# Beyond Leaderboards: A Method for Application-Specific Capability Profiling of Large Language Models

Oleg Grynets
*EPAM Systems*
McLean, Virginia, USA
oleg_grynets@epam.com

Vasyl Lyashkevych
*EPAM Systems*
Lviv, Ukraine
vasyl_lyashkevych@epam.com

Andrii Salyk
*EPAM Systems*
Lviv, Ukraine
andrii_salyk@epam.com

Dmytro Koziuberda
*EPAM Systems*
Kyiv, Ukraine
dmytro_koziuberda@epam.com

Dmytro Kostetskyi
*EPAM Systems*
Lviv, Ukraine
dmytro_kostetskyi@epam.com

Oleh Kaskun
*EPAM Systems*
Lviv, Ukraine
oleh_kaskun@epam.com

Olena Pochernina
*EPAM Systems*
Lviv, Ukraine
olena_pochernina@epam.com

Alona Seletska
*EPAM Systems*
Kyiv, Ukraine
alona_seletska@epam.com

Oleksii Ilchuk
*EPAM Systems*
Lviv, Ukraine
oleksii_ilchuk@epam.com

Viktoriia Tsymbal
*EPAM Systems*
Lviv, Ukraine
viktoriia_tsymbal@epam.com

Daryna Tukalo
*EPAM Systems*
Kyiv, Ukraine
daryna_tukalo@epam.com

Bohdan Saukh
*EPAM Systems*
Lviv, Ukraine
bohdan_saukh@epam.com

Ivan Fedorchuk
*EPAM Systems*
Vinnytsia, Ukraine
ivan_fedorchuk@epam.com

Hryhorii Petrenko
*EPAM Systems*
Kharkiv, Ukraine
hryhorii_petrenko@epam.com

***Abstract*—Public leaderboards rank large language models (LLMs) in general, but an engineer has to choose a model for a specific application, architecture and cost. This paper presents a method for application-specific capability profiling of LLMs and its realization as a containerized testbed. The method treats each application as a unit of extension: an Oracle-to-PostgreSQL migration agent, a documentation-grounded chatbot or an enterprise-support automation agent runs in its own container behind a small contract, brings its own dataset, evaluator and owned attributes, and is onboarded together with its dashboards. Measurements accumulate in a versioned, sparse metamodel of every model. The method separates the model under test from judges and agents, freezes and hashes plans, refuses to compare values from different contexts, diagnoses the instrument before the system, and applies a null-agent check and a noise rule whose operating characteristics are quantified by simulation. In the evaluated automation conditions, model choices produced larger differences on selected observable outcomes than framework choices, although the experimental design does not establish a general causal ranking of these factors. The experiments also show that instruction tuning is deployment-specific and that system-reported success can exceed independently observed success: of 216 migration units checked against a real Oracle database, 108 were demonstrably correct. These measurements establish application-specific observations, not a universal model ranking.**

## I. INTRODUCTION

Choosing a large language model (LLM) for an enterprise application is still, in most teams, an exercise in reading leaderboards and trusting vendor claims. Public benchmarks such as MMLU [1], BIG-bench [2], HELM [3] and Chatbot Arena [4] answer the question "how good is this model in general?", but an engineer who has to migrate a legacy database, operate a documentation assistant or automate a support queue asks a different question: "how good is this model *for this task, in this architecture, at this cost*?" The two questions diverge more than is commonly assumed. In our own experiments a configuration change that improved intent classification by 4.6 percentage points on one deployment lowered it by 3.4 points on a second and had no effect on a third, so that a tuning result measured on a weak deployment described the deployment rather than the prompt. Similarly, in an automation agent that reasons over Jira, Confluence and Outlook events, the choice of the model changed the amount of work actually performed (mean state-changing effects per run) between 0.5 and 1.8, while the choice of the agent framework moved the same quantity only between 0.4 and 0.7.

This paper presents a *method* for application-specific capability profiling of LLMs and the containerised testbed that realises it. The method is designed to make such differences visible, measurable and comparable. It profiles LLMs by running them through *reference applications*: small but complete prototypes of the application classes that matter to the authors' organisation, namely an Oracle-to-PostgreSQL migration agent, a documentation-grounded chatbot and an enterprise-support automation agent. Every application is packaged as a separate container with an evaluation interface. The testbed executes models against datasets and application configurations, records versioned measurements, and exposes them through an analytical interface and through Grafana dashboards. The accumulated measurements form a sparse, versioned *metamodel* of each LLM: a profile of attributes, each measured in the same way for every model, which can be queried to decide which model should do which task at what cost.

The method is organised in eight steps, from defining an application and onboarding it to deciding on a model and monitoring the profile (Section IV). Its central design decision is that an application is a unit of extension of the testbed, not just a workload. When a new application is added, it brings its own dataset, its own evaluator and its own attribute declarations, and onboarding registers the application in the probe registry, in the metric catalogue and in the dashboards. As a consequence the set of LLM features that the testbed can talk about grows with the set of applications it hosts: a chatbot adds retrieval and citation attributes, an automation agent adds tool-use and side-effect attributes, a migration agent adds behavioural-equivalence attributes. Section V describes the mechanism; Section VIII reports what the first three applications have revealed, together with a simulation of the decision rule and a statistical re-analysis of the reported counts.

### A. Problem Statement and Research Questions

We study the following problem. Given a growing set of LLMs and a growing set of application classes, build an experimental infrastructure that (i) measures every model on every application under controlled and fully recorded conditions, (ii) keeps the measuring instrument separate from the object measured, (iii) refuses to report numbers that were produced under different conditions as if they were comparable, and (iv) makes the result usable for model selection. The work is organised around four research questions.

- **RQ1.** What architecture allows heterogeneous reference applications to be added to an LLM-profiling testbed without modifying its core, while preserving reproducibility and cost transparency?
- **RQ2.** Which measurement and diagnostic principles are necessary so that the numbers produced by such a testbed can be trusted, given that agents, judges and applications are themselves LLM-driven, and how should a failure be diagnosed: in the instrument first, or in the system?
- **RQ3.** What do the first three reference applications reveal about the dependence of model behaviour on task, configuration and framework?
- **RQ4.** How can the accumulated measurements be turned into a decision aid for model selection, and which claims must it refuse to make?

### B. Contributions

The paper makes six contributions. First, it states a method for application-specific capability profiling in eight steps (Section IV).

Second, it formalises an LLM metamodel as a sparse partial function from attributes to measurements with explicit context and plan identity, and defines the comparability relation, the evidence levels and a capability taxonomy used throughout (Section III).

Third, it presents the architecture of the testbed with its application-container contract, onboarding algorithm, deterministic plan gate, three-way cost separation and dashboard registration, and it provides a reference implementation of the computational core (Section V).

Fourth, it describes three reference applications and their evaluation stands in sufficient detail to be reproduced (Section VI).

Fifth, it states the measurement and diagnostic principles distilled from defects that the authors found in their own stands, including the instrument-first diagnostic ladder, the degenerate-agent check and a three-state noise rule whose operating characteristics are quantified by Monte Carlo simulation (Section VII).

Sixth, it presents the measured results, their statistical re-analysis, and the resulting evidence and capability maps (Sections VIII and IX).

The methodological contribution is the operational coupling of context-aware measurement classes, evidence-constrained verdicts and instrument-first diagnosis with extensible application probes. The measured evidence is preliminary and does not establish a universal ranking of models or a causal ordering of model and framework effects.

The paper distinguishes measurement validity, repeatability, and generalisability. Some comparisons involve small or single-run samples, and some outcome measures lack complete behavioural ground truth; these boundaries are identified in the experiments and Section IX-J. The study produces conditional capability profiles rather than a context-free ranking of evaluated models.

The remainder of the paper is organised as follows. Section II reviews related work. Section III formalises the problem. Section IV states the method. Section V describes the testbed architecture. Section VI presents the reference applications. Section VII states the evaluation and diagnostic methodology. Section VIII reports experiments and Section IX discusses them. Section X concludes.

## II. Related Work

We reviewed 88 scholarly sources, including peer-reviewed publications and preprints, and organised the literature into six areas: general-purpose benchmarking, application-level evaluation, LLM-as-a-judge, benchmark validity and statistical analysis, cost-aware model selection, and engineering practices for observable machine learning systems. Table I summarises how the testbed relates to each line.

### A. General-Purpose Benchmarks and Holistic Evaluation

Large multi-task suites measure broad knowledge and reasoning. MMLU [1] and AGIEval [5] use examination-style questions, BIG-bench [2] collects hundreds of community-contributed tasks, and HELM [3] argues for holistic reporting over many scenarios and seven metric families, including calibration, robustness and efficiency, rather than accuracy alone. Human-preference platforms such as Chatbot Arena [4] and judge-based protocols such as MT-Bench [6] complement static suites with open-ended comparisons. Surveys [7] show that the evaluation literature is dominated by such general-purpose protocols. Behavioural testing in the CheckList style [8] and dynamic benchmark construction [9] address the fragility of static test sets. These works define *what* to measure in general; the testbed adopts their emphasis on multiple metrics and on explicit scenarios, but replaces the "scenario" with a complete reference application so that the measured quantity is the behaviour of a model inside an architecture.

Two lessons from this line of work shape the testbed more than any specific benchmark. The first is that a single aggregate score hides the structure of model behaviour: HELM reports accuracy next to calibration, robustness, fairness, bias, toxicity and efficiency for the same scenarios precisely so that trade-offs become visible [3], and BIG-bench documents tasks on which scale alone does not produce a smooth improvement [2]. The second is that preference-based and judge-based protocols inherit properties of their instruments. Pairwise voting at scale is informative about open-ended chat quality [4], yet it cannot say whether a model completes a database migration correctly, and a benchmark that ranks models on examination questions says little about their behaviour inside a retrieval pipeline. For these reasons the testbed does not attempt to compete with general leaderboards; it assumes that a team has already shortlisted models from them and asks what happens next.

### B. Application-Level Evaluation: Retrieval, Agents, Code, SQL

Retrieval-augmented generation [10] is evaluated with component-level retrieval metrics on benchmarks such as BEIR [11] and with generation-level criteria such as faithfulness and answer relevance in automated frameworks [12], [13]. RGB [14] diagnoses noise robustness, negative rejection and information integration, and positional effects in long contexts are documented in [15]. Dense retrieval [16] and rank fusion [17] are the building blocks of the retrieval configurations that our chatbot stand compares.

Agent evaluation has moved from single-turn function calling [18], [19] to interactive environments: AgentBench [20], WebArena [21], GAIA [22] and SWE-bench [23] score the end state or the task outcome in realistic environments, and surveys catalogue architectures and evaluation practices [24], [25]. Reasoning-and-acting loops [26], self-reflection [27] and experience-based learning [28] are the mechanisms around which many agents are built, while multi-agent frameworks [29], [30] supply orchestration patterns that we treat as an experimental variable. Our automation stand follows the end-state logic of these benchmarks but adds a distinction that they rarely make explicit: whether a claim of success in the final answer is backed by an observable effect.

For code and data tasks, Spider [31] and BIRD [32] evaluate text-to-SQL by execution, TransCoder [33] and CodeTransOcean [34] address program translation, and EvalPlus [35] shows that weak test suites overstate functional correctness. A study of bugs introduced during LLM-based code translation [36] and two broad surveys of LLMs in software engineering [37], [38] frame the migration problem. Classical text-similarity metrics such as BLEU [39] and chrF [40] are still used as reference-based signals, but our migration results, like [35], indicate that a reference-based or acceptance-based verdict can overstate correctness; execution against the source system is the stronger instrument.

What these application-level works share is that they evaluate a *task* in a fixed *environment*. For retrieval, the environment is the corpus and the index; for agents, a web or code environment; for SQL, a database. The application-level testbed generalises the idea in two directions. It treats the environment as one of the things that can be versioned and registered (the dataset version and the configuration are part of every measurement), and it makes the claim–effect distinction explicit: AgentBench and WebArena score the end state, but an agent that reports success without an observable effect is not separately counted in them, whereas in our automation stand the share of such runs is itself an attribute. The same reasoning applies to code translation: the studies of translation bugs [36] show that models introduce subtle semantic differences, and our finding that a clean install proves little is the database-specific form of the EvalPlus lesson that weak tests overstate correctness [35].

### C. LLM-as-a-Judge

Judge models scale open-ended evaluation [6], [41], [42], yet they exhibit position and verbosity effects [43] and favour their own generations [44]. Our stands therefore use judges only for criteria that cannot be computed deterministically, keep them in a separate role from the model under test, and measure their instability directly: in the migration direction a semantic judge changed its verdict on 16.2% of repeated items at temperature 0. Leakage-resistant evaluation of judges on specifications [45] and execution-judged benchmarks [46] are earlier results from the same group that inform the design of the judge roles.

Judge-based evaluation has a further cost dimension that is rarely reported: the judge's own spend. In the P1 stand, 1,802 combined judge calls cost more than twice as much as the conversion of the same files, which is why the testbed accounts for it separately. Learned judges such as Prometheus [42] and structured prompting such as G-Eval [41] improve agreement with human raters but do not remove the dependence on the instrument, and the evidence of self-preference [44] suggests that the judge should come from a different family than the model under test whenever the comparison is between vendors. In our chatbot stand both the judge and the controller currently use the same model family as one of the compared models, a limitation we record in Section IX-J.

### D. Benchmark Validity, Contamination and Statistics

The testbed reduces contamination exposure through versioned, application-specific test data and isolation of expected outcomes. Synthetic or private datasets alone do not guarantee freedom from data contamination; that remains an evaluation-specific threat to validity.

The statistical literature points in one direction: with few items and few repeats, effects smaller than the noise cannot be established. Dror et al. recommend matching the test to the data and reporting the variability of the evaluation itself [53], and Card et al. show that typical NLP comparisons have too little power to detect the differences they report [54]. Contamination studies add that the same model may have seen the test items, so that a high score can be memory rather than capability [49], [50]. The testbed responds in two ways. Versioned private and synthetic application datasets, together with the separation of ground-truth labels from model inputs, reduce contamination exposure. However, these safeguards do not establish that evaluation data are entirely absent from model training or development histories. For the variability, it declines to publish model numbers until repeats have established the spread, and it applies the coarse noise-floor rule of Section VII-F where a formal test is not applicable.

### E. Model Selection, Routing and Multi-Objective Trade-offs

Cost-aware use of LLMs has been studied as cascades [59] and as quality-aware routing between a cheaper and a stronger model [60]. Selecting among models with several conflicting objectives is a multi-objective problem whose standard solution concept is the Pareto set [61], [62]. The analytical interface of the testbed computes Pareto sets over measured attributes on the server and displays the evidence count behind each claim; our migration agent uses a deterministic complexity score for routing in the spirit of [60], and our token-optimisation study [63] shows that cost and quality must be evaluated jointly.

### F. Engineering Practice for Observable ML Systems

Machine-learning systems accumulate hidden technical debt [64], and rubrics for production readiness explicitly include tests of data, model and infrastructure [65]. A case study of software engineering for ML at scale [66] documents the importance of data management and model evaluation tooling. Programmable guardrails [67] and prompt-programming frameworks [68] address safety and prompt optimisation, respectively. Dashboards for monitoring evolving information systems, including ontology-driven platforms with dashboard-centric analytics, have been investigated by co-author V. Lyashkevych [69]–[73]. The risk perspective on LLM-based development [74], the notion of a software functional state [75] and the orchestration of multi-agent systems with dynamic involvement of new agents [76] inform, respectively, the security attributes, the lifecycle view and the extensibility of the testbed. The metamodel idea itself builds on a unified architecture metamodel for information systems generated by AI [77]. Finally, a comparative analysis of frameworks for the same three application classes [78] was the starting point for the reference applications used here.

### G. Prompt Programming and Reasoning Strategies

The behaviour of a model inside an application depends on how it is prompted. Few-shot prompting [79], chain-of-thought prompting [80] and interleaved reasoning and acting [26] changed what models can do without changing their weights, and declarative frameworks such as DSPy [68] compile pipelines of model calls and optimise their prompts against a metric. The testbed treats the prompt variant as a registered, hashed object, requires a baseline arm next to any compared variant and, in the programme scheme, optimises prompts per probe so that a model is measured under the prompt that suits it and the gain over the baseline prompt is itself recorded. Our measurement that a description rewrite had effects of opposite signs on three deployments is consistent with the prompt-sensitivity literature, and it adds a practical corollary: a prompt optimised against one deployment is a property of the pair, not of the prompt, so the optimisation target must be part of the identity that is stored with the result. DSPy-style optimisation against a metric also inherits the risk that the metric rewards the wrong behaviour; the null-agent check of Section VII-B is the guard that we apply before any loop is pointed at a number.

### H. Specification-Driven Development and Code Regeneration

Specification-driven development treats a specification as the operational artefact consumed by LLM development agents. The authors' earlier studies formalise a specification as a structure of semantic blocks with dependency relations, evaluate it with execution-judged benchmarks that keep the implementer panel fixed and include a no-specification control arm [46], evaluate the evaluators of specifications under sealed ground truth [45], and show on database migration that cost, validity and semantic preservation must be traded off [63]. The generation hook of the testbed reuses this machinery at a smaller scale: the application specification is the artefact from which an adapter is generated, and the smoke test and the control-arm discipline are the checks that the generated adapter must survive. The unified architecture metamodel [77] offers a representation (JSON, PlantUML and Mermaid views of architectural diagrams) that can serve as the structured context for such generation.

### I. Dashboards and Monitoring of Evolving Systems

Operational dashboards for ML systems are often built last, as a view over logs. The earlier work of co-author V. Lyashkevych on intelligent monitoring of evolving information systems treats the dashboard as a first-class element of the architecture, with ontology-driven domain knowledge integrated into dashboard-centric analytics [69], context-aware selection of decision strategies [70], evolution-aware synthesis of monitoring systems by LLMs [71], genetic optimisation of monitoring models for LLM-modified systems [73] and multi-drift predictive monitoring [72]. We adopt this stance: the dashboards of the testbed are registered with the application at onboarding, not added afterwards. A natural next step, discussed in Section IX-D, is to monitor the metamodel itself for drift.

TABLE I. Positioning of the Testbed against Related Lines of Work

| Line of work | Typical unit | What the testbed adds |
|---|---|---|
| General benchmarks [1]–[3] | task item | complete application as scenario; per-application attributes |
| Preference platforms [4], [6] | pairwise vote | execution-based and deterministic metrics |
| RAG evaluation [12]–[14] | answer / retrieval | model as variable under fixed retrieval; failure localisation |
| Agent benchmarks [20], [21], [23] | task outcome | claims versus observed effects; framework as variable |
| Code and SQL benchmarks [31], [32], [34] | snippet / query | execution against source system; fix ledger |
| Validity and statistics [47], [53], [54] | guidelines | enforced in code: comparability, noise floor, fail-closed |
| Routing and cascades [59], [60] | query | measured profile of each model as routing evidence |

Entries paraphrase the cited works; the right-hand column states the contribution of the present testbed.

### J. Intelligent Diagnostics, Knowledge Bases and Testing

The diagnostic stance of this paper has an older root in the co-authors' earlier research on intelligent diagnostics of computer facilities, which treated diagnosing as a knowledge-based task with its own infrastructure: an ontology of the diagnostic subject domain [81], a software tool for building the knowledge base of the diagnosing process [82], retrieval of diagnostic information from microprocessor devices [83], the search for features of diagnostic information and the development of a search service for it [84], [85], criteria for setting correspondence between the terms of different knowledge bases [86], and logic-textual and neural approaches to information search [87]. Four ideas from this line carry over to the profiling of LLMs. First, diagnosis needs an explicit subject-domain vocabulary; in the testbed it is the attribute catalogue, in which every attribute has an owner, a unit and a direction. Second, diagnostic knowledge should be kept as a knowledge base separate from the diagnosed system; the fix ledger of the migration direction and the failure taxonomies of the other directions play this role (Section VII-K). Third, terms must be matched across knowledge bases with explicit criteria; the same quantity appears as recall at K in one probe and as a recall metric in another, and the catalogue must decide when two names denote one attribute. Fourth, retrieval of similar earlier failures is part of diagnosis, which is the practical reason for storing per-unit outcomes and not only aggregates.

Testing under uncertainty has a second root in the quantitative treatment of software risks, where Monte Carlo simulation is used to evaluate the values of risk attributes [88] and where the functional state of software through its life cycle has been formalised [75]. We use the same technique in Section VIII-I to study how often the decision rule of the testbed declares an effect when there is none and how large a real difference must be to be detected. A third relevant observation comes from literature on agent orchestration reviewed in the programme: a subtask graph fixed at design time can raise retry cost, because a failure upstream forces the re-execution of everything downstream, whereas executable control flow that isolates the failing subtask retries only that subtask. The group-level repair loop of the migration agent and the triage of the probe agent follow the same principle of isolating the failure before retrying.

Table II condenses the lessons drawn from the reviewed literature into design decisions. Each row names the decision, the works that motivate it and the mechanism in which it is realised, so that the paper's architecture can be read as an application of the literature rather than a collection of ad hoc choices.

TABLE II. DESIGN LESSONS DRAWN FROM THE REVIEWED LITERATURE

| Lesson | Motivating works | Realisation in the testbed |
|---|---|---|
| report several metric families, not one score | [2], [3], [7] | per-application attribute sets; operational set common to all probes |
| behavioural tests expose what accuracy hides | [8], [9] | behavioural differential against real Oracle; scenario-based chatbot checks |
| execution is stronger evidence than text similarity | [31], [32], [35], [36] | verification levels; demonstrably correct units; effects of tool calls |
| retrieval and generation must be evaluated separately | [10]–[13] | retrieval-only runner; six metric groups; causal gating |
| noise robustness and negative rejection matter | [14], [15] | citation, refusal and empty-retrieval attributes |
| agents are scored on end state, not on prose | [20]–[23] | frozen world state; expected tool sequences and final states |
| tool use is a distinct capability | [18], [19], [26] | tool selection, argument validity, hallucinated-call rate |
| judges are biased and must be separated | [6], [43], [44] | separate roles; judge spend counted apart; judge churn measured |
| benchmarks can fail as instruments | [47], [48] | null-agent check; evidence levels; fix ledger |
| contamination inflates scores | [49]–[52] | private or synthetic datasets; sealed ground truth |
| significance claims are fragile | [53], [54] | noise-floor rule; repeats before publication |
| reproducibility needs identity | [55], [56] | plan hash; prompt identity; frozen metric libraries |
| document the context of every result | [57], [58] | dataset version, configuration, warnings stored with each run |
| cost and quality are traded off | [59]–[62] | Pareto view with evidence counts; cost separation |
| ML systems accumulate hidden debt | [64]–[66] | tests without network; pinned metric ruler; orphan recovery |
| agent frameworks supply orchestration patterns | [24], [29], [30] | framework as experimental variable behind one contract |
| experience can improve agents, but must be validated | [27], [28], [68] | self-learning as hypotheses; approval and immutable snapshots |
| guardrails are programmable but measurable | [67] | guardrail precision, recall and false-block rate measured offline |

In summary, existing work provides the measuring instruments and the warnings. What is missing, and what this paper offers, is an extensible infrastructure in which every application brings its own instruments, all measurements share one identity and comparability discipline, and the result is a profile of each model rather than a rank in a single list.

## III. PROBLEM FORMULATION AND REQUIREMENTS

This section fixes the terminology and the formal objects used in the rest of the paper. The notation is summarised in Table III.

### A. Formal Model

Let $M$ be a set of models and $A$ a set of reference applications. An application $a \in A$ is a tuple

$$a = (D_a, K_a, E_a, \Sigma_a), \tag{1}$$

where $D_a$ is a family of versioned datasets, $K_a$ is a configuration space (retrieval settings, framework, prompt variant, model roles, evaluation method), $E_a$ is the evaluator that produces measurements, and $\Sigma_a$ is the set of attributes that the application declares and owns. A *probe* $p_a$ exposes the application to the testbed: it names the modes in which the application can be run, the prompt slots that can be varied and the attributes it measures.

A *run* fixes a model $m$, an application $a$, a dataset version $d$, a configuration $\kappa \in K_a$ and a plan identity $h$, the hash of the frozen plan that produced the run. A *measurement* is the tuple

For a declared comparison, partition the recorded configuration into the evaluated treatment (for example, the tested model or prompt variant) and an evaluation context κE containing the evaluator and metric-library versions, scoring protocol, fixed decoding and application settings, and any other factors required to be held constant. A plan hash h identifies the canonical execution plan; it does not itself establish equality of evaluation contexts.

$$\mu = (m, \sigma, v, n, \kappa, d, h), \sigma \in \Sigma_a, \tag{2}$$

with value $v$ computed from $n$ samples. The *metamodel* of model $m$ is the partial function

$$\Psi(m)\colon \sigma \mapsto \{(v_i, n_i, \kappa_i, d_i, h_i)\},$$

$$\text{defined only on measured } \sigma. \quad (3)$$

Partiality is essential. An attribute that has not been measured is *absent*; it is never replaced by zero or by an average, and each view of the metamodel lists the attributes that are missing. Two values of the same attribute are *comparable* if they were produced in the same context:

$$v_1 \sim v_2 \Leftrightarrow \sigma_1 = \sigma_2 \wedge \kappa E(\kappa_1) = \kappa E(\kappa_2) \wedge d_1 = d_2. \quad (4)$$

The analytical interface applies (4) before it draws any comparison and displays a warning when a selection mixes contexts (Section V-H). Model selection is expressed over a set of objectives $O = \{(o_1, \delta_1), \ldots, (o_k, \delta_k)\}$ with directions $\delta \in \{\max, \min\}$. Model $m_1$ dominates $m_2$ on $O$ if it is at least as good on every objective and strictly better on one, and the recommended set is the non-dominated set:

The treatment dimension is declared in advance: models or prompt variants may differ between arms, but the evaluation context, dataset, scoring protocol and other fixed factors must match. Equality is tested on registered identifiers and evaluator versions; a changed metric library starts a new comparison class.

$$P(O) = \{\, m \in M' : \neg\exists\, m' \in M', m' >_O m \,\}, \quad (5)$$

where $M'$ is the set of models that satisfy the hard constraints of the task (for instance a minimum accuracy threshold) and for which every objective is measured in a comparable context.

TABLE III. NOTATION

| Symbol | Meaning |
|---|---|
| $M$, $A$ | set of models, set of reference applications |
| $a = (D, K, E, \Sigma)$ | application: datasets, configurations, evaluator, owned attributes |
| $p_a$ | probe that exposes application $a$ to the testbed |
| $\sigma \in \Sigma_a$ | attribute owned by application $a$ |
| $\kappa$ | run context: mode and configuration identifier |
| $h$ | hash of the frozen plan (identity of a measurement) |
| $\mu$ | measurement: model, attribute, value, sample count, context, dataset, plan |
| $\Psi(m)$ | metamodel of model $m$ (partial function over attributes) |
| $\sim$ | comparability relation |
| $P(O)$ | Pareto set for objectives $O$ |
| $C_t$, $C_j$, $C_g$ | spend of tested model, judge and agent model |

## B. Formal Properties

Three simple properties follow from the definitions and justify the mechanisms of Section V. They are stated with short proofs because the mechanisms rely on them.

**Lemma 1 (comparability). On measurements of one attribute, the relation (4) is an equivalence relation. Proof.** It is defined by equality of σ, the evaluation-context projection κE(κ), and d; equality is reflexive, symmetric and transitive, and a conjunction of equivalence relations is an equivalence relation. Consequently measurements fall into context classes, and any comparison, ranking or Pareto set is well defined within one class and meaningless across classes. **Lemma 2** (plan identity). Let the plan hash be computed over the list of runs after sorting by a total order on runs. Then two plans that contain the same runs in any order have the same hash, and a rerun that replays the stored plan has the hash of the original. Proof. Sorting maps every permutation of a multiset of runs to the same list, and a hash of the same list is the same value. Replay reads the stored row, which is the sorted list. **Lemma 3** (non-emptiness). If the set M′ of eligible models is finite and non-empty, the Pareto set P(O) of Equation (5) is non-empty. Proof. Dominance is irreflexive and transitive, hence a strict partial order; a finite non-empty set has a maximal element under a strict partial order, and a maximal element is not dominated.

**Lemma 1** applies within a declared comparison design whose treatment and evaluation-context coordinates have been registered in advance. **Lemma 3** concerns exact Pareto dominance only. Uncertainty bands are reported as supplementary pairwise annotations and do not replace the transitive dominance relation used to construct P(O); approximate ties need not be transitive.

## C. Evidence Levels and Demonstrable Correctness

The formal model above says what a measurement is; it does not say how much a measurement proves. Experience with the migration direction showed that the same word, "accepted", can denote evidence of very different strength, so we attach an *evidence level* λ to every verdict. The levels are ordered by what the instrument could in principle detect (Table IV): a statement by the system itself, a static check of the output, installation or execution in a target environment, comparison with a human reference, and comparison with the behaviour of the real source system. A claim made at level λ is only as strong as the weakest level at which every unit that contributes to it was checked, and a report must state the denominator at each level.

TABLE IV. EVIDENCE LEVELS

| Level | Instrument | Detects | Misses |
|---|---|---|---|
| L0 | system's own report | nothing independent | everything |
| L1 | static check of the output | syntax, lint, structure | unresolved references; semantics |
| L2 | install or execute in the target | load and runtime errors | wrong but valid behaviour |
| L3 | reference comparison (text, tree, judge) | divergence from a reference | behaviour equal to a different valid reference |
| L4 | behaviour of the real source system | behavioural differences | inputs not exercised |

L0–L4 generalise the verification levels of the migration agent (CHECKED_STATIC, VERIFIED_INSTALL, VERIFIED_BEHAVIOR) to the other applications.

Let UL4 ⊆ U contain only units for which the prescribed Oracle differential checks were actually executed. For u ∈ UL4, let acc(u) indicate the agent's own acceptance verdict, con(u) a contradiction detected on the exercised L4 cases, and exp(u) a failure that reproduces the source-system error under the recorded test.

Operational limitation: in the underlying migration stand, "demonstrably correct" denotes the following L4-screened operational count, not a proof of semantic equivalence over all possible inputs. A candidate can remain uncontradicted on the exercised Oracle tests while still diverging on untested cases. The count and the tested denominator must therefore accompany every claim.

$$N_{dc} = |\{u \in U_{L4} : [\mathrm{acc}(u) \wedge \neg\mathrm{con}(u)] \vee [\neg\mathrm{acc}(u) \wedge \mathrm{exp}(u)]\}|. \quad (6)$$

so that a unit counts both when it is accepted and not contradicted, and when it fails in exactly the way the source system fails. The definition applies only to UL4 and reports agreement on the executed checks, not exhaustive equivalence; for units that the stand cannot test the verdict is "not testable", which is a separate category rather than a failure. The equation also shows why the agent's own count of accepted units can rise while Ndc rises less: some accepted units are contradicted, and some failures are correct.

### *D. A Capability Taxonomy for Mapping*

The word *capability* is used loosely in the LLM literature; for a testbed it must be operational. We define a capability as a named behaviour of a model that is observable in at least one reference application, is measured by at least one declared attribute at a stated evidence level, and can differ between models. The capability map of a model is then the set of its capabilities with the measured values and evidence levels, which is exactly the metamodel $\Psi(m)$ read through the taxonomy. Table V lists the ten capabilities that the first three applications make measurable, with the attributes through which they are observed and the largest variation that the reports show. The taxonomy is not exhaustive and is not a claim about how model capabilities are organised internally; it is a bookkeeping device that lets us state which questions the testbed can currently answer.

TABLE V. CAPABILITIES MADE MEASURABLE BY THE REFERENCE APPLICATIONS

| Capability | Observed through | Application | Level | Largest variation reported |
|---|---|---|---|---|
| C1 behavioural fidelity of translation | demonstrably correct units; defects; faithful raises | migration | L4 | 108 of 216 units; 25 of 114 accepted wrong or unverifiable |
| C2 whole-file output completeness | refusals; file efficiency; abbreviation of large files | P1 | L3 | about 69 large files abbreviated by one model; refusals 8.32% |
| C3 grounded answering | faithfulness; citation validity and compliance | chatbot | L1–L3 | correctness 0.977–0.995 across finalists |
| C4 retrieval use | recall, precision, MRR under fixed retrieval | chatbot | L1 | sparse recall 0.083 against 1.000 for dense (retrieval side) |
| C5 tool routing | tool selection; forbidden calls; router model effect | chatbot, automation | L1–L2 | 4 routing failures of 10 scenarios; 2 fixed by a router change |
| C6 tool-argument validity | argument correctness; hallucinated tool calls | chatbot, automation | L1 | two scenarios still failing on arguments after routing fix |
| C7 action honesty | unbacked success claims; effect rate | automation | L2 | 0% to 55% of effect-free runs claim success |
| C8 instruction robustness | effect of an instruction rewrite | intent classification | L1 | +4.6, −3.4, 0.0 pp on three deployments |
| C9 cost efficiency | tokens per run; cost per evidence | all | L1 | 1,953 to 9,465 tokens per automation run |
| C10 run-to-run stability | variance over repeats; judge churn | chatbot, P1, judges | L1 | SD 0.013 to 0.037; judge churn 16.2% |

Evidence levels as in Table IV; values from the experiments of Section VIII. C4 is a property of the retrieval configuration and is listed because the model-side effect is held fixed in those experiments.

Two points about the taxonomy matter for the rest of the paper. First, several capabilities are only visible in combination: a model's grounded answering (C3) cannot be separated from the retrieval it is given (C4), which is why the chatbot stand uses a retrieval-only runner and holds the retrieval fixed when the model is varied. Second, the evidence level is part of the capability: C1 is measured at L4, but C3 is largely measured at L1 to L3 because no hard oracle exists for the quality of a free-text answer, in the sense of the generalisation caveat of the score-model decision record. Where a capability cannot be verified, we report it as such and do not substitute a judge and call it measured.

### *E. Requirements*

Starting from the questions in Section I and from the defects described in Section VII, we derived ten requirements. Table VI lists them with the mechanism that satisfies each one and the section in which it is described.

### *F. Scope and Assumptions*

The testbed measures models, not applications: the application is held fixed and the model is varied, or the application configuration is varied while the model is held fixed. The comparison of application architectures themselves is the subject of a separate study [78]. All model calls go through a single gateway, so that the models are reached in a uniform manner and cost can be attributed to the caller.

The programme that this work belongs to plans two further stages, requirement analysis with specification generation tailored to a chosen model, and specification decomposition with decision making about which model handles which part. They depend on stable profiles and are outside the scope of this paper.

TABLE VI. REQUIREMENTS AND THE MECHANISMS THAT SATISFY THEM

| ID | Requirement | Mechanism | Sec. |
|---|---|---|---|
| R1 | Add an application without changing the core | probe contract; container adapter | IV |
| R2 | Every attribute has an owner and a unit | metric catalogue owned by probes | IV |
| R3 | Reproduce a measurement exactly | frozen, hashed plans; prompt identity | IV |
| R4 | Tested model, judge and agents costed apart | three-way spend accounting | IV |
| R5 | No metric is produced by a model | agents propose, code computes | IV, VI |
| R6 | Missing is not zero | partial metamodel; missing list | III |
| R7 | Never compare across contexts | comparability check in the UI | IV |
| R8 | Outage is not a verdict | INCONCLUSIVE / absent verdict | VI |
| R9 | Aggregate must beat the degenerate agent | null-agent check | VI |
| R10 | Observable by non-authors | dashboards, registration at onboarding | IV |

## IV. The Profiling Method

The method answers a practical question: given a shortlist of LLMs, which of them is suited to a particular application, in a particular architecture, at what cost, and how sure can the team be of that answer? It does so by turning each application into an instrument and each model into a profile. The method has eight steps, M1 to M8 (Fig. 1, Table VII). The steps are ordered, but the method is iterative: a new application re-enters at M1, a new model at M3, and a diagnosis at M5 that blames the instrument sends the work back to M2 or M4.

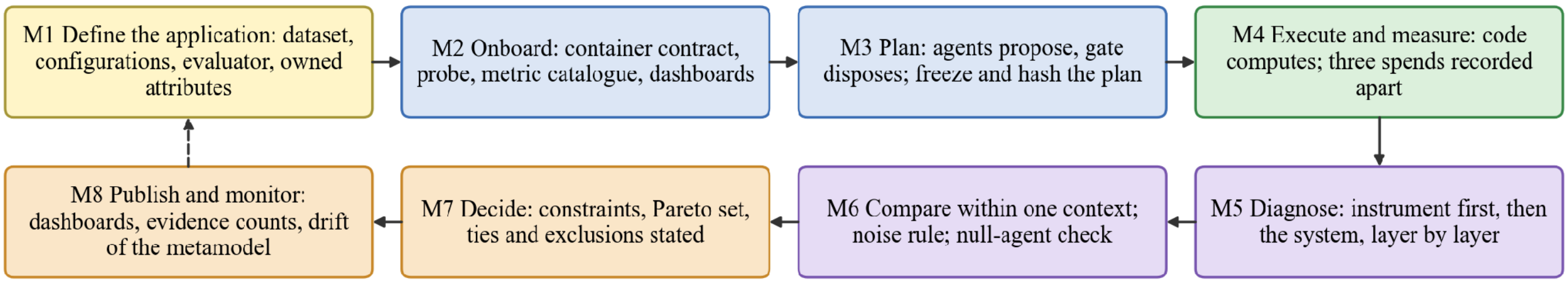

Fig. 1. The eight steps of the profiling method. Solid arrows: the normal order; the dashed arrow: a new application or a new model re-enters the method.

TABLE VII. Steps of the Method, Their Outputs and the Guards Against Known Failures

| Step | Activity | Output | Guard | Section |
|---|---|---|---|---|
| M1 Define | fix the dataset, the configuration space, the evaluator and the owned attributes with unit and direction | application specification | every attribute has an owner, a unit and a direction | VI |
| M2 Add probe | package as a container, expose the contract, pass the smoke test, register probe, catalogue entries and dashboards | probe identifier; dashboards | a generated adapter must pass the smoke test | V-E |
| M3 Plan | agents propose runs; the gate validates, canonicalises, hashes and freezes the plan | frozen plan with hash | scope re-applied by code; baseline arm required | V-D |
| M4 Execute | run under the caller's key; code computes the metrics; tested model, judge and agents are costed apart | measurements with provenance | prompt identity; fail-closed reporting | V-G |
| M5 Inspect | test the instrument before the system, then locate the first failing layer | diagnosis; verdict from a fixed taxonomy | an outage is not a verdict; the null agent must lose | VII-J |
| M6 Compare | compare only values of one context class; apply the noise rule and evidence levels | effect, suggestive, or no result | comparability; per-axis noise floor | VII-F |
| M7 Decide | apply hard constraints, then the Pareto set with ties and exclusions | recommendation with evidence counts | missing is not zero; context warning | VIII-M |
| M8 Publish | show the profile on dashboards; monitor the profile for drift | views; alerts | dashboard variables fix the context class | V-H |

Sections give the place where the mechanism behind each step is described.

Three properties distinguish the method from a benchmark run. *The unit of evaluation is the application, not the item.* A leaderboard item is a question; a unit here is a complete behaviour in an architecture, such as the migration of a group of Oracle objects, a chatbot turn with its retrieval and tool calls, or the handling of an enterprise event. *The artefact is a profile, not a rank.* The result of the method is the metamodel of a model: its measured attributes with their contexts, and an explicit list of what was not measured. *The guards are part of the method.* Each step carries a guard that exists because a stand once misled its authors without it (Table VII), and a result produced without its guards is reported with a warning instead of a number. Table VIII contrasts the method with a leaderboard on the dimensions that matter to a team that has to choose.

TABLE VIII. Leaderboard Versus Application-Specific Profiling

| Dimension | Leaderboard | Profiling method |
|---|---|---|
| unit of evaluation | item or task | application in its architecture |
| question answered | how good is the model in general | how good is the model for this application, configuration and cost |
| artefact | rank or score | sparse metamodel with contexts and missing list |
| comparability | same benchmark version | same attribute, context and dataset version, checked by code |
| variables | model | model, prompt variant, framework, retrieval, configuration |
| failure handling | item scored wrong | failure diagnosed in instrument or system; outage not a verdict |
| uncertainty | often single number | per-axis noise floor; evidence counts; 'no result' allowed |
| cost | rarely reported | three spends, reported separately |
| extension | new benchmark | new application brings new attributes and dashboards |

### A. Re-Entry Rules

What must be repeated when something changes is as important as what to do the first time (Table IX). The rules follow from the comparability relation: anything that changes a component of the context class opens a new class, and results from different classes are never merged.

TABLE IX. RE-ENTRY RULES OF THE METHOD

| Change | Re-enter at | Consequence for existing results |
|---|---|---|
| new model | M3 | existing results unchanged; new profile with absent attributes filled by runs |
| new application | M1 | new attributes in the catalogue; existing profiles list them as not measured |
| new dataset version | M3 | new context class; no comparison with older versions without bridging runs |
| new configuration or prompt variant | M3 | new context class; baseline arm required for a prompt comparison |
| metric library upgrade | M2 | new ruler; old and new results not comparable |
| diagnosis blames the instrument | M2 or M4 | affected runs marked inconclusive; fix and rerun |
| drift of a hosted model detected | M3 | re-run a frozen plan; compare as a new context class |

A new model is run on the existing applications and fills the absent attributes of its own profile. A new application adds attributes to the catalogue and, until models are run on it, every profile lists them as not measured. An upgrade of a metric library changes the ruler itself, so the testbed freezes the libraries to the versions that produced the results and treats an upgrade as the opening of a new context class.

## V. TESTBED ARCHITECTURE

This section answers RQ1. We first give the overview and the design principles, then describe the metamodel and the programme scheme from which the testbed originates, the planes of the core, the life cycle of a goal, the application-container contract with its onboarding procedure, the specification-driven generation hook, the accounting of cost and prompts, and finally the dashboards. The core, called MAS Core (metamodel extraction pipeline), was implemented by D. Koziuberda, with the shared agent-integration layer contributed by H. Petrenko; the analytical interface and the three reference applications were built by the programme teams, as indicated in Section VI.

### A. Overview and Design Principles

Figure 2 shows the testbed. Reference applications live in their own containers on the left. MAS Core in the centre receives a *goal* (a statement of what to measure), plans runs with the help of agents, validates and freezes the plan with deterministic code, executes it through probes and writes measurements to a single store. Models under test, judges and the models that drive the agents are all reached through one LLM gateway, which keeps the cost of every role attributable. The store feeds the analytical interface, which is deterministic and contains no LLM in its analytics pages, and Grafana dashboards that are registered when an application is onboarded.

Five principles, listed in Table X, run through the whole design. They were not postulated in advance; each was introduced after a defect in an earlier stand showed what happens without it (Section VII).

TABLE X. DESIGN PRINCIPLES OF THE TESTBED

| Principle | Why it matters | Where it is enforced |
|---|---|---|
| The instrument is not the subject | judge or agent cost and errors must not become the model's | three spends reported separately; distinct model roles |
| Agents decide, code computes | a model must not write its own score | aggregator schema has no numeric field; metrics are deterministic code |
| Reproducibility | numbers are comparable only if produced alike | plans frozen and hashed; prompts identified by hash; one change per experiment |
| Missing is not zero | averaged or zero-filled values hide gaps | partial metamodel; explicit list of unmeasured attributes |
| An outage is not a verdict | infrastructure failure says nothing about the model | INCONCLUSIVE status; absent verdict when an agent is unreachable |

The five design principles in Table X capture the implementation safeguards introduced after faults in the evaluation infrastructure were identified experimentally.

### B. Programme Scheme and Metamodel

The proposed framework supports a three-stage research programme (Figs. 3 and 4). The implementation and experiments in this paper concern Stage 1: constructing a capability metamodel from application-specific probes. The two subsequent stages—task-aware specification generation and model assignment to specification components—are proposed consumers of the profile, not evaluated functions of the present testbed. A task is represented as T = {agents, operations, environments, modules, context, …}, consistent with the architectural metamodel [77].

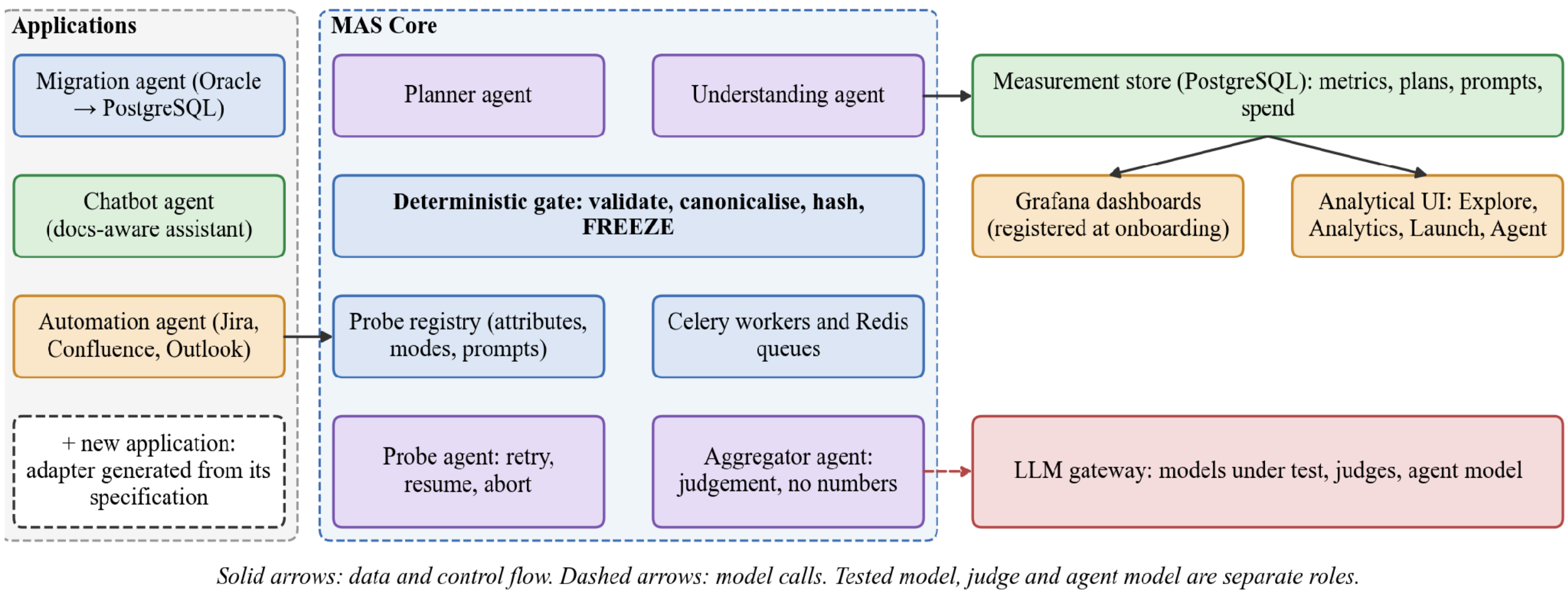

Fig. 2. Overview of the testbed: reference-application containers, MAS Core, measurement store, dashboards and the LLM gateway. Tested model, judge and agent model are separate roles.

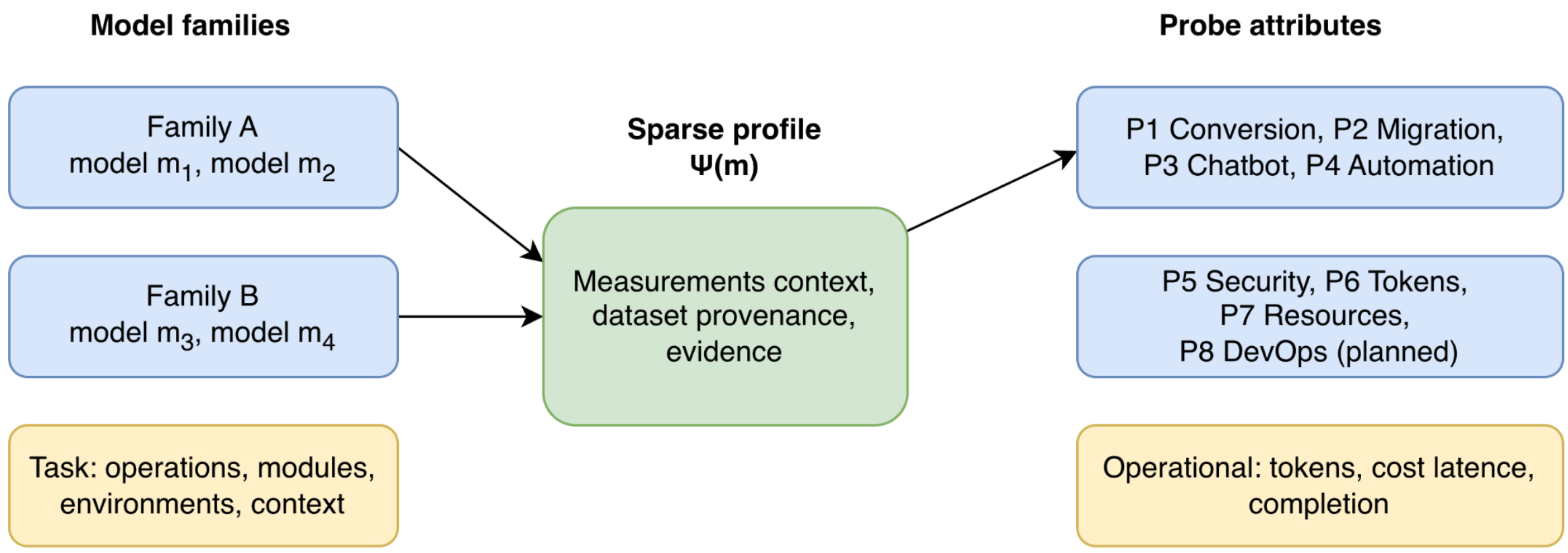

Fig. 3. Model families and their sparse capability profiles, linked to application-owned probe attributes and operational measurements; T denotes structured task context.

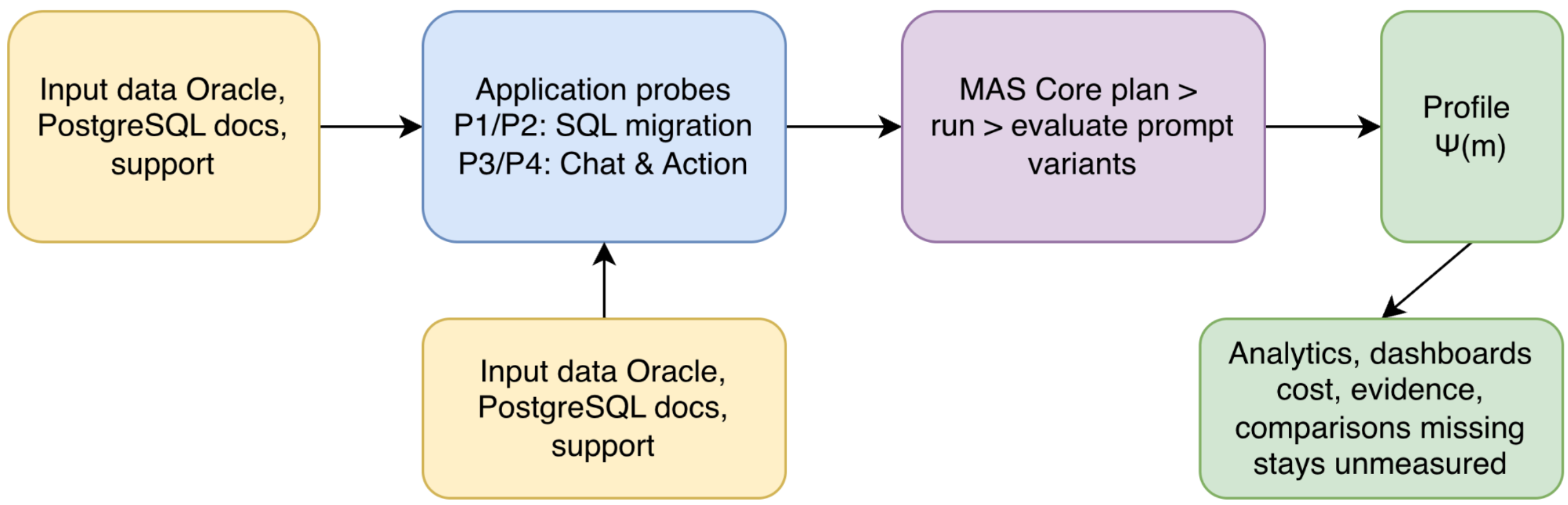

Fig. 4. Application probes and controlled prompt variants produce versioned capability measurements, which are aggregated into model profiles and exposed to analytical queries.

A metamodel is a sparse record per model, grouped by model family, whose attributes are the quantities that probes measure. Table XI lists the attribute groups of the scheme and the probes that own them. The numbering follows the scheme (P1 conversion, P2 migration, P3 chatbot, P4 automation); in the registry of the implementation the probes have identifiers such as p1_o2pg and p2_chatbot, which is a naming difference only. Attributes are owned by probes: conversion accuracy is an attribute of P1 and not a universal attribute of every model, and only a small operational set (tokens, cost, latency, completion) is collected identically for every probe. Figure 5 shows the resulting structure.

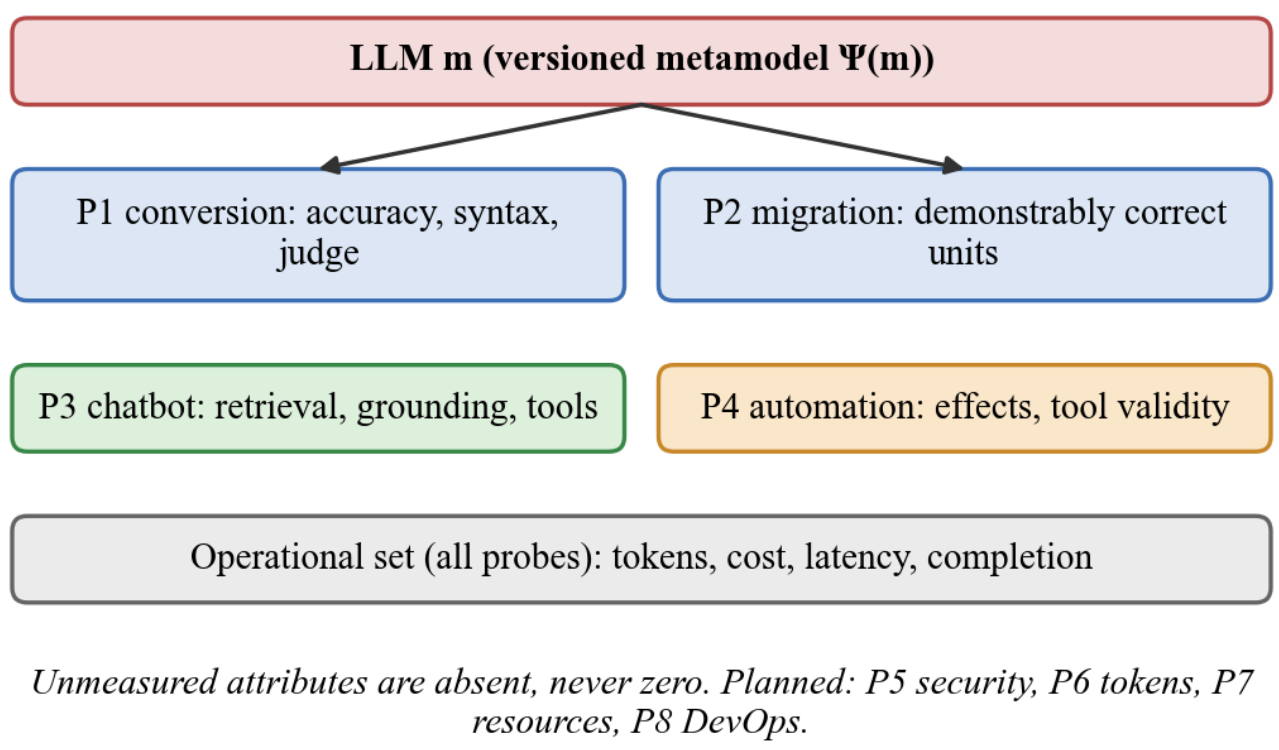

Fig. 5. Structure of a model's metamodel. Probe-owned attributes (P1–P4 implemented or integrated; P5–P8 planned) are measured in the same way for every model; the operational set is common.

TABLE XI. ATTRIBUTE GROUPS OF THE METAMODEL AND THEIR OWNERS

| Attribute group | Probe | State |
|---|---|---|
| Conversion accuracy (Oracle, PostgreSQL) | P1 | measures |
| Migration accuracy (Oracle, PostgreSQL) | P2 | own stand; integration planned |
| Knowledge and retrieval, chatbot | P3 | integrated (p2_chatbot) |
| Knowledge and retrieval, automation | P4 | own stand; integration planned |
| Security treatment | P5 | planned |
| Token efficiency | P6 | planned |
| Index of required resources | P7 | planned |
| DevOps | P8 | planned |
| Volume, quantity of parameters | — | facts about the model |
| Tokens, cost, latency, completion | all | operational set |

Probe integration status and attribute ownership evaluated in October 2026; P5–P8 denote planned extensions.

### C. MAS Core: Planes and Components

MAS Core is organised in four planes (Fig. 6). The *control plane* is a FastAPI service with endpoints for goals, models, probes, prompts, metamodels and cost; it hosts the registries (models and families, probes, attributes, datasets, techniques, prompt variants), the planner and understanding agents and the deterministic gate. The *data plane* is a PostgreSQL blackboard with registries, goals, frozen plans, runs, measurements, agent prompts and agent spend, plus a Redis broker with no persistence, no host port and a password. The *execution plane* consists of Celery workers with two queues, a probe agent that triages failures, a runner that executes the probe and computes attributes deterministically, and an aggregator agent. The fourth plane is the measurement stand of the first probe, P1, which is a copy of the ConversionEvaluator adapted to the probe contract; it has its own deployment and database. The responsibilities of each MAS Core component are distinguished in Table XII.

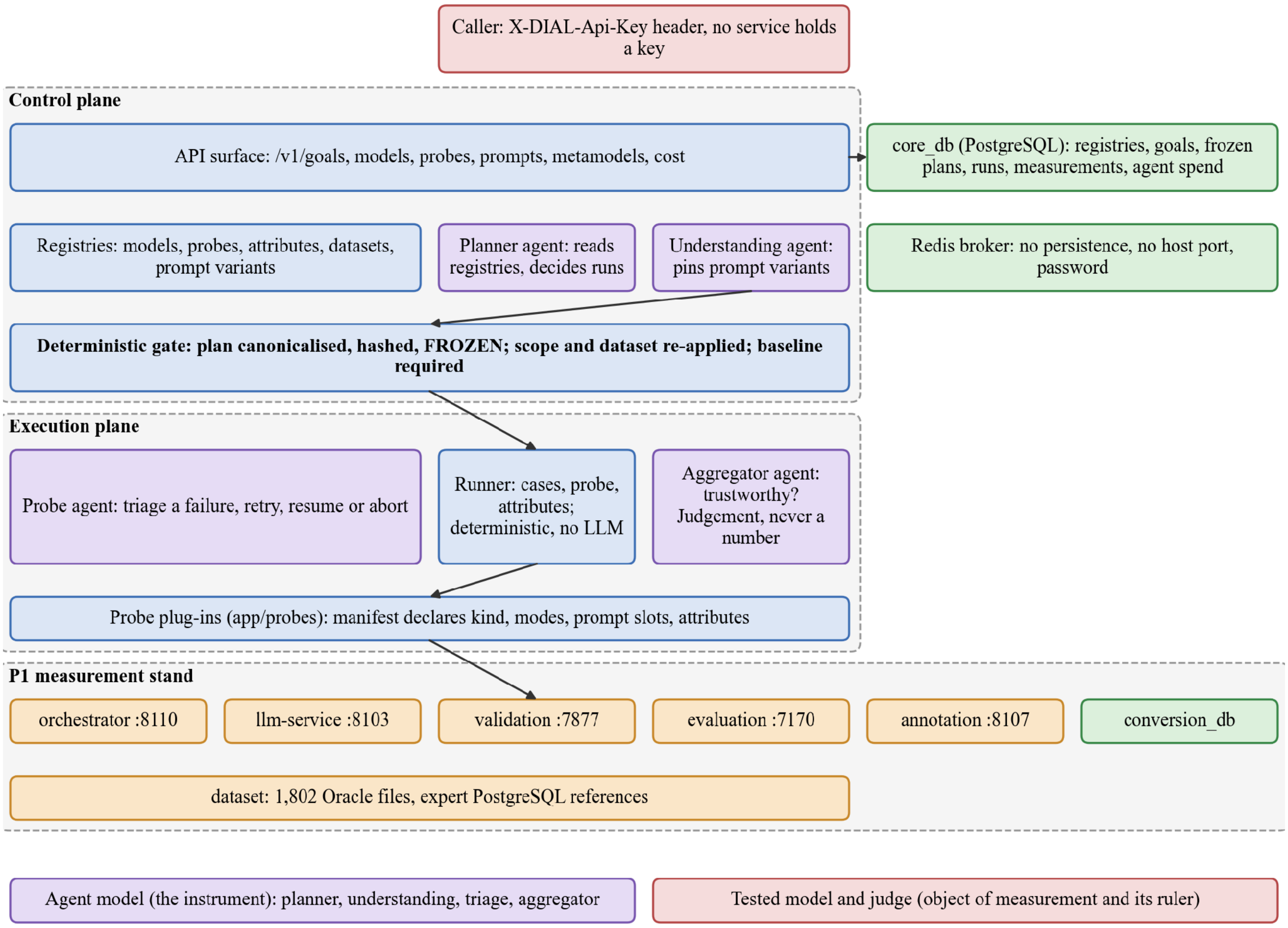

*Solid: the model under test and the judge that scores it. Dashed paths (agents that decide what to measure) never meet it.*

Fig. 6. MAS Core architecture as implemented, redrawn from its documentation: control plane, execution plane, data stores and the P1 measurement stand. The agent model (the instrument) and the tested model with its judge (the object of measurement and its ruler) are separate roles.

TABLE XII. COMPONENTS OF MAS CORE AND THEIR RESPONSIBILITIES

| Component | Plane | Responsibility |
|---|---|---|
| API surface | control | goals, models, probes, prompts, metamodel views, cost |
| Registries | control | models, probes, attributes, datasets, prompt variants |
| Planner agent | control | reads registries via read-only tools, proposes runs |
| Understanding agent | control | pins a prompt variant to every slot before freezing |
| Deterministic gate | control | validates, canonicalises, hashes and freezes the plan |
| Blackboard (PostgreSQL) | data | single write path into measurements |
| Broker (Redis) | data | queues; task messages carry the caller's key |
| Probe agent | execution | triage of a failed run: retry, resume or abort |
| Runner and probe plug-ins | execution | execute cases; compute attributes without an LLM |
| Aggregator agent | execution | judges whether a goal's result is trustworthy; no numbers |
| P1 stand (5 services) | stand | conversion, validation, evaluation, annotation, gateway |

A new probe is a package with a manifest plus registry rows; the core is not edited.

### D. *Life Cycle of a Goal*

A request is a *goal*, for example "profile this model on Oracle-to-PostgreSQL conversion with the baseline prompt on a small subset". Figure 7 and Algorithm 1 show how it is processed.

**Algorithm 1.** Goal planning, freezing and settlement.

**Input:** goal $G$, caller key $k$, registries *Reg*
**Output:** settled goal with measurements and assessment
$S \leftarrow$ ModelSession($k$)  // one client per unit of work
*draft* $\leftarrow$ Planner($G$, read-only tools over *Reg*, $S$)
*draft* $\leftarrow$ Understanding(pin prompt variant per slot, $S$)
**if** dry_run **then return** preview(*draft*)  // nothing is stored
*plan* $\leftarrow$ Gate(*draft*)  // reject on any violation:
  undeclared mode; variant of another probe; duplicate run;
  budget overrun; comparison without baseline arm
re-apply goal scope and dataset version by code
*plan* $\leftarrow$ canonicalise(sort runs); $h \leftarrow$ SHA-256(*plan*); freeze(*plan*, $h$)
**for each** run **in** *plan* **enqueue**(run, $k$)  // idempotent by run id
**on** run end **do** request settlement  // success and failure alike
**if** infrastructure failure **then** status $\leftarrow$ INCONCLUSIVE
**if** no measurements **then** settle without a model call
$A \leftarrow$ Aggregator(results, $S$)  // judgement; unreachable $\rightarrow$ absent
store $A$ beside the goal, never inside a measurement

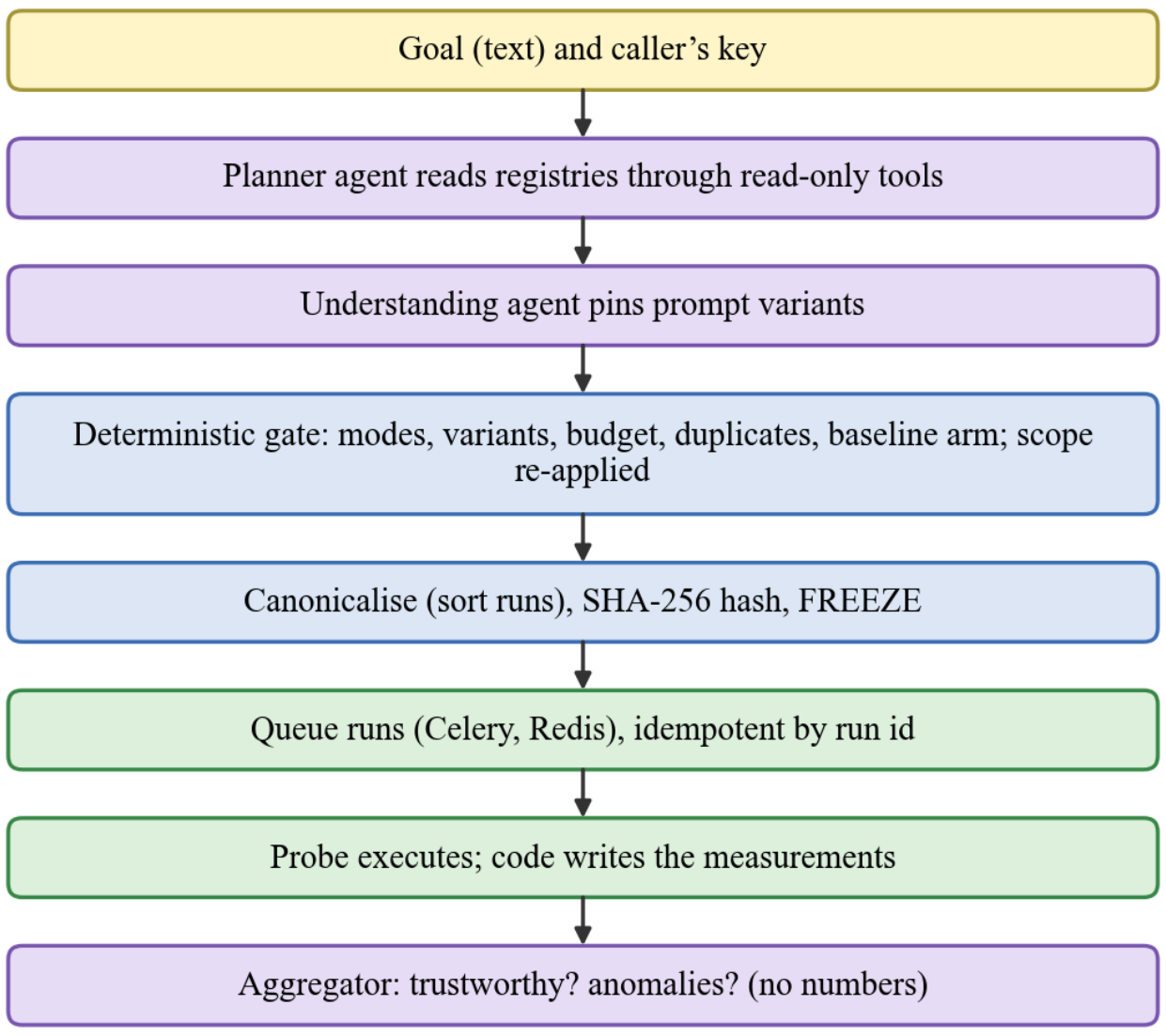

Fig. 7. From a goal to a settled result: the agents propose, deterministic code disposes, and the plan is frozen and hashed before anything is spent.

Four agents take part, declared in the same way: an *AgentSpec* names the agent, its system prompt and its response schema; a shared *run_agent* call performs the structured model call, converts any vendor failure into a single domain error and optionally records spend and the prompt hash; and a *ModelSession* holds the caller's key and builds one model client per unit of work, so that the planner and the understanding agent of one goal share one client. Table XIII summarises the agents and what happens when one of them is unreachable.

TABLE XIII. Agents of MAS Core

| Agent | Response schema | If unreachable |
|---|---|---|
| Planner | PlanDraft | planning fails; no plan stored |
| Understanding | VariantChoice | planning fails; single candidate needs no call |
| Probe agent (triage) | TriageDecision | decision is abort, never a guessed retry |
| Aggregator | GoalAssessment (4 fields) | verdict absent, not "no" |

The aggregator schema has four fields (trustworthy, summary, anomalies, caveats); no field can hold a number.

The gate deserves emphasis because it is where an LLM in the control flow is made safe. The planner is told to pass the goal's subset through, but "the agent was told to" is not a control: in an early run a dropped subset would have launched the full 1,802-file dataset, hours of work and an unapproved bill. The scope and the dataset version are therefore re-applied by code after planning.

Table XIV lists the checks. Because LLM planning cannot be reproducible by determinism, reproducibility is obtained by materialisation: a rerun replays the stored plan instead of asking the agent again, and the hash is the identity of a measurement. Plans with identical canonical representations produce the same SHA-256 hash. Matching hashes are treated as plan identity under the assumption of negligible cryptographic collision probability; strict content identity can be verified by comparing the stored canonical plans. Different hashes do not alone imply incomparable conditions, because controlled experimental arms may intentionally differ. Comparability additionally requires the same evaluation protocol, data and controlled context.

TABLE XIV. Checks Applied by the Deterministic Gate

| Check | Rejects |
|---|---|
| Mode declared | a mode the probe does not declare |
| Variant ownership | a prompt variant belonging to another probe |
| Uniqueness | duplicated runs |
| Budget | plans exceeding the stated budget |
| Baseline arm | a prompt comparison without the baseline on the same model |
| Scope | dropped subset or changed dataset version (re-applied) |

### E. Application Container Contract and Onboarding

Reference applications are not linked into the core. Each one runs as a separate container and is reached over HTTP, so a new application can use its own language, dependencies and data. The chatbot application is the first to follow this pattern end to end: MAS Core reaches the Chatbot Evaluation API through a configured stand address and a shared container network, queries its datasets and configurations through a probe-options endpoint, submits an evaluation and persists the returned measurements. Table XV abstracts this into a contract. The application boundary and its onboarding data flow are illustrated in Fig. 8.

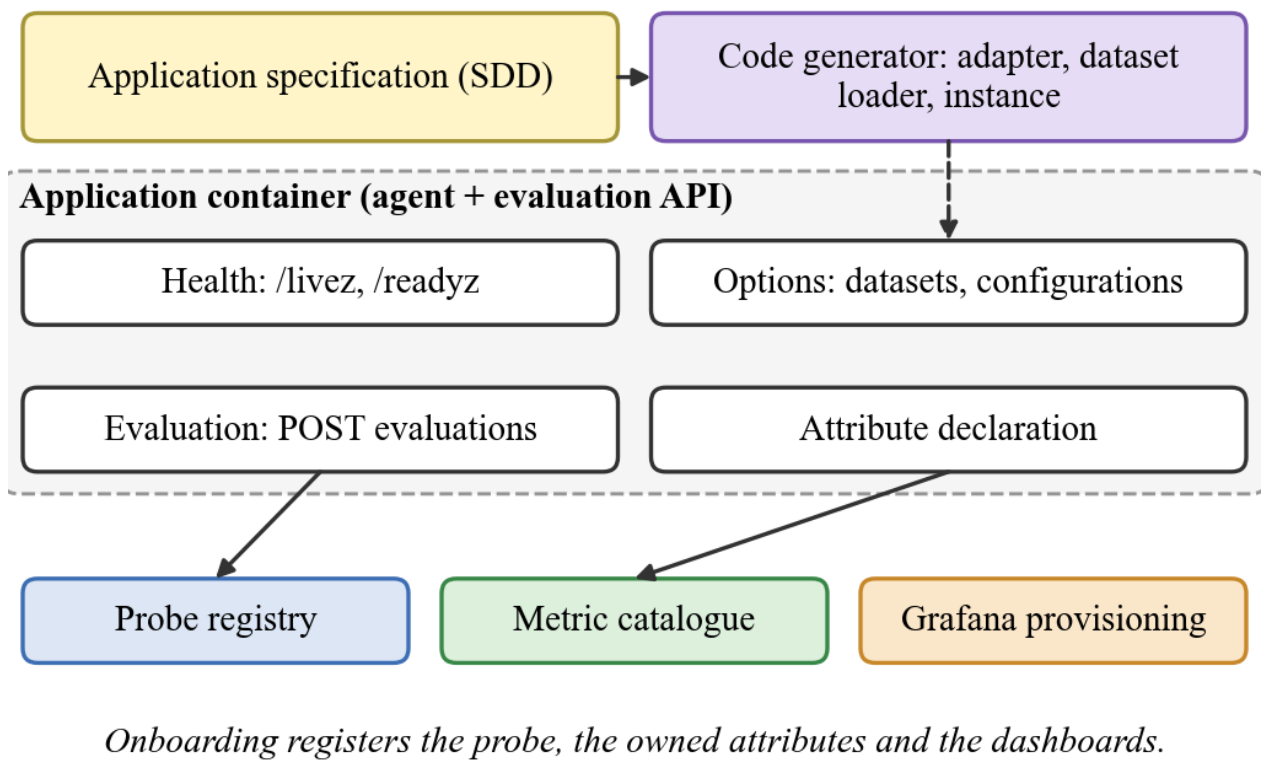

Fig. 8. Application container contract and onboarding. A specification drives the code generator; the container exposes health, options, evaluation and attribute declarations; onboarding registers the probe, the metrics and the dashboards.

Algorithm 2 gives the onboarding procedure. Its two outputs are what turns an application into new knowledge about models: the attributes are added to the metric catalogue with an owner, a unit and a direction, and every model that is subsequently run on the application acquires measured values of them.

TABLE XV. Application Container Contract

| Element | Content |
|---|---|
| Health | liveness and readiness endpoints (/livez, /readyz) |
| Options | datasets, configurations and modes the application offers (GET probe options) |
| Evaluation | submit an evaluation for model(s), dataset version, configuration; return per-case and aggregate measurements |
| Attributes | declaration of owned attributes: name, unit, direction, category |
| Provenance | dataset version, configuration identifier, prompt identity, warnings |
| Variables | stand address, network, dataset, application, framework, model-set and evaluation configuration identifiers |

The run parameters observed for the chatbot are app_config_id, framework_config_id, pinned_model_set_id, dataset_version and evaluation_config_id.

**Algorithm 2.** Onboarding a reference application.

**Input:** application container *C*, specification *Spec*, datasets *D*
**Output:** registered probe, catalogue entries, dashboards
**if** *C*.adapter missing or variables reference unsupported dataset/instance
  *C*.adapter ← Generate(*Spec*, variables)  // specification-driven
**if not** SmokeTest(*C*) **then reject**  // health, options, one case
Σ ← *C*.attribute_declaration()
**for each** σ ∈ Σ: assert unique name, owner = probe, unit, direction
register probe(*C*.modes, *C*.prompt_slots, *D*) in registry
add σ ∈ Σ to metric catalogue with owner; bump metamodel schema
provision Grafana: data source, per-application dashboard,
  panels derived from Σ (direction-aware, unit-aware)
link probe to existing models; list missing attributes per model
**return** probe id

Generating Grafana dashboard definitions from attribute declarations is part of the onboarding design; successful import into a live Grafana server was not measured in this study.

TABLE XVI. Effects of Onboarding a New Application

| Registry | Effect |
|---|---|
| Probe registry | probe with modes, prompt slots and dataset versions |
| Metric catalogue | new attributes with owner, unit and direction |
| Model metamodels | attributes appear as “not measured” until a run completes |
| Analytical UI | use case, metrics and evaluations are listed without UI changes |
| Grafana | data source and dashboard panels for the new attributes |

Chatbot observation: the chat.* metrics appear on the Metrics page, owned by p2_chatbot, with measurement counts.

The metric catalogue associates each chat.* measurement with its owning probe, unit, direction, context and count; unmeasured attributes remain explicitly absent rather than taking a numerical value of zero (Table XVI).

### F. Specification-Driven Generation of Agent Code

Each container is parameterised by template variables (Table XV). In most cases a variable only selects among existing assets: another dataset version, another application configuration, another model set. For the remaining cases the testbed relies on the specification-driven techniques developed in the authors' earlier work, in which a technology-neutral specification is the operational artefact from which code is regenerated [45], [46], [77]. When a variable refers to a dataset or an instance of the reference application for which no adapter exists, for example a new chatbot instance over a different documentation corpus, the generation hook produces or modifies the adapter code from the application specification. The generated adapter must pass the smoke test of Algorithm 2 before it can be registered, and the measurements it produces carry the dataset version and configuration identifier like any other. We stress that this paper evaluates the contract and the measurement machinery; the quality of generated adapters is not evaluated here and is listed among the open items in Section IX-J.

### G. Prompt Identity, Cost Accounting and Credentials

Three properties concern what is recorded about every call. *Prompt identity.* The inherited P1 stand called its templates directly and recorded nothing about the text sent to the model, so every earlier run was, strictly, unreproducible. Templates are now accepted as source plus the identity under which they were registered; the stand verifies that the source hashes to that identity and binds only variables the mode provides, then records the variant identifier and the hash on the run. Omitting a prompt records the built-in under its own name and hash. A regression test asserts that the new path renders byte-identically to the old, so earlier results remain comparable. *Cost separation.* The tested model, the judge and the agents are counted as three numbers that are never summed, because the first is a property of the model and the other two are the price of measuring it (Table XVII).

TABLE XVII. Three-Way Cost Separation

| Spend | Role | Belongs to |
|---|---|---|
| $C_t$ | model under test | property of the model |
| $C_j$ | judge scoring the output | price of measuring |
| $C_g$ | planner, understanding, triage, aggregator | price of measuring |

The three cost components are reported separately; their sum is used only when explicitly labelled as total experimental cost, as in Equation (10), and is not treated as a property of the model under test.

The judge's spend was found to be a substantial fraction of the tested model's, and it had been invisible because the stand discarded it. *Credentials.* No service stores a gateway key. Each request carries the caller's own key, so every model call is billed to whoever asked for it; the Launch page of the interface states that the key is kept in memory and that the core stores only a fingerprint. The key crosses a process boundary inside the task message and is mitigated operationally (a broker without persistence, host port or default password) but not yet cryptographically.

### H. Dashboards: Grafana Registration and the Analytical Interface

Two complementary analytical views consume the common measurement store. Per-application Grafana panels show observed metric series and execution health, while a deterministic API provides filtered measurement and comparison data. Testbed metrics include conversion scores, judge criteria, static validation, file completion, refusals, latency, tokens and separate cost components. The work implements dashboard definitions from attribute declarations, but does not measure end-to-end automated provisioning effort. The architecture of the resulting measurement and dashboard views is shown in Fig. 9.

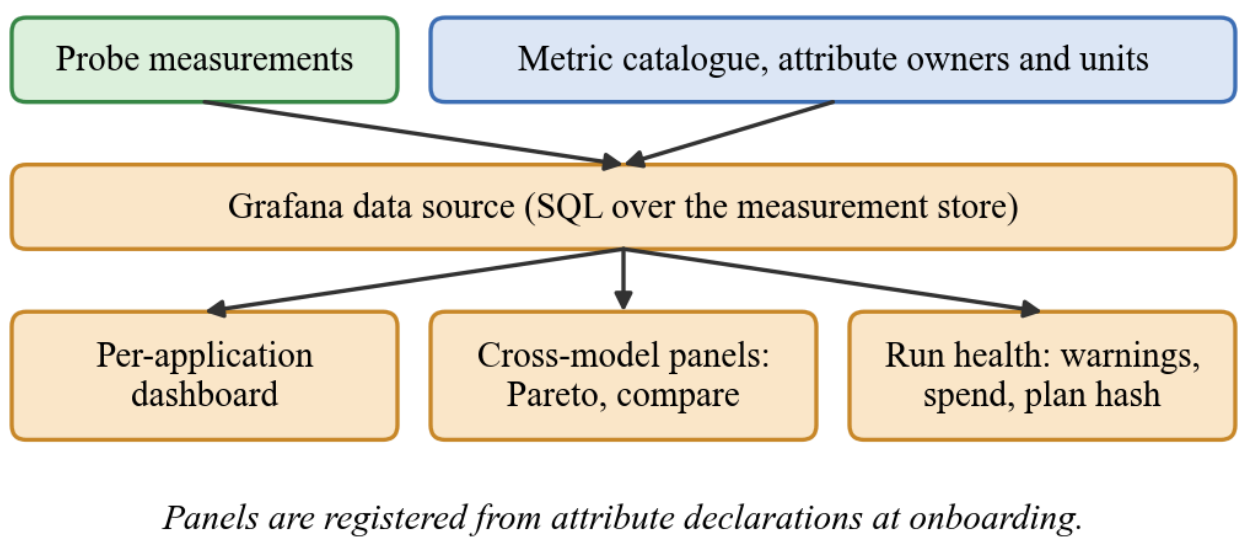

Fig. 9. Measurement store and attribute catalogue determine context-aware dashboard definitions and execution-health views; live deployment of generated dashboard definitions was not evaluated.

The registration rests on a small set of rules that map the declaration of an attribute, its unit, direction and category, to a panel (Table XVIII, Algorithm 3). The dashboard variables are the run context and the dataset version, so that a panel can only ever show values of one context class (Lemma 1); an attribute that is not yet measured is shown as an explicit “not measured” panel instead of an empty graph or a zero. These are design rules of the testbed. Their implementation for a new application, and the effort that it saves, have not been evaluated quantitatively in this work.

**Algorithm 3.** Provisioning dashboards from attribute declarations.

**Input:** probe $p$, attribute set $\Sigma_p$, data source $S$
**Output:** dashboard $B_p$ registered in Grafana
create data source $S$ if absent (read-only role on the measurement store)
$B \leftarrow$ new dashboard titled by $p$; variables $\leftarrow$ {run context κ, dataset version $d$, model}
**for each** $\sigma \in \Sigma_p$
  *panel* ← Template(σ.unit, σ.direction, σ.category)   // Table of panel rules
  query ← SELECT value, n, run FROM measurements WHERE attribute = σ AND context = \$κ AND dataset = \$d
  **if** no rows **then** *panel* ← "not measured"
  add *panel* to $B$
add cost panel (three spends, reported separately) and run-health panel
register $B_p$; return identifier

Implementation status: The dashboard provisioning rules describe the intended onboarding contract. Their fully automatic deployment to a live Grafana instance, and the effort saved relative to manual onboarding, were not evaluated quantitatively.

TABLE XVIII. RULES MAPPING ATTRIBUTE DECLARATIONS TO DASHBOARD PANELS

| Attribute type | Panel | Rule |
|---|---|---|
| ratio or percentage, higher is better | gauge and per-run trend | thresholds from the declaration; direction-aware colour |
| count or iterations, lower is better | stat and trend | direction reversed |
| latency (ms) | time series; percentile bars | median and tail |
| cost (USD), tokens | stacked bars by spend role | tested, judge and agent costs shown separately |
| categorical verdict | state timeline; distribution | PASS, FAIL, degraded; ACCEPTED, FAILED, BLOCKED |
| text evidence | link to run report | text is evidence, not a value |
| not measured | explicit "not measured" panel | never zero, never blank |
| cross-model view | Pareto scatter; comparison table | only within one context class; evidence count shown |
| run health | warnings, plan hash, prompt hash | provenance beside every result |

Design rules of the onboarding step; not evaluated quantitatively.

The analytical interface is a single-page application over a deterministic backend. Scores and comparisons originate from server-side aggregations; client-side computation is limited to presentation. The main interface areas and their analytic functions are defined in Table XIX.

TABLE XIX. AREAS OF THE ANALYTICAL INTERFACE

| Area | Content | Nature |
|---|---|---|
| Explore | models, use cases, metrics, evaluations; filters; raw JSON parameters | read-only catalogue |
| Analytics: Summary | KPI tiles, leaderboard for a chosen metric, breakdown charts | deterministic |
| Analytics: Model performance | model passport: gauges per metric, evaluation reports | deterministic |
| Analytics: Compare models / evaluations | per-category scores; per-metric delta; context warning | deterministic |
| Analytics: Pareto | objectives with direction; frontier with evidence counts | server-side |
| Launch | choose probe, model(s), dataset, configuration; preview plan; start | calls the core |
| Agent | conversational recommendation using the analytical API as tools | LLM-backed |

Technology: React, Vite, TanStack Query, React Router, Tailwind with Radix primitives, Recharts; a typed API client surfaces network, HTTP and validation errors through shared components.

The conversational recommendation component is architecturally separate from the deterministic measurement and Pareto selection procedures, and its interactive behaviour is not used as experimental evidence. The comparison API enforces the context constraints of Equation (4): measurements generated using different datasets, evaluation protocols or controlled configurations cannot be combined as if directly comparable. Incomplete model usage, missing semantic scores and unavailable cost records are returned with explicit warnings rather than converted to zero-valued measurements.

### *I. Data Model and API Surface*

The blackboard holds a small number of entities, and every API path maps onto them (Table XX). *Registries* describe what exists: models and their families, probes, attributes, datasets, techniques and prompt variants. *Goals* carry the request, the caller's key fingerprint and the state of settlement. *Plans* are frozen and keyed by hash. *Runs* point to a plan, a model, a mode and a prompt variant and own their *measurements*. *Agent prompts* and *agent spend* record the text and the cost of every agent call. Because measurements have a single write path, the metamodel views are projections that can be recomputed, and they do not contain information that the underlying rows lack. Two views deserve attention. The metamodel of a model lists the measured attributes together with an explicit list of what was not measured. The comparison view prevents direct comparisons between measurements obtained under different evaluation contexts, thereby enforcing the comparability constraints defined in Equation (4).

TABLE XX. PRINCIPAL ENTITIES AND INTERFACES

| Entity or path | Role |
|---|---|
| POST /v1/goals (dry_run option) | plan a goal; with dry_run only a plan preview is returned and no goal is written |
| GET /v1/goals/{id}, /cost | state, runs, assessment; three-way cost |
| GET /v1/models, /v1/probes, /v1/prompts/{id} | registries; variant ids are readable coordinates and may contain slashes |
| GET /v1/probes/{id}/options | datasets and configurations that a probe offers (chatbot integration) |
| GET /v1/metamodels | per-model profile with an explicit list of missing attributes |
| analytical API: GET /models, /use-cases, /metrics, /evaluations | raw catalogue with filters and pagination |
| analytical API: POST /analytics/* | breakdown, compare-models, compare-evaluations, Pareto |
| Blackboard tables | registries, goals, frozen plans, runs, measurements, agent prompts, agent spend |

The listed endpoints describe the implemented control and analytical interfaces.

### *J. Operational Properties*

Automated verification comprises 93 tests for the MAS Core, 21 for the P1 evaluation service and 232 backend tests for the integrated chatbot evaluation interface. Metric-library versions are pinned for a given evaluation context because changing the scoring implementation changes the measurement instrument.

## VI. REFERENCE APPLICATIONS

The three reference applications provide distinct, implemented evaluation environments: Oracle-to-PostgreSQL translation with differential behavioural verification, documentation-grounded question answering with retrieval and tool-use diagnostics, and enterprise workflow automation with observable side effects. Table XXI summarises their experimental configurations; Sections VI-A–VI-C describe the associated task and measurement contracts.

TABLE XXI. Overview of the Reference Applications

| | Migration | Chatbot | Automation |
|---|---|---|---|
| Purpose | Oracle PL/SQL → PostgreSQL | documentation-grounded assistant | support tasks over Jira, Confluence, Outlook |
| Probe | P2 (own stand), P1 (conversion) | P3, id p2_chatbot | P4 (own stand) |
| Orchestration | LangGraph repair loop | LangGraph | 4 interchangeable frameworks |
| Dataset | 1,448 tasks; 216 units; 1,802 files (P1) | mvp_scenarios v6, 10 scenarios | pilot.v4, 49 samples |
| Ground truth | real Oracle run; expert references | expected sources, tools, answers | expected tool operations, final states |
| Judge | P1 judge; semantic judge in depth study | semantic_judge_v1 | none (effects observed) |
| Integration | own stand; MCP and API | integrated in MAS Core | own stand |

Experimental configurations and dataset sizes used in this study.

## A. Migration Agent: Oracle PL/SQL to PostgreSQL

The migration agent converts a directory of Oracle SQL and PL/SQL files into PostgreSQL files and, for every database object, reports how it was converted, which checks were completed and what could not be verified. The pipeline and routing were designed by A. Seletska, the integrations by O. Ilchuk and the measurement of quality by D. Kostetskyi. It does not connect to an Oracle database; it works only with the files it is given. The migration sequence from parsing to validation and bounded repair is shown in Fig. 10.

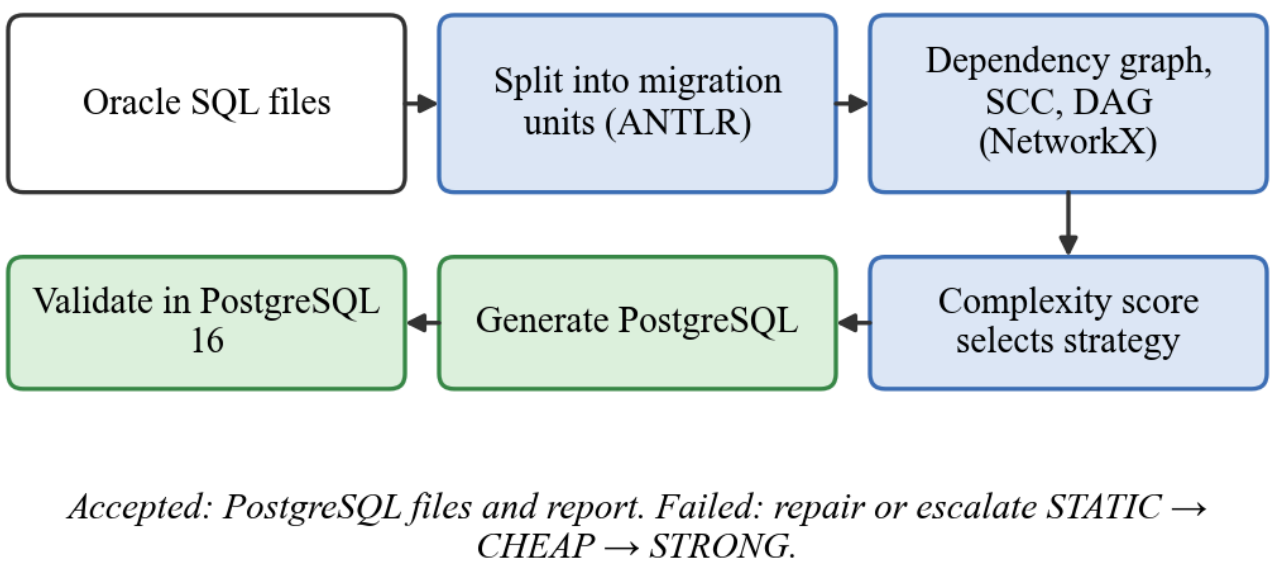

Fig. 10. Migration pipeline: files are split into migration units, ordered by a dependency graph, routed by a complexity score, generated, validated in PostgreSQL 16, and repaired or escalated on failure.

*Decomposition.* The parser splits each file into *migration units* (table, view, sequence, function, procedure, package, trigger, query or data-changing statement) while preserving the original source text. A package remains one unit because its members share state. Each unit carries a parse confidence: full, partial, lexical-only or failed, and a low confidence excludes the deterministic converter. *Ordering.* A dependency graph places required objects before their users; mutually dependent objects form one migration group (a strongly connected component), and the condensation is a directed acyclic graph that is processed in topological order. A failing group blocks its dependants while independent branches continue. The same decomposition idea underlies the earlier study of task decomposition for specification generation by the same team.

*Routing.* A deterministic weighted score in [0, 1] estimates the difficulty of each group from source size, control flow, data access, Oracle-specific features, dependencies, cycles, package size and parse quality. Scores up to 0.30 select the STATIC strategy (Ora2Pg and deterministic rules, no model tokens), scores above 0.30 and up to 0.65 select a cheaper model, and higher scores a stronger model (Table XXII). Large packages go directly to the strong model, and low analysis confidence promotes a unit to a stronger strategy. Every routing decision records a reason. The individual feature families of the complexity score are detailed in Table XXIII. The bounded migration, validation, and repair procedure is formalised in Algorithm 4, which defines the iterative correction process, strategy escalation, and termination conditions under predefined execution and retry budgets.

TABLE XXII. Migration Strategies

| Complexity | Strategy | Mechanism |
|---|---|---|
| 0 – 0.30 | STATIC | Ora2Pg plus exact rules (e.g. NVL → COALESCE); no tokens |
| > 0.30 – 0.65 | CHEAP | less expensive configured model |
| > 0.65 | STRONG | more capable model; large packages; escalations |

Models in the breadth study: gpt-5.4-nano, gpt-5.4-mini, gpt-5; in the depth study: DeepSeek-V4-Flash (cheap) and DeepSeek-V4-Pro (strong).

TABLE XXIII. Inputs of the Complexity Score

| Feature family | Role in routing |
|---|---|
| source size | longer sources raise the score |
| control flow | branching and loops raise the score |
| data access | cursors, bulk operations, dynamic SQL |
| Oracle-specific features | constructs without a PostgreSQL equivalent |
| dependencies and cycles | coupling, mutually dependent groups |
| package size | large packages go directly to the strong model |
| parse quality | low confidence moves a unit to a stronger strategy |

Weighted formula with YAML-configured weights; not a trained classifier.

**Algorithm 4.** Bounded migrate–validate–repair loop for one migration group.

**Input:** group *g*, dependency signatures, budget (2 repairs/strategy, 2 escalations, 5 attempts, 900 s)
**Output:** status, verification level, PostgreSQL text
s ← Route(score(g), parse confidence, type, size); sql ← Generate(s, g, signatures)
repeat while attempts < 5 and elapsed < 900 s
  attempts ← attempts + 1; r ← Validate(sql)
  // static → install → structure → fixed scenarios
  **if** *r*.pass **then return** ACCEPTED, level(*r*)
  **if** *r*.infra_error **then return** INCONCLUSIVE
  if repairs(s) < 2 then sql ← Repair(sql, r.diagnostics); repairs(s) += 1
  else if escalations < 2 and next(s) exists then s ← next(s); sql ← Generate(s, g, signatures); escalations += 1
  **else return** FAILED
return FAILED // attempts or time exhausted; dependants become BLOCKED

*Generation and validation.* The model receives the Oracle source, all members of a mutually dependent group, the signatures of dependencies already migrated, the names of missing dependencies and conversion guidance; it returns structured PostgreSQL and notes. It does not choose the order, compute its own complexity or decide whether its output is correct. Validation runs against a PostgreSQL 16 database kept for the run: static analysis, installation, structural checks, generated behaviour scenarios and derivable invariants. Scenarios are created once per group and reused after repairs so that tests cannot drift to fit a candidate. The strongest completed check is recorded as the verification level (Table XXIV), and a group takes the result of its least successful member.

TABLE XXIV. VERIFICATION LEVELS AND UNIT STATUSES

| Term | Meaning |
|---|---|
| CHECKED_STATIC | static checks passed; object could not be installed |
| VERIFIED_INSTALL | PostgreSQL installed or accepted the object |
| VERIFIED_BEHAVIOR | installed and generated behaviour checks passed |
| ACCEPTED | passed every check available for the unit |
| FAILED | attempted but still failing |
| BLOCKED | an upstream unit failed; never tried |
| INCONCLUSIVE | infrastructure unavailable; not a verdict on the SQL |
| UNSUPPORTED | could not be attempted |

*Interfaces.* The same service layer is exposed through a command line, an HTTP API with asynchronous jobs and an MCP server, so that an orchestrating model can discover tools and follow a run through structured resources. Provider credentials stay in server-side configuration and are never tool arguments. An optional integration generates a repair agent with EvoAgentX in an isolated project that communicates through a small versioned JSON protocol; it only produces candidates, and installation and scenario execution in the main process remain authoritative.

*Evaluation corpora.* Two corpora answer two questions. The *breadth* corpus is a grammar-clean subset of paired Oracle fragments and human-written PostgreSQL references: 1,000 pairs drawn with seed 42 produced 1,448 migration tasks. The *depth* corpus consists of 216 migration units from 34 complete Oracle PL/SQL example files from a public repository, each executed on a real Oracle instance (Oracle 26ai Free in Docker), so that a translation can be compared with what Oracle printed and raised. The first measures how often the agent's own checks pass and how close the output is to a reference; the second measures whether the translation behaves like Oracle.

## B. Chatbot Agent: Documentation-Grounded Assistant

The chatbot is a developer-support assistant that answers questions over a body of documentation, retrieving passages, calling tools and citing sources. Its evaluation framework measures how well different LLMs and chatbot configurations work for documentation-based question answering and where a failure occurs: in execution, tool routing, retrieval, grounding or answer generation. It runs the real chatbot with selected retrieval and model configurations, scores it against a versioned dataset and saves the evidence. Ground truth is never passed to the chatbot; expected sources, tool behaviour and answer content are used only by the evaluation engine. Its execution and evaluation stages are summarised in Fig. 11.

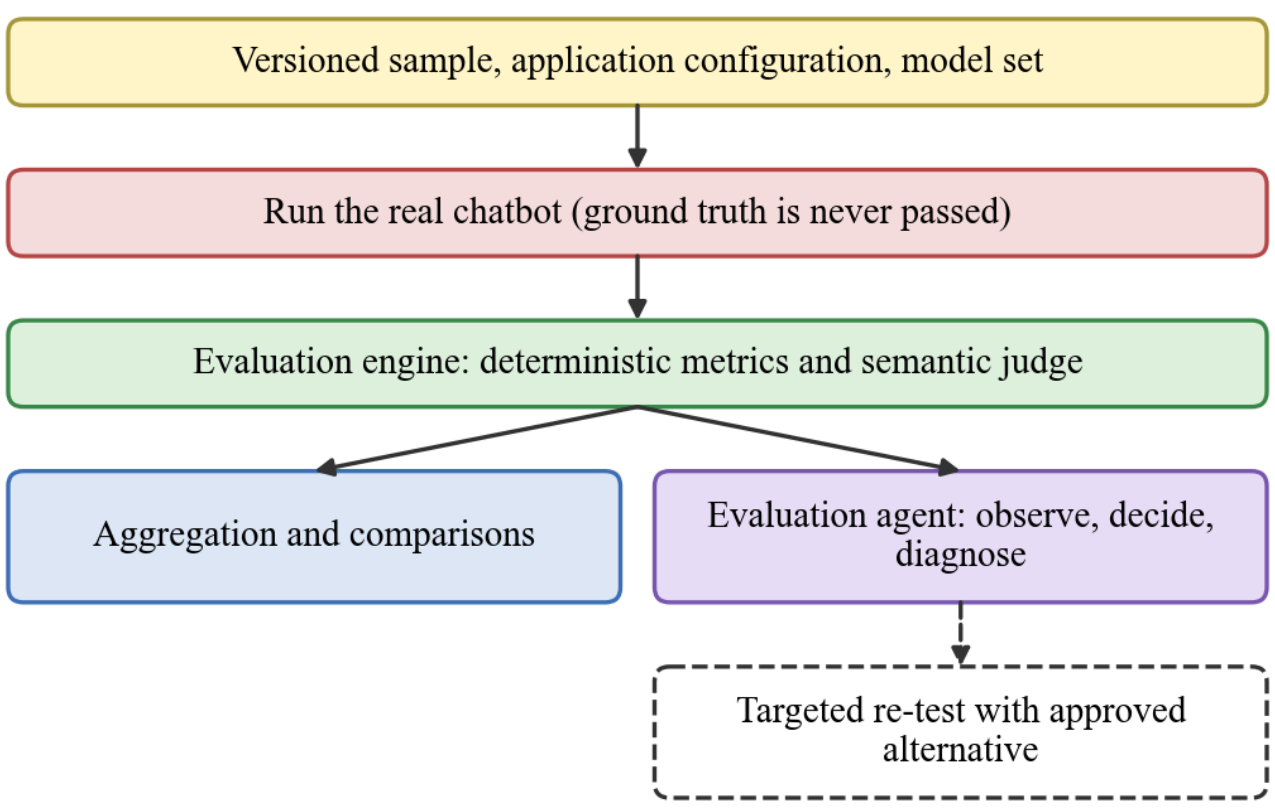

Fig. 11. Chatbot evaluation workflow. The real chatbot is executed on a versioned sample; deterministic metrics and a semantic judge produce results that feed aggregation and an evaluation agent.

Application, model, evaluation and experiment configurations are stored separately, and every run saves the configuration actually used. Six metric groups evaluate the components separately (Table XXV); two methods are used. Deterministic checks cover execution, tools, retrieval and citation structure. A judge, semantic_judge_v1, scores faithfulness, answer correctness, answer relevance and task completion with fixed criteria; its scores support analysis and comparison but do not by themselves decide PASS or FAIL. The judge is an evaluation component built on an existing LLM, not a separately trained model.

TABLE XXV. CHATBOT METRIC GROUPS

| Group | Measurements | Method |
|---|---|---|
| Execution | execution success, tool execution success | deterministic |
| Agent / tools | tool selection, tool arguments, forbidden calls | deterministic |
| Retrieval | recall@K, precision@K, R-precision, MRR, metadata | deterministic |
| Grounding | citation compliance, validity, completeness, source match; faithfulness | deterministic; judge |
| Answer quality | correctness, relevance, task completion | judge |
| Operational | latency, tokens, retries, iterations, LLM calls | deterministic |

The metric groups are registered as chat.* attributes in the metamodel, with explicit ownership, measurement context, and provenance.

*Execution contract.* The execution layer runs one conversation through the existing LangGraph chatbot under a selected configuration and returns a factual result to the evaluation layer. It does not rebuild the chatbot, does not see ground truth and does not compute metrics. A run plan expands every combination of sample, application configuration, model set and repeat, records the outcome of each execution and hands each result across a framework-neutral boundary (Table XXVI). Per-run settings (chat model, router model, decoding parameters, retrieval strategy, Top-K, reranker state, enabled tools) are injected at run time, which was verified in focused tests and by baseline and alternative-model smoke runs; model selection is therefore explicit and not merely recorded as metadata. Each execution starts from a fresh thread and state and uses a dedicated evaluation principal with memory isolation, and 21 execution tests cover, among others, the mechanical rejection of ground-truth fields, the preservation of retrieved chunks after serialisation and the capture of successful and failed tool attempts.

TABLE XXVI. EXECUTION CONTRACT OF THE CHATBOT EVALUATION

| Object | Content |
|---|---|
| ExecutableSample | sample id, dataset version, ordered user turns only (no ground truth) |
| AppConfiguration | retrieval strategy, Top-K, reranker state, enabled tools, identifier |
| ModelSet | chat model, router model; optional temperature, top-p, token limit, structured output |
| RunPlan | dataset version, configurations, model sets, repeats, concurrency, failure handling |
| ChatbotAdapter | invokes the chatbot turn by turn; converts runtime data into a RunResult |
| RunResult | answer, structured citations, ordered tool calls and results, retrieval evidence, statuses, typed errors |
| RunResultConsumer | boundary through which the result is handed to the evaluation layer |

Chatbot evaluation execution contract and interfaces.

*Preflight and datasets.* Before evaluation, the runner verifies readiness of the database, prompt registry, LLM

gateway and configured tool services. Experiments use the versioned mvp_scenarios v6 dataset. Expected answers, sources and tool actions are held in the evaluator, and never supplied to the tested chatbot. Some expected labels have not been independently validated through execution, so corresponding measurements are restricted to their declared evidence level.

*The evaluation agent.* A stateful LangGraph agent observes the computed metrics, decides on a diagnostic action, updates its state and decides again until it can name the failing layer. It does not change metric values or verdicts. Its five diagnostic tools (Table XXVII) are internal functions and not tools of the chatbot. The agent applies *causal gating*: it examines execution first, then routing and tools, retrieval, grounding and answer quality, and the first failing layer is the root cause (Fig. 12). For example, if the documentation-search tool was never called, routing is the cause and retrieval, which never ran, is not assigned a separate cause. A deterministic fallback replaces the agent's controller if the model is unavailable.

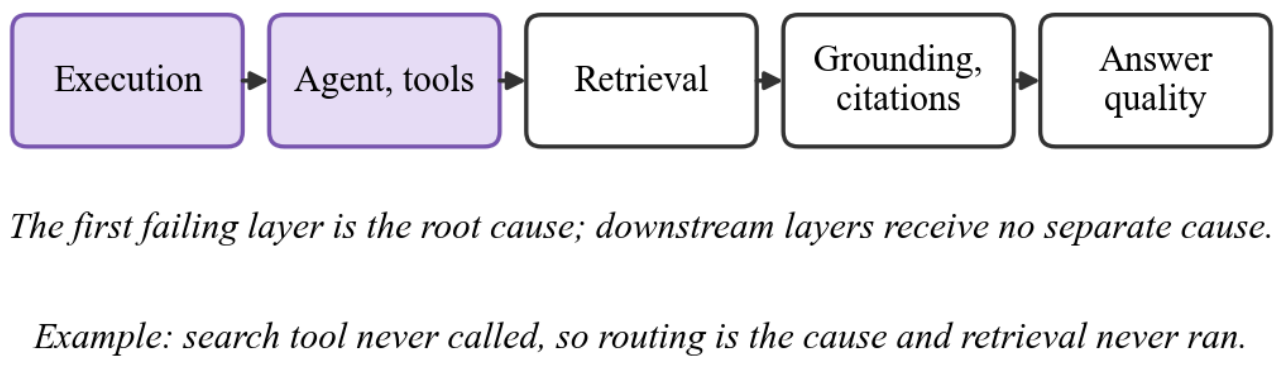

Fig. 12. Causal gating: layers are examined in dependency order.

TABLE XXVII. Diagnostic Tools of the Evaluation Agent

| Tool | Analyses | Example finding |
|---|---|---|
| diagnose_execution | runtime and tool failures | error during a tool call |
| diagnose_agent_tools | tool selection and arguments | required search tool not called |
| diagnose_retrieval | sources and ranking | a required document not found |
| diagnose_grounding | source support, citations | one of two source groups cited |
| diagnose_answer_quality | correctness, relevance, completion | required information missing |

**Algorithm 5.** Causal-gated diagnosis with optional targeted re-test.

**Input:** run result *R* with metrics, policy *Π* of allowed actions, candidates *Cand*
**Output:** diagnosis, optional accepted candidate
L ← [execution, tools, retrieval, grounding, answer_quality]; root_cause ← none
**for each** layer *l* **in** *L* **in order**
  **if** *R*.passes(*l*) **then continue**
  *d* ← diagnose_*l*(*R*)  // action chosen by controller from *Π*
  root_cause ← l; evidence ← d; break  // downstream layers not assigned
if root_cause ≠ none and approved candidate c ∈ Cand exists
  rerun same scenario with *c*; compare target metrics with baseline
  if PASS is restored or a prespecified target-metric improvement is confirmed then mark c as a candidate for validation
return root_cause, evidence, candidate status; do not mutate the live system

TABLE XXVIII. Model Roles in the Chatbot Evaluation

| Role | Configuration | Purpose |
|---|---|---|
| Chat model | GPT-5.6 Terra (selected) | generates answers |
| Semantic judge | semantic_judge_v1, GPT-5.6 Sol reasoning | scores semantic quality |
| Agent controller | GPT-5.6 Sol reasoning | selects the next permitted diagnostic action |
| Router model | baseline-models (variable) | decides tool routing inside the chatbot |

Judge and controller have separate roles and instructions although both use the same model in the current configuration. Vendor and deployment designations are reproduced as recorded by the gateway; internal aliases are not verified product identifiers. The explicit model-role configuration is documented in Table XXVIII.

*Integration.* The chatbot evaluation API executes the configured chatbot, collects structured execution evidence and writes measurements into the MAS Core store. The integrated evaluation path passed an end-to-end validation, including deterministic retrieval and tool-call checks. Semantic metrics are emitted only when the specified judge is available, while missing token usage and incomplete citations generate explicit warnings. These integration checks establish the measurement path but do not substitute for application-quality evaluation.

### C. Automation Agent: Enterprise Support Tasks

The automation system handles inbound Jira, Confluence or Outlook events with a single LLM agent that reasons in a loop and acts through fifteen canonical business operations, five per system. Every execution starts from a fresh state, carries no memory from earlier executions, reads knowledge without changing it and is bounded to ten steps. Failures are typed outcomes, never hidden or fabricated. The system has five layers (Fig. 13): entry points; strategy selection; prompt orchestration, model-call interception and tool management; business integrations, which are live systems in production and a frozen world state in evaluation; and infrastructure for configuration, retries, evidence and authentication.

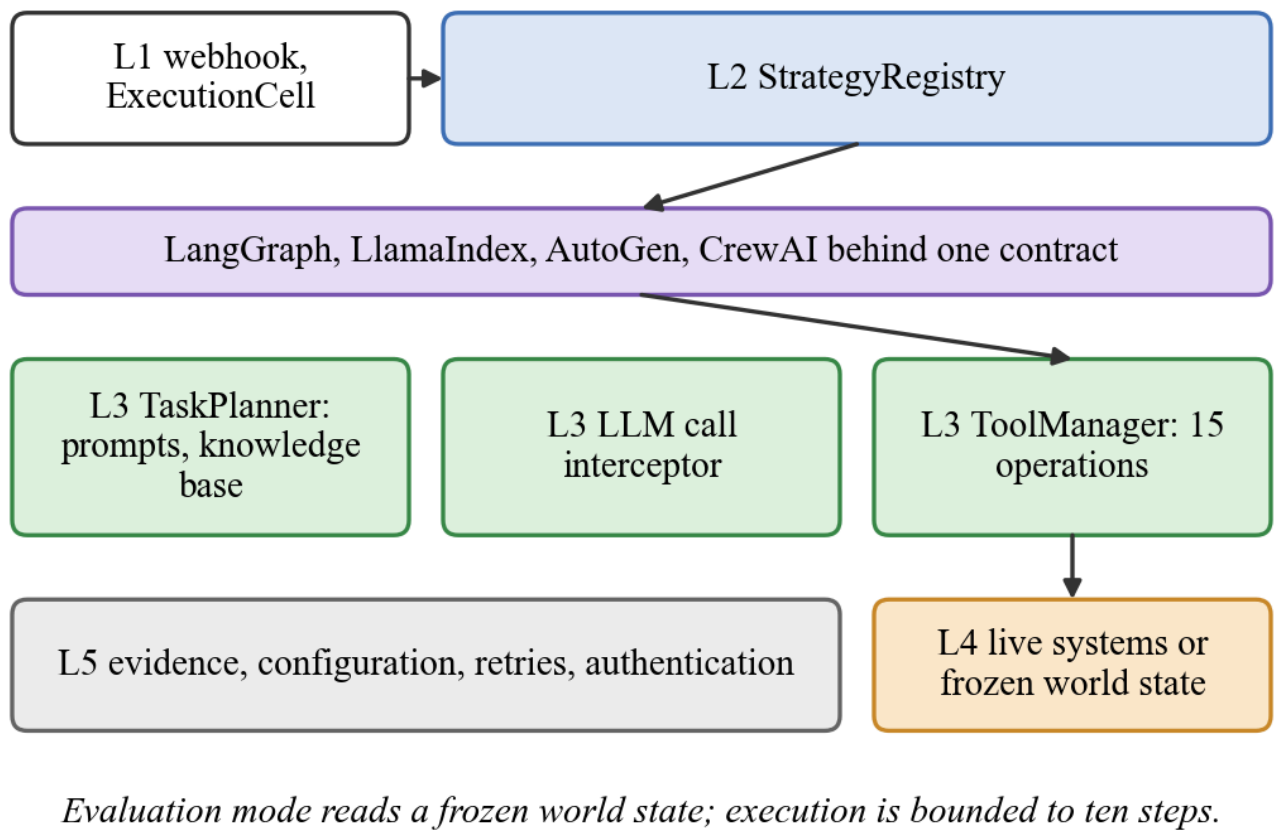

Fig. 13. Layered architecture of the automation agent. In evaluation mode the tool manager reads a frozen world state.

*Interchangeable frameworks.* A framework-neutral strategy contract lets a factory build the configured implementation once. LangGraph (compiled graph) and LlamaIndex (event-driven workflow) apply the step limit per execution and support model-call retries; AutoGen (group chat) and CrewAI (agent, task, crew) fix the step limit at start-up and do not support retries. All adapters share the same prompts, tools, retry handling and evidence, which keeps results comparable; adding a framework takes one adapter and one registry entry. The framework is thereby a variable of the experiment, in line with the framework-level comparison in [78]. Experimental-design limitation: Model and framework sweeps are analysed separately here. Their association statistics and effect-rate ranges do not estimate an interaction term in a crossed model-by-framework factorial design and should not be interpreted as a general causal superiority of model choice.

*Configuration schema.* An agent identity specification fixes the framework, model, prompt, chain-of-thought and few-shot choices [79], [80]; a system configuration adds sampling (temperature, maximum tokens), knowledge-base retrieval settings and retry budget. The configuration examples comprise six identities and eleven configurations: five baseline model conditions (GPT-5.4, GPT-5.4 Mini,

Claude Opus 5, Gemini 3.5 Flash, o4-mini) and six controlled variations around GPT-5.4 (temperature 0.3 and 0.7, chain of thought, knowledge base, retry budgets 1 and 2); the candidate models were checked through the gateway for native capability rather than assumed from their names.

*Sample model.* Running against live Jira, Confluence and Outlook cannot provide an evaluation because live systems mutate between runs, hold no authored ground truth and would produce hundreds of real tickets and emails under a configuration, task and repeat fan-out. The dataset is therefore a closed simulation: the structure matches real payloads so that the model sees production-shaped data, the content is synthetic and the world is frozen. A sample has three layers (Table XXIX); the expected layer is designed backward from the evaluation axes so that every axis is computable and nothing is decorative. Tool names in the expected steps are bound to the operations that the tool manager really exposes (Table XXX); any other name that an agent requests is by definition what the hallucinated-call rate catches. The dataset identifier is immutable after first use, and the same body of tasks is partitionable into learning, validation and locked-test roles without leakage, so that evidence gathered for self-learning cannot inflate scores on the locked test.

TABLE XXIX. LAYERS OF AN AUTOMATION SAMPLE

| Layer | Purpose | Feeds |
|---|---|---|
| task_input | the event that starts the run (production-shaped, synthetic) | input |
| world_state | frozen snapshot that tools read during the run | determinism |
| expected | authored ground truth: steps, parameters, valid tools, final state, completion signal | all correctness and reasoning axes |

Source: dataset schema of the automation direction.

TABLE XXX. CANONICAL OPERATIONS OF THE THREE TOOLS

| Tool | Operations |
|---|---|
| Jira | create_issue, update_issue, search_issues, add_comment, get_issue |
| Confluence | create_page, update_page, search_pages, get_page, add_comment |
| Outlook | send_email, list_emails, read_email, reply_email, list_calendar_events |

Fifteen operations; add_comment is shared and takes exactly one target (issue key or page id).

*Two-phase tool lifecycle.* A tool call passes through three events: a request, emitted before validation with the tool name and the arguments exactly as requested; a start, emitted after validation and before dispatch; and a finish, emitted on completion or rejection. A request rejected before dispatch produces a finish without a start. A tool call is counted from a unique agent-requested action, including requests rejected before dispatch, so that an invented tool is counted even if it was never executed; transport retries and event-delivery retries are not new calls. This convention is what makes the hallucinated-call rate and the argument-validity rate well defined across frameworks that differ in step granularity.

*Dataset.* The pilot dataset, identified as pilot.v4, contains 49 generated task samples across six task classes (jira_create, confluence_lookup, email_compose, incident_triage, cross_system_sync and escalation). Each sample has a production-shaped input, a frozen world state and authored expected tool operations and final states. All 49 samples pass schema validation and the repository tests (35 passing). The dataset is synthetic and has known gaps: fifteen samples require a contact address that no tool can read, so a completed run on them can still mean that no email was sent. The resulting eleven measurable automation axes are defined in Table XXXI.

TABLE XXXI. AUTOMATION EVALUATION AXES

| Metric | Measures |
|---|---|
| Task completion rate | share of executions achieving the expected outcome |
| Tool selection accuracy | requested tool equals expected tool at the position |
| Tool sequence order accuracy | complete expected sequence matched |
| Tool argument validity rate | structurally valid, contextually appropriate arguments |
| Mean agent loop steps | execution units, framework-specific |
| Mean total LLM calls | model invocation attempts incl. retries |
| Mean total tokens | input and output tokens |
| Error retry rate | executions triggering retries |
| Hallucinated tool call rate | requests naming unavailable tools |
| Premature termination rate | limits, timeouts, incomplete responses |
| Unnecessary loop rate | loops not warranted by the observations presented (trace analysis) |

Eleven evaluation axes are specified and listed. Assessments use recorded execution evidence, not inferred reasoning; unavailable values remain explicitly missing.

*Self-learning.* The system treats execution experience as evidence and generated lessons as hypotheses, in the spirit of experience-based agents [27], [28]. Candidate procedures are generated from failed executions, validated one at a time on separate tasks, and published only after demonstrated improvement and human approval, as immutable knowledge snapshots selected explicitly by later configurations. This mechanism is designed but not part of the measurements reported here.

*Self-learning pipeline and its measurement.* As suggested by experience-based agent research [27], [28], candidate procedures may be abstracted from successful and failed trajectories, then compared with an unchanged baseline on independent validation tasks. The testbed provides the separate evaluation, control-arm, evidence and noise-floor mechanisms required by such a loop. The generation and long-term retention of learned procedures are future applications of this measurement method; no learning gains are claimed in this study.

**Algorithm 6.** Validation of one candidate procedure in the self-learning loop.

**Input:** failed executions $F$, successful examples $S$, snapshot $K_0$, partitions (learning, validation, locked test)
**Output:** published snapshot $K_1$ or rejection
1 $c$ ← Generate($F$, $S$) // contrast failures with successes; narrow procedure
2 if ¬Relevant(c) ∨ ¬Safe(c) then reject // before any execution
3 $arm_{ctrl}$ ← run validation tasks with $K_0$; $arm_{sys}$ ← run with $K_0 + c$
4 $\Delta$ ← score($arm_{sys}$) − score($arm_{ctrl}$) // per axis; repeats ≥ 5
5 if repeats are insufficient or $|\Delta| \le$ range(armsys) + range(armctrl) + $\tau$ then return “no result”
6 if $\Delta$ does not improve the direction-adjusted target metric, or any outcome axis regresses, then reject // compliance never instead of outcome
7 **if** human approves **then** publish $K_1 = K_0 + c$ (immutable)
8 **else** defer; a revision is a new explicit attempt
9 never touch the locked-test partition during steps 1–8

*Experimentation strategy.* Every configuration choice is an explicit experimental variable. Stage 0 screens acting models against applicability thresholds with all else at baseline; Stage 1 sweeps behavioural factors one at a time on survivors; Stage 2 sweeps retrieval only if a knowledge base proves useful; a final bounded stage probes a few interactions. Repeats quantify variation, and an effect smaller than the repeat spread is not reported as an improvement. Table XXXII lists the axes. The applicability thresholds (hallucinated-tool-call rate at most 5%, completion at least 60% on the screening subset) are placeholders pending team agreement, so Stage 0 is not executable yet.

TABLE XXXII. AUTOMATION EXPERIMENT AXES

| Axis | Bound field | Disposition |
|---|---|---|
| Acting model | identity model id | swept (primary) |
| Chain of thought | prompt chain-of-thought flag | swept |
| Temperature | system model temperature {0.0, 0.3, 0.7} | swept |
| Knowledge base | knowledge base enabled | swept |
| Retrieval | strategy, top-k, max context tokens | if knowledge base helps |
| Retry budget | max retries {0, 1, 2} | swept |
| max_tokens, framework, prompts | various | fixed at baseline |

Representative axes from the experimentation strategy.

### D. What the Applications Contribute to the Metamodel

Taken together, the applications contribute attributes that no general benchmark provides (Table XXXIII). The migration agent contributes behavioural equivalence to a real system, the notion of a demonstrably correct unit and a ledger of fixes; the chatbot contributes retrieval and grounding attributes and the localisation of failures by layer; the automation agent contributes the distinction between claimed and effective action. Each of these is available to every model that is run on the corresponding application, which is the mechanism by which the set of measurable LLM features grows with the set of applications.

TABLE XXXIII. ATTRIBUTES CONTRIBUTED BY EACH APPLICATION

| Application | New attributes (examples) | Question answered |
|---|---|---|
| Migration | units demonstrably correct; defects; verification level; fix confirmations | does the output behave like Oracle? |
| Chatbot | retrieval recall/precision/MRR; citation validity; faithfulness; tool-argument correctness | where does grounding fail? |
| Automation | effects per run; unbacked success claims; hallucinated tool calls; premature termination | does the model act, and does it say so honestly? |
| P1 conversion | NLP aggregate; judge criteria; linter error rate; refusals | how complete is whole-file conversion? |

## VII. EVALUATION AND DIAGNOSTIC METHODOLOGY

This section answers RQ2. Its content is not a list of good intentions but a set of rules that were introduced because a stand misled its authors. We first summarise the defects, then state the rules they led to, and finally the diagnostic procedure that connects them (step M5 of the method).

### A. Defects Found by Running the Stands

Execution of the testbed revealed six infrastructure defects (Table XXXIV), including invalid planning output, omitted dataset restrictions and inconsistent settlement of queued tasks. Independent Oracle-to-PostgreSQL differential checks uncovered evaluator failures rather than migration failures (Table XXXV). Fixing these defects changed the number of units that could be reliably assessed, demonstrating why an evaluation pipeline must be diagnosed before blaming the model.

TABLE XXXIV. DEFECTS FOUND IN MAS CORE BY RUNNING IT

| Defect | Consequence | Resolution |
|---|---|---|
| strict JSON-schema output unusable for open maps | no plan could be parsed | tool-calling instead |
| settlement scheduled once on a countdown | goal stayed “running” | run requests settlement |
| orphan recovery swept queued runs | healthy work cancelled | only “running” rows, on worker start |
| registry populated only in API process | worker could not run any probe | lazy discovery |
| dropped subset not enforced | would run all 1,802 files | scope re-applied by code |
| variant ids contain slashes | prompt endpoint never matched | fixed |

Defects found during MAS Core integration tests and execution.

TABLE XXXV. DEFECTS IN THE MIGRATION TEST STAND (NOT IN THE AGENT)

| Defect | Effect | State |
|---|---|---|
| rolled-back install of a unit that raises on a test input | about 32 of 41 BLOCKED units | fixed |
| unreachable provider scored as bad translation | 109 FAILED units from one outage | fixed; now INCONCLUSIVE |
| test input crashed validation database memory | 2–4 units per run never tried | fixed |
| dependent shown wrong upstream signature | 3 callers invoked wrong procedure | fixed |
| COMMIT/ROLLBACK blocks in per-attempt transaction | 9 units not evaluable | fixed (disposable copy) |
| decomposer merges statements in 9 of 34 files | units not checkable per unit | open |

Migration evaluator defects observed during differential validation; an earlier resolver error blocked 21 dependency groups, representing 57% of the affected cascade.

### B. The Degenerate-Agent Check

The most consequential rule is simple: before trusting any aggregate metric, score the agent that does nothing and see whether it wins. Let $f$ be a candidate headline metric, $a_\varnothing$ the null agent (an empty migration, an empty answer, a run without tool calls) and $a_{\text{ref}}$ a reference agent. A metric is *admissible* as a headline only if

$$f(a_\varnothing) + \varepsilon \le f(a_{\text{ref}}),\ \ \varepsilon > 0, \tag{7}$$

and a loop that optimises against $f$ (a prompt-tuning or self-learning loop) must not be pointed at a metric that fails (7). The check caught two different metrics in the migration direction. “PostgreSQL accepted the SQL” rewards wrapping statements in a function body in which nothing is resolved, and mean behavioural fidelity rewards silence, because the similarity of two empty sequences is 1.0 and Oracle itself is silent on 11 of the first 20 curated units. An empty migration outscored the agent on every headline metric that had been published up to that point. The rule has a practical corollary used throughout the testbed: every aggregate is accompanied by the numerator and denominator and by the evidence level at which it was obtained. The four acceptance levels and their observed values are summarised in Table XXXVI.

### C. Tiers of Acceptance and the Behavioural Differential

What “accuracy” means for a migration turned out to depend on the tier at which it is measured (Table XXXVI). The reference loads on a clean database for only 35.93% of the units in isolation, 40.26% with dependency closure and 49.50% as a union. The agent's output loads less often than the reference by 6.08 percentage points ($p \approx 5 \cdot 10^{-11}$). But about 78% of every "ok" verdict proved nothing, because a function body in PostgreSQL is accepted without resolving the references inside it; restricted to verdicts that actually resolved their references, both headline acceptance differences collapse to nulls (−6.15 pp becomes −0.45 pp, $p = 0.39$). Finally, execution-judged against a captured real Oracle run, no non-trivial exact behavioural match was found on 20 curated units until the fifth prompt fix. A prompt loop pointed at raw acceptance would have learned to hide statements inside function bodies; this is exactly what separated two real prompt versions early on.

TABLE XXXVI. Four Tiers of "Accuracy" in the Migration Direction

| Tier | Measures | Headline |
|---|---|---|
| Corpus filter | does the reference load | 35.93% / 40.26% / 49.50% |
| Candidate vs reference | does the agent's output load | −6.08 pp, $p \approx 5 \cdot 10^{-11}$ |
| Acceptance | what "PostgreSQL accepted it" proves | ≈78% of verdicts proved nothing |
| Behavioural differential | does it behave like Oracle | 0 non-trivial exact matches on 20 units, until fix 5 |

Observed migration acceptance measurements; the three percentages describe reference installation in isolation, with dependency closure and as their union.

In the depth study the verdict is therefore read through three channels (Table XXXVII). The agent's own status says whether a unit installed and ran. A differential of errors says whether the translation raises an error Oracle does not, or misses one Oracle raises. A per-unit output comparison replays each translated unit in order on PostgreSQL and compares with what Oracle printed. Every FAILED unit and every disagreement between channels is then read by hand, and where a judgement depends on Oracle's behaviour it is executed on a real Oracle instance rather than assumed. Two consequences follow. A translation that correctly raises the same error Oracle raises must not be scored FAILED; and when the agent was taught Oracle's SELECT … INTO semantics, ACCEPTED went *down*, since two translations that had silently returned NULL now raised Oracle's error.

TABLE XXXVII. Reading a Migration Verdict Through Three Channels

| Channel | Question | Instrument |
|---|---|---|
| Status | did it install and run? | agent's own report |
| Errors | extra or missing errors relative to Oracle? | behaviour differential tool |
| Output | does each unit print what Oracle printed? | per-unit transcript replay |
| Manual | is the judgement Oracle-dependent? | execute on real Oracle |

Observed differential verification channels used in the migration experiment.

### D. The Score Record, the Source Hierarchy and Control Arms

The programme fixed what a score is in a decision record before the stands were built, and the testbed inherits it. A score is a record, not a number: it carries the axis, the value and its type, the source, the arm, the number of samples, an explicit and non-blank denominator, the run and the requested model identifier and a timestamp (Table XXXVIII). The measurement μ of Equation (2) is a projection of this record. Two details were learned from defects. A denominator that is present is not a denominator that is stated: a whitespace string once passed a length check and produced a record that was technically complete, said nothing and was incomparable with anything. And the requested model identifier, not the one the gateway answered with, keys the record, because the gateway may return a prefixed variant of the same deployment; the returned identifier belongs in the trace as evidence of what actually served the call, which makes a silent fallback visible.

TABLE XXXVIII. Fields of a Score Record

| Field | Rule |
|---|---|
| axis, value, value_type | boolean, categorical, numeric or text; text is evidence, never the value |
| source | one of four levels (Table XXXIX) |
| arm | mandatory: system under test or control with the component removed |
| n, denominator | explicit and non-blank; validated on content, not presence |
| run_id, model_id | model_id is the requested deployment; the returned id is traced |
| timestamp | time of the measurement |

Fields established by the implemented measurement schema and evidence policy.

Scores come from four sources, ordered by what they can verify (Table XXXIX). A lower level never overrides a higher one on the same question; where levels disagree the higher level is the verdict and the disagreement is recorded, because it is usually the most informative thing in the run. A judge may triage, prioritise and explain; it may not decide, and it is never an acceptance gate. The rule has a numerical justification in the migration programme that preceded the testbed, which kept LLMs out of its judging loop: executability moved from 72.0% to 97.3% under a specification, while behavioural exact match on the same artefacts was 2 to 6 of 20, and 42 structured probes against a live Oracle instance returned 19 correct results in both the specification and the control arm [46]. A judge reading the generated code would have seen well-formed PostgreSQL in both arms and scored them close; the difference was not in the text.

TABLE XXXIX. Source Hierarchy of Scores

| Level | Source | Use |
|---|---|---|
| 1 | verifiable outcome: execution, differential comparison with a reference system, deterministic check against labelled ground truth | decides |
| 2 | human annotation | small-n; only source for intent |
| 3 | static programmatic checks (string, structure, schema, compliance) | objective about the artefact, not its behaviour |
| 4 | LLM-as-a-judge | advisory; never an acceptance gate |

Correspondence to evidence levels: level 1 ≈ L4/L2; level 3 ≈ L1; level 4 ≈ L3 when the judge compares with a reference.

Two further rules complete the model. A score without a control arm is not reportable: one axis scored 100% in every arm and carried no information about the system at all, and where a control arm is impossible the score is labelled "uncontrolled" and cannot support a claim of improvement. And compliance is reported beside outcome, never instead of it: in one experiment adding a specification raised the compliance axis it targeted while the number of invalid declarations rose from 0 in the control to 10, and a composite score would have shipped that release as a win. Finally, offline measurements with a fixed dataset, a control arm and thresholds per axis are the only ones that may gate a change or support a claim, whereas online traces without ground truth detect drift and nominate candidates for the offline set. The price of this discipline is evaluation cost: control arms double generation and thresholds need repeats, which the decision record estimates at two to three times the cost of a naive evaluation, and we accept it.

### *E. Fail-Closed Reporting and the Verdict Taxonomy*

A metric that cannot be computed is reported as an error, not as zero. In the chatbot smoke run the semantic metrics requested by the dataset were reported as errors because no judge was configured; in the interface a run with incomplete token usage is shown with a warning rather than a zero cost. An infrastructure failure is not a verdict on the model: a network outage once turned into 109 FAILED migration units and is now recorded as INCONCLUSIVE; an unreachable triage agent aborts instead of guessing a retry; an unreachable aggregator yields an absent verdict. Each stage has a distinct vocabulary (Table XL).

TABLE XL. VERDICT TAXONOMY BY STAGE

| Stage | Verdicts | Outage handling |
|---|---|---|
| Migration unit | ACCEPTED, FAILED, BLOCKED, INCONCLUSIVE, UNSUPPORTED | INCONCLUSIVE |
| Fix cycle | CONFIRMED, PARTIAL, FAILED, INCONCLUSIVE | INCONCLUSIVE |
| Chatbot run | PASS, FAIL, degraded | error, not zero |
| Goal assessment | trustworthy: yes, no, absent | absent |
| Triage | retry, resume, abort | abort |

### *F. Noise Floor, Repeats and Comparability*

Repeated runs of an identical plan do not return identical numbers. Two identical goals over 20 cases of the P1 stand differed by 0.056 in conversion accuracy; a semantic judge changed its verdict on 16.2% of repeated items at temperature 0; and in the specification benchmark of the same group arm deltas across identical re-runs had a median spread of 14.4 percentage points and a worst case of 32.7 [46]. A judge that grades identical retrieved context twice disagrees with itself too: on a retrieval-augmented evaluation with identical context in two databases, answer relevancy came out as 0.4923 and 0.4111, a measured noise floor of up to 0.081, which turned an earlier reported gap of 0.834 against 0.519 on six cases into noise. Such variability, a known obstacle to reproducible AI research [55], [56] and to meaningful significance claims [53], [54], has two consequences for the testbed. Numbers from the first P1 test runs were deliberately not published, because a figure in a report tends to be quoted as a statement about the model. And a difference is reported as an effect only if it exceeds the spread of the repeats of *that axis*:

$$\mathrm{report}(\Delta) \Leftrightarrow |\Delta| > s + \tau,\ \ s = \mathrm{range}(v_1) + \mathrm{range}(v_2),\ \ v_1 \sim v_2, \qquad (8)$$

Here, Δ denotes the difference between the mean scores of the two experimental arms, $\mathrm{range}(v_i)$ is the difference between the maximum and minimum scores obtained from independent repetitions of arm i on the same fixed evaluation dataset, and τ is a numerical tolerance introduced to account for floating-point representation. The two arms must satisfy the comparability relation defined in Equation (4), with the investigated treatment being the only intended difference between their experimental configurations.

The criterion is evaluated separately for each metric rather than using a single global threshold. A difference is considered reportable only when its magnitude exceeds the combined within-arm ranges and the numerical tolerance. The sum of the two ranges provides an upper bound on the variation of pairwise differences under arbitrary re-pairings of the repeated measurements. If the observed difference does not exceed this threshold, the coarse rule does not establish an effect. This outcome should not be interpreted as evidence that the two configurations are equivalent.

The need for a metric-specific threshold is illustrated by the observed experiments. A nominal improvement of 18.7 percentage points was not retained because the corresponding repeat spread reached 20.3 percentage points, whereas an improvement of 17.4 percentage points exceeded the 12.3-point spread observed for another metric. In a separate case, the estimated effect changed sign between nominally identical repetitions, from +23.3 to −9.3 percentage points. Pairing stochastic repetitions solely by their positions can also underestimate variability: one observed pairing yielded five identical differences of +2.86 percentage points, while alternative pairings produced a spread of up to 5.71 percentage points.

A zero observed spread must not be interpreted as proof of high precision, particularly when only a few repetitions are available. If the difference and the spread are both zero, no improvement or regression is reported. The numerical tolerance prevents floating-point rounding near the decision boundary from generating a spurious result. Comparisons with fewer than two independent repetitions per arm are classified as unreplicated according to Equation (9). The decision rule is deliberately conservative and does not replace a calibrated statistical test when sufficient independent measurements are available.

The same principle appears in the automation experimentation strategy, where an effect smaller than the repeat spread is not reported as an improvement, and the paired comparisons of the migration depth study use exact tests where a count is available (for example a McNemar test, $p = 0.208$, for whole-file versus split migration in Section VIII). Equation (8) is deliberately conservative, and the simulation of Section VIII-I shows how conservative: with five repeats per arm it needs a difference of about 5.8 per-run standard deviations to be detected with 80% power, and, because the range of a sample grows with its size, it becomes *less* sensitive as repeats are added. We therefore use it as the guard against false effects and add a second tier for the cases in which the repeats allow a calibrated test. The verdict has three states: *effect* when Equation (8) holds; *suggestive* when it does not hold but a Welch test of the repeats rejects equality at level $\alpha = 0.05$; and *no result* otherwise,

$$\mathrm{verdict} = \text{unreplicated if } r < 2;\ \text{effect if } r \geq 2 \wedge |\Delta| > s + \tau;\ \text{suggestive if } r \geq 2 \wedge |\Delta| \leq s + \tau \wedge pW < \alpha;\ \text{no result otherwise.} \qquad (9)$$

Here r is the number of repeated independent runs per arm, and pW is the Welch p-value. The "suggestive" category is advisory and cannot gate a change; with small r its calibration must be treated cautiously. A single run is reported as "unreplicated", not as evidence of an effect, even if its nominal range is zero. The reference implementation described in Section VIII-K illustrates these rules; no single-run claim is promoted to a confirmed improvement.

**Algorithm 7.** Noise-aware comparison and Pareto selection.

**Input:** models $M$, task constraints $H$, objectives $O$ with directions, metamodels Ψ
**Output:** Pareto set, ties, exclusions with reasons
$M' \leftarrow \varnothing$; $Excl \leftarrow \varnothing$
**for each** $m \in M$
  **if** some $o \in O$ or constraint in $H$ unmeasured **then** $Excl$ += ($m$, "not measured"); **continue**
  if evaluation contexts differ outside the declared treatment then Excl += (m, "other context"); continue
  **if** violates $H$ **then** $Excl$ += ($m$, constraint); **continue**
  $M'$ += $m$
P ← exact non-dominated set of M′ using direction-adjusted recorded values
for each pair and objective: if both have ≥2 repeats and |Δo| ≤ range1 + range2 + τ, annotate "uncertain difference"
if repeats are unavailable, annotate "unreplicated"; never treat uncertainty annotations as transitive equalities
return P, uncertainty annotations, exclusions and evidence counts

### G. Gated Fix Cycles

In the migration direction fixes are not tuned against a score, because every score that was tried rewarded the wrong behaviour at some point; an empty translation once outscored the agent. Each fix instead passes through a gated cycle (Fig. 14, Algorithm 8). Evidence comes from a real run: a named Oracle construct, the units it breaks, verified on both databases. The change and three tiers of checks are pre-registered before the run. Tier A is the gate, a deterministic check on the generated SQL, for example that no DATE type survives translation. Tier B, reported only, asks whether the target units improved. Tier C explains every other unit that changed status. A run then yields an honest verdict, and failed fixes are kept in the ledger.

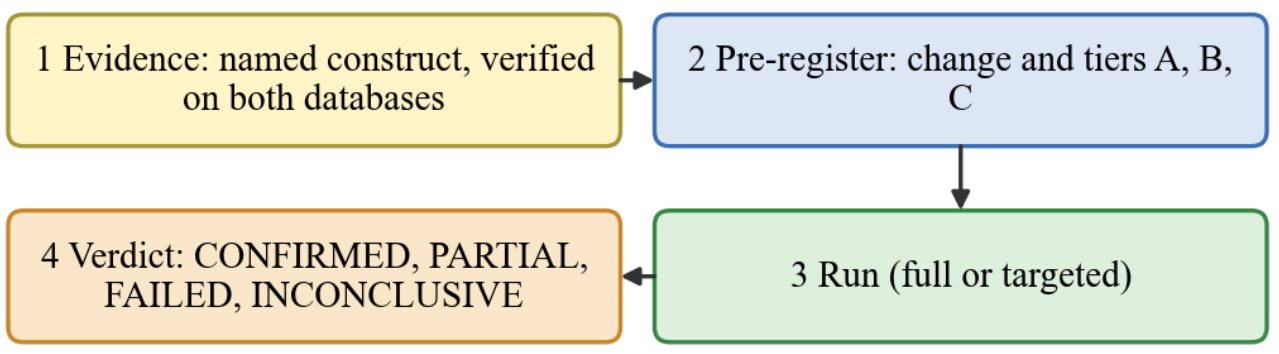

*Tier A: deterministic gate on generated SQL. Tier B: target units, reported. Tier C: every other status change explained. Failed fixes stay in the ledger.*

Fig. 14. Gated fix cycle: evidence, pre-registration, run, verdict.

**Algorithm 8.** Gated fix cycle for one construct.

**Input:** evidence $E$ (construct, affected units, both-database verification)
**Output:** verdict recorded in the ledger
pre-register change $\Delta$ (prompt section or code), tier A gate, tier B targets, tier C scope
commit pre-registration // before the run
$run \leftarrow$ full or targeted run with $\Delta$
**if not** TierA($run$) **then return** FIX_FAILED
**if** infra error **then return** INCONCLUSIVE
report tier B (targets improved?); explain every tier C status change
**return** verdict ∈ {FIX_CONFIRMED, FIX_PARTIAL, FIX_FAILED} from tiers B and C
failed fixes remain in the ledger

The ledger of fixes from FIX-011 to FIX-056 contains 46 cycles: 26 confirmed, 8 partial, 3 failed, 1 inconclusive and 8 from the last two days of the observation period whose Tier A check holds and whose verdict awaits the next full run. One example shows why checking each unit against Oracle matters: a regression caused by an earlier fix (the cheaper model began to keep :name bind placeholders in dynamic SQL, most likely after the NULL-concatenation fix) retained its ACCEPTED status and was visible only by comparing with what Oracle printed.

### H. Cost Model and Cost per Evidence

Because the cost of measuring is itself a result, we define it explicitly. For a run of model $m$ on application $a$ the three spends of Table XVII are accumulated separately, and the cost of the measurement is the vector ($C_t$, $C_j$, $C_g$), never its sum. For decision purposes a scalar is sometimes needed; the natural one is the cost per unit of *evidence* at a chosen level, for example per demonstrably correct unit, Application-specific experimental costs are listed separately in Table XLI.

$$c_{dc} = (C_t + C_j + C_g) / N_{dc}, \tag{10}$$

where the sum is taken explicitly and labelled as the cost of the experiment, not a property of the model. With the reported figures, a full production-routed migration run on the 216-unit corpus costs about USD 3 to 4, and 108 units were demonstrably correct, which gives a derived cost of about USD 0.03 to 0.04 per demonstrably correct unit for the experiment as a whole; the P1 first cycle cost about USD 4 for 1,802 files (about USD 0.002 per file) of which the judge accounted for about three quarters. These derived numbers are included to show the form of the computation; they are not comparable across applications, because the evidence levels differ.

TABLE XLI. Reported Costs of the Experiments

| Experiment | Reported cost | Derived |
|---|---|---|
| P1 first cycle, 1,802 files | conversion ≈ USD 1.2; judge ≈ USD 3; total ≈ USD 4 | ≈ USD 0.002 per file; judge ≈ 2.5 × conversion |
| Migration depth, full run, 216 units | ≈ USD 3–4 | ≈ USD 0.03–0.04 per demonstrably correct unit (108) |
| Migration breadth, 1,448 tasks | 10.0 M tokens (agent) | 3.0% fewer tokens than the cheap-model baseline |
| Intent classification, one repeat | 72 model calls; ≈ 37.4 k tokens | mean 1,385 ms per call |
| Shared proxy key (several users) | ≈ USD 331 of USD 500 rolling 30-day budget on 2026-09-23 | quota may decide before price |

Measured costs and arithmetically derived unit costs; approximate totals reflect source rounding.

### I. Statistical Treatment

The stands report counts with denominators and use exact tests where the data are paired counts. The comparison of whole-file against split migration used a McNemar test on the discordant pairs of exact matches ($p = 0.208$), and the acceptance-tier comparison reported a difference with $p \approx 5 \cdot 10^{-11}$ that collapsed to $p = 0.39$ once unresolved verdicts were removed. Two practices are worth stating. First, an aggregate is accompanied by its numerator and denominator and its evidence level, so that a reader can see that 284 of 1,448 tasks, not 1,330, reached execution in PostgreSQL. Second, results that depend on a single run are labelled as such, and no confidence interval is claimed where the number of repeats is one. The coarse noise-floor rule of Equation (8) is a substitute for a test, not a stronger version of it: it is applied when a repeat spread exists and the number of repeats is too small for a formal test, which is the case for most of the application experiments.

### J. Instrument-First Diagnosis

When a number is surprising, the question is where the cause lies. The measured defects of Tables XXXIV and XXXV show that a stand can mislead for a long time before the model is at fault: a resolver bug held up 57% of a dependency cascade, a network outage turned into 109 failed units, a test input crashed the validation database and left the following units untried. The diagnostic procedure of the method therefore tests the instrument before the system (Fig. 15, Algorithm 9).

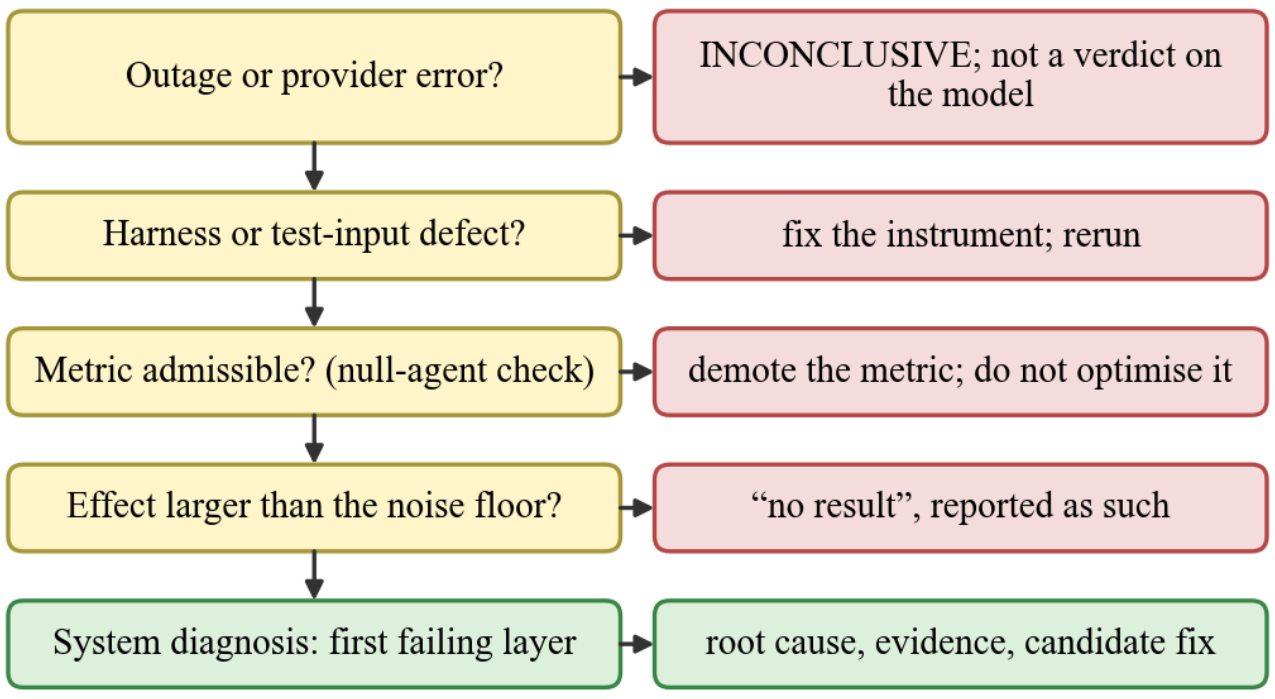

*Left: question asked; right: outcome if the answer is yes. A no answer moves down.*

Fig. 15. Instrument-first diagnostic ladder: each question has an outcome if the answer is yes; a no answer moves down to the next question.

**Algorithm 9.** Instrument-first diagnosis of a surprising result.

**Input:** result $R$, run log, metric $f$, repeats, application layers $L$
**Output:** diagnosis with evidence and action
**if** outage or provider error in log **then return** INCONCLUSIVE; rerun
**if** harness or test-input defect reproduces without the model
  **then return** instrument defect; fix instrument; rerun // e.g. rollback, memory crash
**if not** admissible($f$) // null agent must lose, Eq. (7)
  **then return** demote $f$; do not optimise it
**if** verdict($R$) = no result **then return** no result; add repeats
**for each** layer $l$ **in** $L$ **in causal order**
  **if not** passes($l$) **then return** root cause $l$, evidence, candidate fix
**return** no defect found; record as pass

It asks, in order, whether an outage or provider error explains the result (then the verdict is inconclusive and the units are rerun), whether a harness or test-input defect explains it (then the instrument is fixed and the run repeated), whether the headline metric passes the null-agent check of Equation (7) (then it may be optimised), whether the difference exceeds the noise floor (else it is reported as no result), and only then diagnoses the system itself, layer by layer, in the causal order of Fig. 12. The order is not arbitrary. An instrument defect can masquerade as any system defect, whereas a system defect cannot masquerade as an instrument defect, so ruling out the instrument first is the only ordering that does not confound the two.

### *K. Diagnostic Knowledge Base and Symptom Taxonomy*

Diagnosis is only repeatable if the knowledge about causes is kept apart from the diagnosed system, which is the lesson of the knowledge-based diagnostics of computer facilities [81], [82]. In the testbed this knowledge takes two forms. The first is the symptom taxonomy of Table XLII: for a symptom that was actually observed, the candidate causes, the evidence that separates them and the resulting action. The second is the fix ledger of the migration direction, an append-only record of 46 cycles in which every entry states a named construct, the units it breaks, the predicates that gate it, the verdict and the cost. A search service over such a base, with explicit criteria for matching the terms of different entries [84]–[86], is what turns a collection of reports into a diagnostic instrument: a new failure is first matched against known symptoms before anyone reads a trace.

### *L. Negative Controls*

A negative control is an instrument that must produce a null result; if it does not, the instrument is broken. The stands use several (Table XLIII). The null agent tests whether a metric rewards emptiness. A keyword or rule baseline as the control arm of a classifier shows what an instrument reports when the system contributes nothing; in the intent experiment it scored 0.000, which satisfied the control formally and carried almost no information. A no-specification arm shows what the models do without the component under study. And in the exploratory analysis of the Oracle script dataset, a random control column was added to the features of a random-forest model; a real feature is taken to carry signal only if its importance exceeds that of the random column, which confirmed that the core SQL and PL/SQL counts carry genuine predictive weight rather than noise. The same device applies to any attribute whose predictive value is in question: add a random attribute and require it to lose.

TABLE XLII. SYMPTOM TAXONOMY OBSERVED IN THE STANDS

| Application | Symptom | Candidate causes | Separating evidence | Action |
|---|---|---|---|---|
| P1 | low recall, many refusals, short outputs | truncation by output cap; voluntary abbreviation | completion length against the cap (maximum 7,548 of 32,768 tokens) | report abbreviation as model behaviour; add a truncated status |
| migration | FAILED unit | defect; faithful raise; invalid test input; harness limitation; no-op | execute on the real Oracle instance | reclassify; fix defect or harness |
| migration | BLOCKED unit | upstream failure; upstream correct but raising by design | trace to the upstream and its failure stage | keep a correct upstream installed for dependants |
| migration | mass of FAILED after one event | provider outage; model regression | provider status; token usage of failed calls | mark INCONCLUSIVE; rerun |
| migration | ACCEPTED that Oracle contradicts | silently wrong translation | per-unit output comparison; second model disagrees | fix construct; do not count as correct |
| chatbot | FAIL scenario | execution; routing; retrieval; grounding; answer quality | first failing layer in causal order | targeted re-test with an approved alternative |
| automation | no effect but success stated | unbacked claim; candid minimalism | final answer against recorded effects | report claim–effect gap |
| automation | tool error on an invented identifier | hallucinated argument | argument against the frozen world state | count in hallucinated-call rate |
| judge | different scores on identical context | judge noise | repeat on identical context | report noise floor; do not gate |
| comparison | two systems return identical results | true equivalence; comparison that cannot differ | does the experiment exercise the index or path under test? | redo at realistic scale |

Symptoms are derived from the execution and validation observations analysed in Sections VII and VIII.

TABLE XLIII. NEGATIVE CONTROLS USED IN THE STANDS

| Control | What it must show | Observed |
|---|---|---|
| null agent (empty migration, empty answer) | headline metric does not reward emptiness | empty migration outscored the agent on published metrics |
| keyword or rule baseline (control arm) | what the instrument reports with no system | 0.000 on intent; 100% on an axis in every arm carried no information |
| no-specification arm | effect attributable to the component studied | executability 72.0% → 97.3%; agreement 83.1% → 83.8% |
| random control feature | feature importance exceeds noise | core SQL and PL/SQL counts exceeded the random column |
| exact scan against index | comparison can come out differently | identical rankings were guaranteed at 31 chunks |
| identical repeat | noise floor of the instrument | judge relevancy 0.4923 against 0.4111 on identical context |

Negative controls used in the evaluation protocols; see Sections VII-B, VII-F and VIII-G.

## M. Cross-Model Disagreement as a Defect Detector

A property that is specific to a testbed that runs many models on the same application is that disagreement between models is itself a signal. If two models produce different behaviours on the same unit under comparable evaluation conditions, the disagreement identifies a candidate for further verification using the stronger evidence levels defined in Table IV. Behavioural differences do not necessarily imply that either model is incorrect, since multiple valid outputs may satisfy the same task requirements. In the migration direction the first silently wrong accepted translation was found this way: a second model, tried for cost reasons, disagreed with the baseline on one unit. The testbed therefore stores, for every application context, the per-unit outcomes of every model, so that a disagreement query is a join and not a new experiment. The signal is asymmetric: agreement is weak evidence of correctness, because models share training data and defaults, whereas disagreement is strong evidence that something deserves a look. This asymmetry agrees with the observation that the quality of a unit test suite determines how much an apparently correct generation can be trusted [35], and it motivates the use of verified outcomes (levels 1 and 2 of Table XXXIX) to settle the disagreement.

## N. Separation of Roles and Leakage Hygiene

The model under test is kept apart from the models that judge and decide. In the chatbot direction the judge and the controller have separate roles, ground truth is never passed to the chatbot, and in the specification-evaluation work of the same group ground truth is mechanically sealed from inference and checked with canary and label-permutation tests, with judges compared against trivial baselines such as specification length [45]. These practices follow the recommendations on contamination [49], [51], [52] and on judge bias [43], [44] and are the reason why the testbed prefers deterministic computation wherever one exists. A further consequence concerns cost: because the judge's spend can be a substantial fraction of the model's, a cross-vendor judge to control same-vendor self-preference is still an open item of the P1 stand.

# VIII. EXPERIMENTS AND RESULTS

This section answers RQ3 using the recorded experimental measurements from the three application-specific evaluation environments. Dataset versions, configurations, denominators and evidence levels are retained with each result. Because experiment designs differ between applications, cross-application numerical comparisons are not interpreted as a single model ranking. Values for which only one run is available are identified explicitly. The full comparison set and experimental denominators are summarised in Table XLIV.

TABLE XLIV. SUMMARY OF THE EXPERIMENTS REPORTED

| ID | Application | Data | Models / variables | Runs | Section |
|---|---|---|---|---|---|
| E1 | P1 conversion | 1,802 Oracle files | gpt-4.1-nano; judge gpt-4.1-mini | 1 full cycle | VIII-A |
| E2 | Automation | 49 samples (pilot.v4) | 4 frameworks, gpt-4o | 1 per framework | VIII-B |
| E3 | Automation | 49 samples | 6 models, LlamaIndex | 1 per model | VIII-B |
| E4 | Automation (intent) | 40 frozen cases | 3 deployments, 2 arms | 5 repeats | VIII-C |
| E5 | Chatbot | mvp_scenarios v6; 6 / 5 / 10 scenarios | retrieval, Top-K, reranker, temperature, 9 models | 1; 15 for finalists | VIII-D |
| E6 | Chatbot evaluation agent | 10 scenarios | v1 diagnostic; v2 adaptive | 1 campaign each | VIII-D |
| E7 | Migration breadth | 1,000 pairs, 1,448 tasks | agent, cheap LLM, rules, whole-file | 1 per configuration | VIII-F |
| E8 | Migration depth | 216 units, 34 files | production routing; 2 prompt versions | 2 full runs | VIII-F |
| E9 | Fix cycles | FIX-011…056 | prompt sections and code | 46 cycles | VIII-F |

Experiments, sample sizes and evaluation protocols included in this study.

## A. P1 Conversion: First Full Cycle

The P1 stand (Fig. 16) is split into two independently triggered phases: dataset conversion (dataset, prompt, model, database) and conversion evaluation (stored results, validation, metrics, judge, annotators, report). Because evaluation runs entirely from the database, a no-judge evaluation completed in about 7 s with the conversion backend stopped; this property allows the expensive GPU box planned for generation to be powered off afterwards. The GPU service does not exist yet; its slot is filled by the gateway, and the swap boundary is a single environment variable and a documented generation contract.

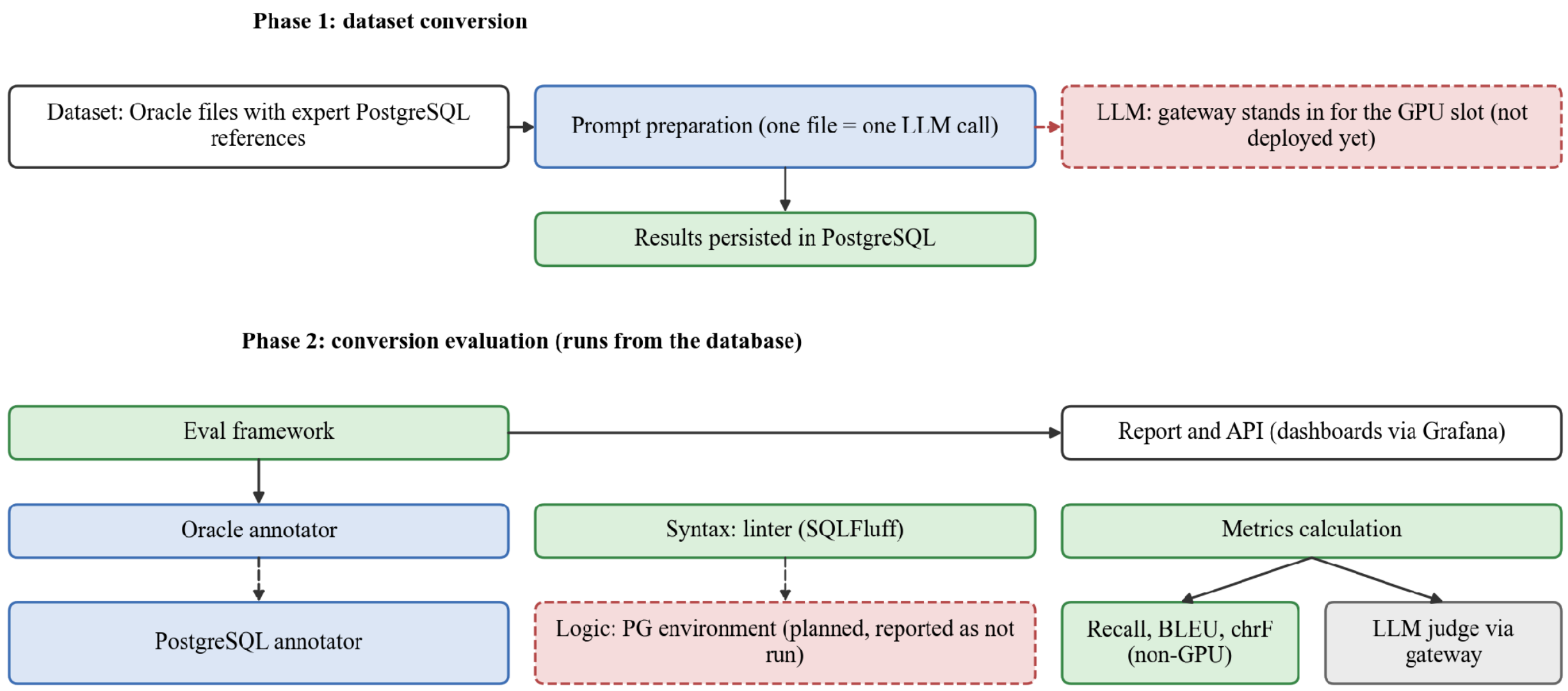

Fig. 16. Two-stage conversion evaluation: generation and subsequent evaluation from stored outputs; the live PostgreSQL logic check is outside the completed P1 experiment.

The first full cycle converted 1,802 of 1,802 files with gpt-4.1-nano in 48 min with zero skips, and the evaluation took 14 min 38 s (Table XLV). The whole cycle cost about USD 4 on the caller's own key, of which the judge, gpt-4.1-mini, accounted for about USD 3 and the conversion for about USD 1.2. The aggregate NLP score was 0.513 against 0.736 on the previous stand that used chunked retrieval; the linter-based syntax error rate improved from 14.87% to 11.65%, and refusals rose from 2.94% to 8.32%. This is not a regression of the stand but the finding: without chunking, the model does not truncate large files mid-token, it abbreviates them voluntarily. About 69 files with references above 5 KB received under 35% of the expected output; one 59.6 KB reference was answered with 269 bytes and a remark that complete conversion required thorough rewriting. The short outputs parse cleanly, which explains the better syntax figure, and fail recall, efficiency and the refusal detector, so the metrics punish the behaviour fairly. Direct mode therefore measures a capability that the chunked flow could not: honest whole-file conversion. The distribution of evaluator scores for this completed run is shown in Fig. 17.

TABLE XLV. P1 First Full Cycle with gpt-4.1-nano

| Quantity | Value |
|---|---|
| Coverage | 1,802 / 1,802 converted, 0 skipped, 48 min |
| Tokens, conversion | 4.67 M input, 1.81 M output; about USD 1.2 |
| Output ceiling hit | 0 files (maximum completion 7,548 tokens) |
| NLP vs references | Recall 0.725; BLEU 0.249; chrF 40.44; aggregate 0.513 |
| Judge (0–10) | correctness 7.50; functional equivalence 7.36; syntactic validity 9.02; clarity 7.59 |
| Linter | NSER 11.65% (210 files, 214 errors); 9,402 warnings |
| Annotators | mean file efficiency 63.07; refusals 8.32% (150 files) |
| Evaluation time | 14 min 38 s; judge cost about USD 3 |

Single run; judge gpt-4.1-mini over all 1,802 items with 0 failures. BLEU [39] and chrF [40] are reference-based.

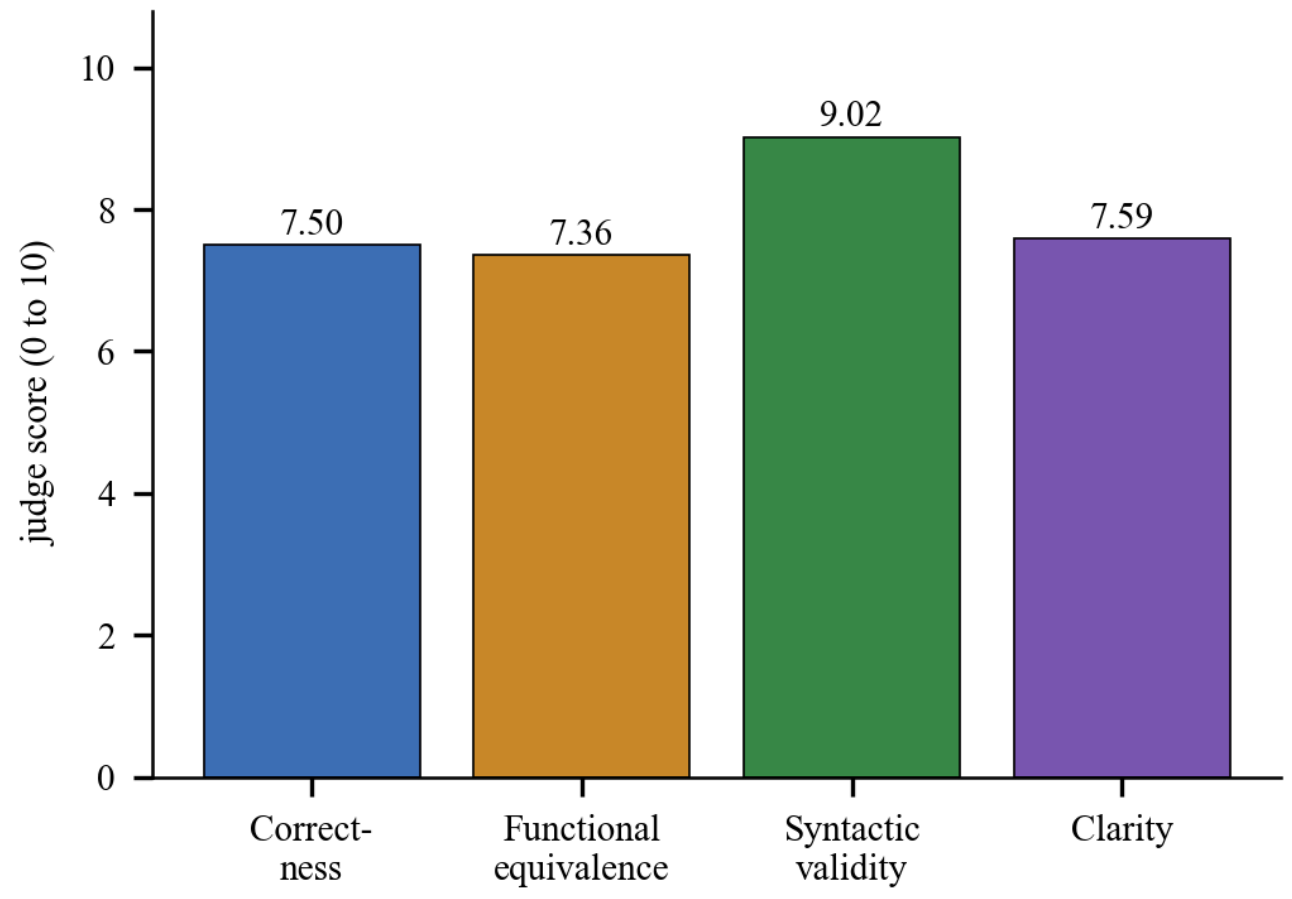

Fig. 17. Judge scores for the P1 first cycle (single run, 1,802 items).

A reliability incident belongs to the methodology rather than to the model. The default metric set initially contained translation edit rate, whose block-shift edit distance is quadratic in file length; it ran for more than 57 minutes without finishing on the full dataset, held the service's CPU, and the orchestrator's one-hour timeout reported a misleading "service unreachable". The metric was removed from the default set (it remains available on request) and the failure message was rewritten to name the actual cause.

### B. Automation: Frameworks and Models

In E2 the 49 pilot samples were executed once per framework with gpt-4o through the gateway, a ten-step limit, the knowledge base off and no retries. The results (Table XLVI, Fig. 18) cover execution outcomes only, without ground-truth scoring. Completion status does not separate the frameworks: every run returned a final answer, none reached the step limit and no tool call failed. LangGraph, LlamaIndex and AutoGen behave almost identically; they usually perform one read and then answer, and on cross-system synchronisation and most escalation samples they write nothing while many of their answers claim the work was done. CrewAI does the most work, producing the most writes and the fewest unbacked claims, at about four times the token cost, which comes mostly from a larger per-call prompt. LlamaIndex is the cheapest and matches LangGraph's effects almost exactly.

TABLE XLVI. Framework Comparison (gpt-4o, 49 Samples, One Run Each)

| Framework | LLM calls | Tool calls | Tokens | Effects | No effect | Unbacked |
|---|---|---|---|---|---|---|
| LangGraph | 2.1 | 1.1 | 2,260 | 0.5 | 27 | 14 |
| LlamaIndex | 2.0 | 1.0 | 2,055 | 0.5 | 27 | 14 |
| AutoGen | 2.0 | 1.0 | 2,210 | 0.4 | 29 | 16 |
| CrewAI | 2.5 | 1.5 | 8,942 | 0.7 | 21 | 5 |

All 49 runs completed in each framework. “Unbacked”: run produced no effect but the final answer states the work was done (heuristic; may over-count when an answer quotes past events). Means per run.

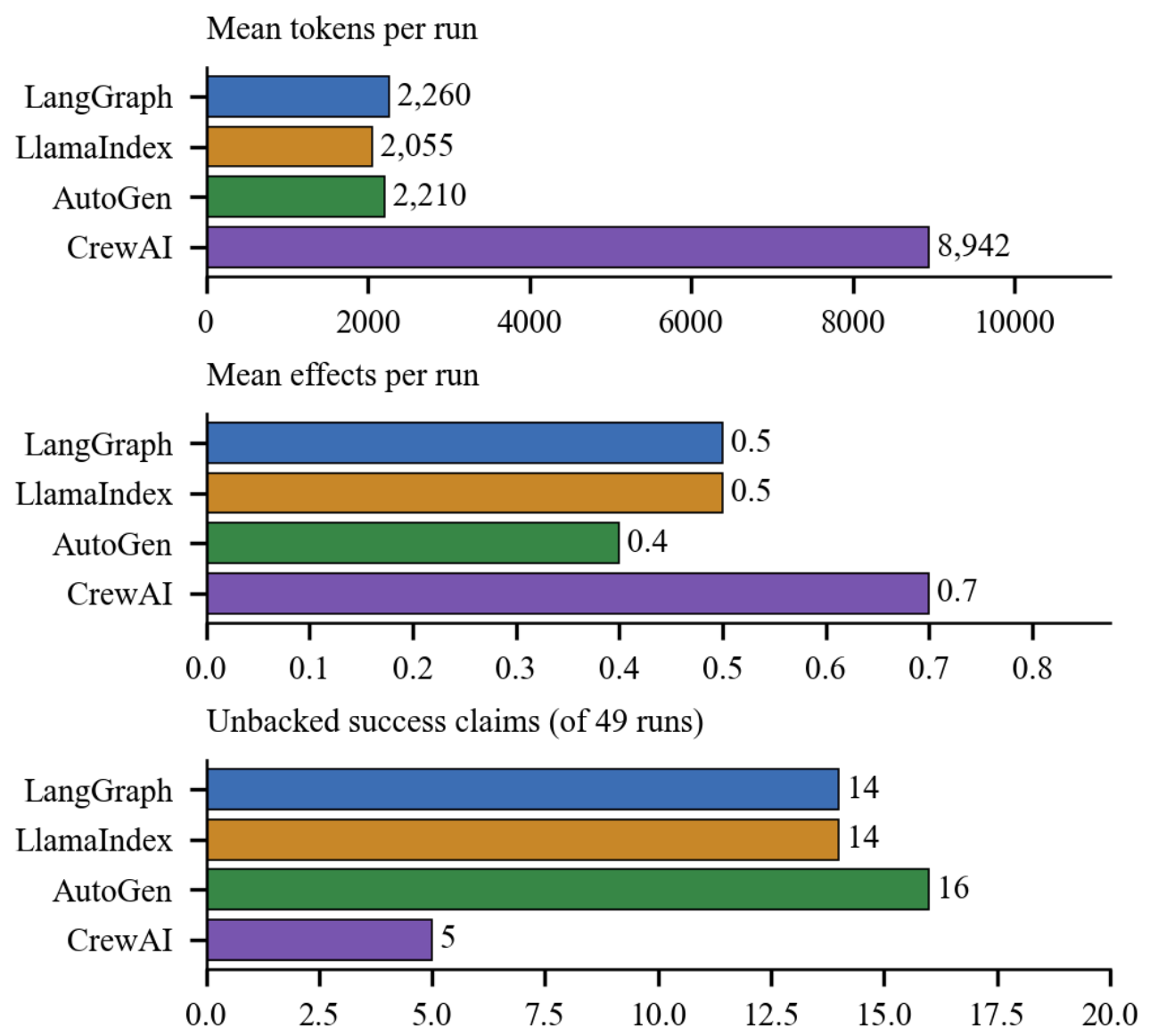

Fig. 18. Framework comparison: tokens, effects and unbacked success claims.

In E3 the framework was fixed to LlamaIndex and only the model was varied (Table XLVII, Fig. 19). The models fall into two groups. gpt-4o, gpt-5.5, Haiku 4.5 and Gemini 3.1 Pro do the minimum, with about one tool call and 0.5 effects per run. Sonnet 4.6 and Gemini 2.5 Pro carry out the multi-step synchronisation, escalation and incident-triage tasks. Unbacked success claims vary by model: gpt-4o, Haiku 4.5 and Gemini 3.1 Pro often report work that was not done; gpt-5.5 does as little but mostly says so; Sonnet 4.6 and Gemini 2.5 Pro have none. Sonnet 4.6 is the most thorough and the most expensive at about 9.5 k tokens per run; it also takes actions nobody asked for, such as 13 new Confluence pages, and its three tool errors come from invented page identifiers.

Gemini 2.5 Pro is cheap but sent no email on any email_compose sample, and its unusually small reported first-call prompt should be checked before its numbers are trusted. Haiku 4.5 costs more than its output suggests, since the Claude deployments report a larger first-call prompt, and it never adds Jira comments.

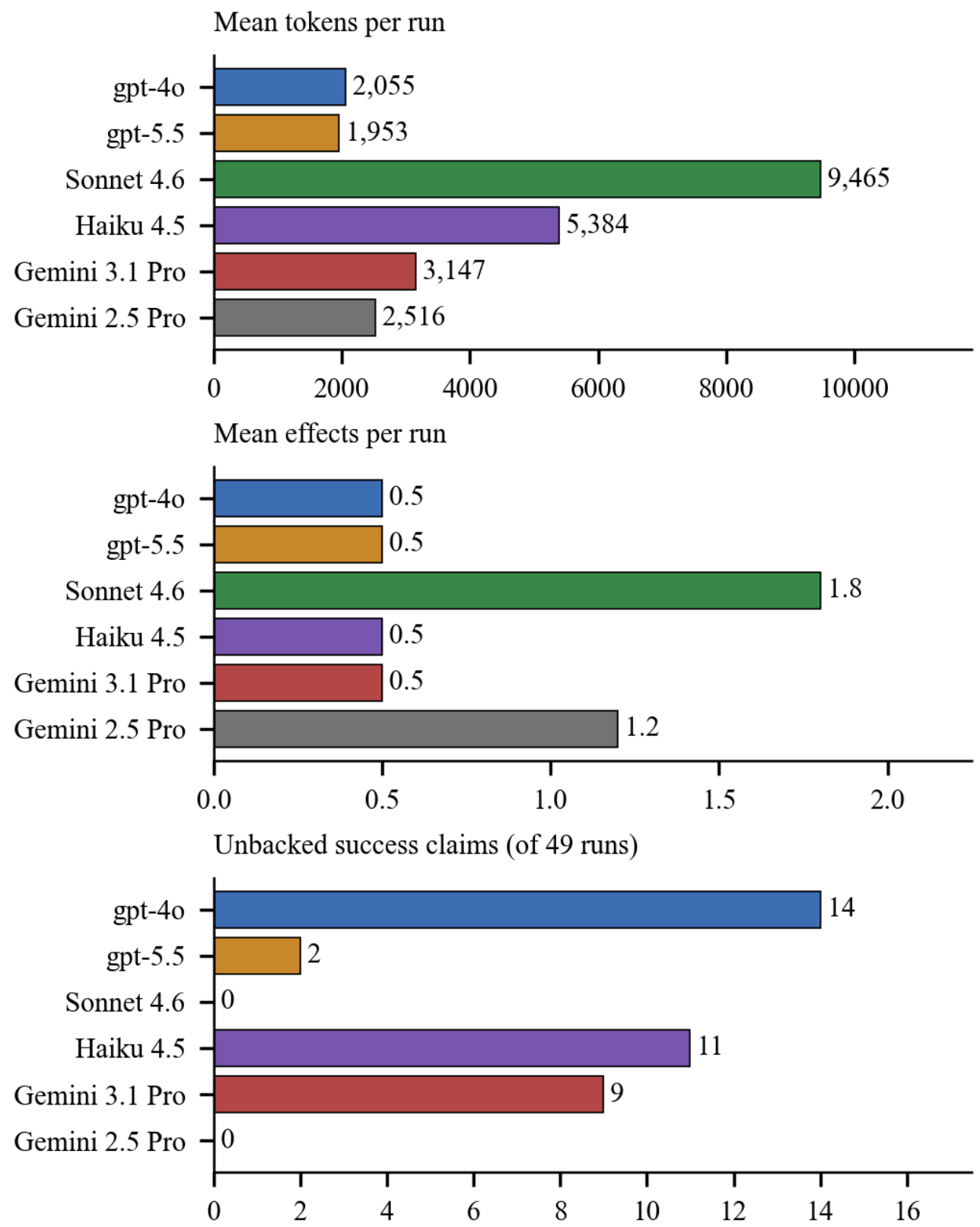

Fig. 19. Model comparison under one framework: tokens, effects and unbacked claims per run.

TABLE XLVII. Model Comparison (LlamaIndex, 49 Samples, One Run Each)

| Model | LLM calls | Tool calls | Tool errors | Tokens | Effects | No effect | Unbacked |
|---|---|---|---|---|---|---|---|
| gpt-4o | 2.0 | 1.0 | 0 | 2,055 | 0.5 | 27 | 14 |
| gpt-5.5 | 1.8 | 0.9 | 0 | 1,953 | 0.5 | 28 | 2 |
| Sonnet 4.6 | 2.8 | 3.2 | 3 | 9,465 | 1.8 | 13 | 0 |
| Haiku 4.5 | 2.1 | 1.1 | 0 | 5,384 | 0.5 | 24 | 11 |
| Gemini 3.1 Pro | 2.4 | 1.5 | 0 | 3,147 | 0.5 | 27 | 9 |
| Gemini 2.5 Pro | 2.8 | 1.9 | 1 | 2,516 | 1.2 | 17 | 0 |

Token counts come from each provider’s own tokenizer and are only roughly comparable across families.

The central observation is the size of the effects. Mean effects per run range from 0.5 to 1.8 across the six models and from 0.4 to 0.7 across the four frameworks. In these separate, unreplicated sweeps, observed outcomes vary more across model conditions than framework conditions; this is not a causal model-by-framework comparison. Under the noise rule (Equation 8) these single-run numbers are descriptive: no repeat spread was measured, so a difference such as 0.5 versus 0.4 is not reported as an effect, and the larger range of 0.5 to 1.8 is also descriptive until run-to-run variability is measured. More importantly, the effect count is not a quality score. Whether the additional writes of Sonnet 4.6 and Gemini 2.5 Pro are correct requires ground-truth scoring, which is pending, and the screening stage cannot run until the applicability thresholds are agreed.

Two derived quantities make the claim–effect structure visible (Table XLVIII, Fig. 20). The *effect rate* is the share of the 49 runs with at least one state-changing effect, and the *unbacked share* is the fraction of runs without any effect whose final answer nevertheless states that the work was done. Both are computed here from the reported counts, using the same heuristic as the source and with its caveat that it may over-count when an answer quotes past events from the data. The models separate along two nearly independent axes. Sonnet 4.6 and Gemini 2.5 Pro have both a high effect rate and no unbacked claims; gpt-5.5 has a low effect rate but is almost always candid about it (7.1% unbacked share); gpt-4o, Haiku 4.5, AutoGen, LangGraph and LlamaIndex combine a low effect rate with a share of unbacked claims between 46% and 55%. A single score such as task completion would place the second and third groups together, whereas the plane of Fig. 20 separates the candid minimalist from the confident one. This is the quantitative form of the claim that a system's own report of success overstates it.

TABLE XLVIII. Derived Claim–Effect Measures (49 Runs per Arm)

| Arm | Effect rate | Unbacked / no effect | Unbacked share |
|---|---|---|---|
| LangGraph (gpt-4o) | 44.9% | 14 / 27 | 51.9% |
| LlamaIndex (gpt-4o) | 44.9% | 14 / 27 | 51.9% |
| AutoGen (gpt-4o) | 40.8% | 16 / 29 | 55.2% |
| CrewAI (gpt-4o) | 57.1% | 5 / 21 | 23.8% |
| gpt-5.5 | 42.9% | 2 / 28 | 7.1% |
| Claude Sonnet 4.6 | 73.5% | 0 / 13 | 0.0% |
| Claude Haiku 4.5 | 51.0% | 11 / 24 | 45.8% |
| Gemini 3.1 Pro | 44.9% | 9 / 27 | 33.3% |
| Gemini 2.5 Pro | 65.3% | 0 / 17 | 0.0% |

Derived by the authors from the counts of Tables XLVI and XLVII; the gpt-4o/LlamaIndex row is shared by both comparisons.

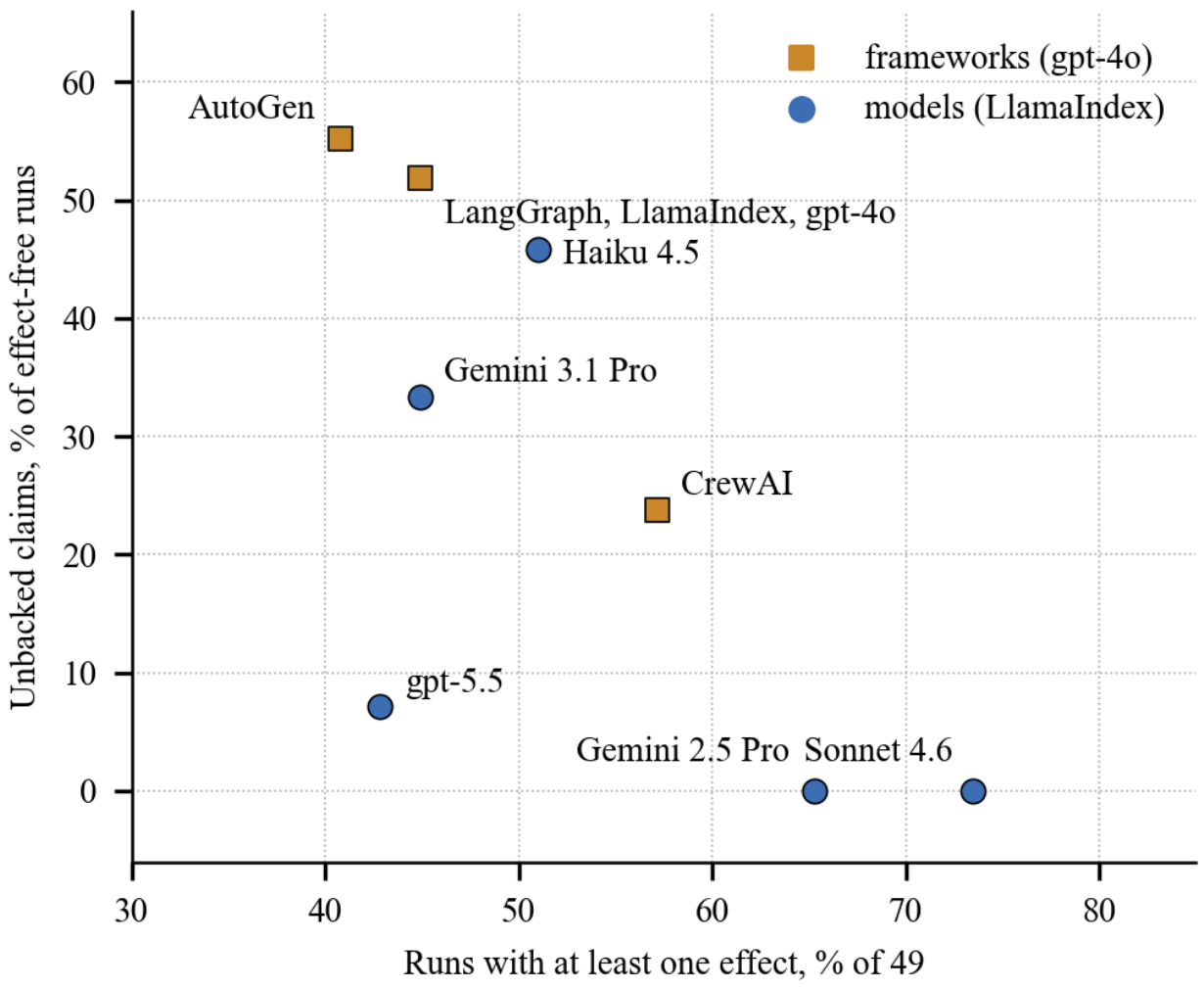

Fig. 20. Claim–effect plane (derived): share of runs with an effect versus share of effect-free runs whose answer claims success.

### C. Model-Specific Behaviour of Tuned Instructions

A component of the automation direction classifies inbound messages into intents using natural-language intent descriptions held in a rule store. On a frozen 40-case set with five repeats of both arms, the same unedited descriptions scored 89.7% on gpt-4.1-nano with a one-case wobble, 94.3% on DeepSeek-V4-Flash with zero variance, and 97.1% on Claude Haiku 4.5 with zero variance. A description rewrite improved classification accuracy by 4.6 percentage points on the weakest deployment, reduced accuracy by 3.4 percentage points on the intermediate deployment, and produced no observed change on the strongest deployment (Table XLIX, Fig. 21). A confusion between two intents that had generated two description rewrites and a claimed specification gap simply did not occur on the other models.

TABLE XLIX. Intent Classification on Three Deployments

| Deployment | Accuracy, unedited | Variance (5 repeats) | Effect of rewrite |
|---|---|---|---|
| gpt-4.1-nano | 89.7% | one-case wobble | +4.6 pp, reportable |
| DeepSeek-V4-Flash | 94.3% | zero | −3.4 pp, no result |
| Claude Haiku 4.5 | 97.1% | zero | 0.0 pp |

Frozen 40-case set; labels self-authored by the intent author, so the system arm is an upper bound; the keyword control arm scores 0.000 and is a floor, not a rival.

The three deployment conditions and their classification counts are presented in Table XLIX; the same comparison is visualised in Fig. 21.

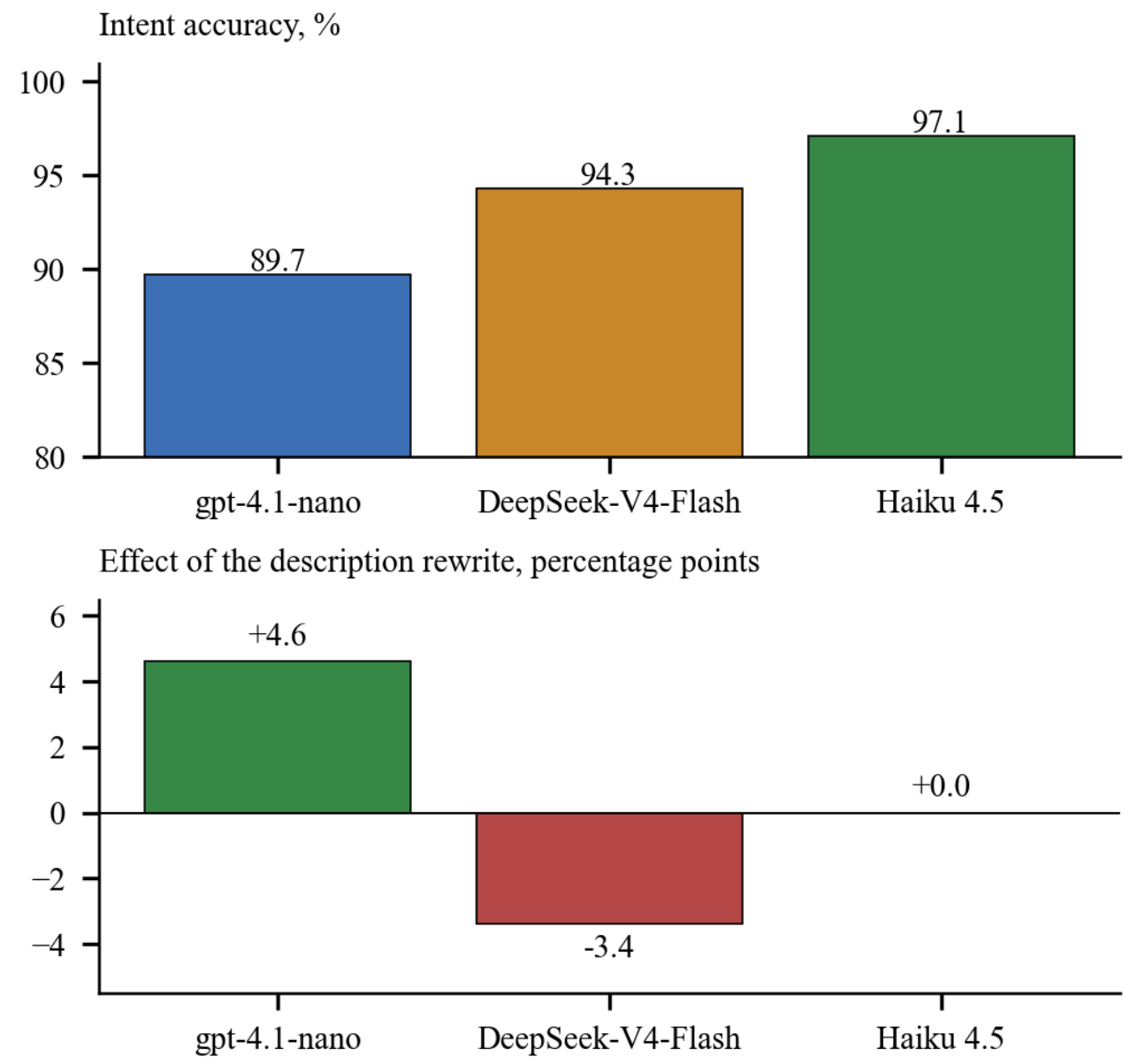

Fig. 21. Intent classification accuracy and the effect of a description rewrite on three deployments.

The transferable claim is that prompt and description engineering measured on a weak deployment measures the deployment, not the prompt. An edit whose benefit shrinks as the model gets better was a workaround for a model that could not follow the instruction already present, not an improvement of the instruction. The cheapest experiment in the whole exercise, changing the model before changing the words, was the one that had been run last. The same study measured a regex-based guardrail on 29 labelled drafts: precision 0.733, recall 0.688 and a false-block rate of 0.308. This is the one result in the study that does not depend on a model, since the guard uses regular expressions and substring matching, so it reproduces offline for anyone; its misses are evasions by surface (a lowercase IBAN) and its false blocks are shape collisions (an order reference shaped like a card number), and the adversarial set describes the shape of the failure modes, not their frequency. Programmable guardrails of this kind [67] are one of the planned attribute groups (P5).

### D. Chatbot: Controlled Experiments and the Evaluation Agent

The chatbot experiments used the mvp_scenarios v6 dataset: most controlled comparisons used six scenarios, the final model comparison five scenarios with three repeats, and the agent evaluation all ten scenarios. A retrieval-only runner isolated retrieval from answer generation and tool routing. Table L and Fig. 22 summarise the controlled experiments on retrieval. Dense and hybrid retrieval [16], [17] reached the same recall, precision and MRR, but dense was faster (487 ms against 634 ms) and was selected. Sparse retrieval found almost none of the required sources; its higher precision does not indicate better overall quality. Retrieving more chunks did not improve recall or MRR and reduced precision (Top-K of 4 against 8 and 12), so K = 4 was selected. A reranker raised precision slightly (0.375 to 0.417) but left MRR unchanged and raised latency (the 2,960 ms includes model-loading cold start), so it was left off. With the retrieval fixed, temperature 0.0 gave the highest correctness (0.952) and task completion (0.963) for GPT-5 mini.

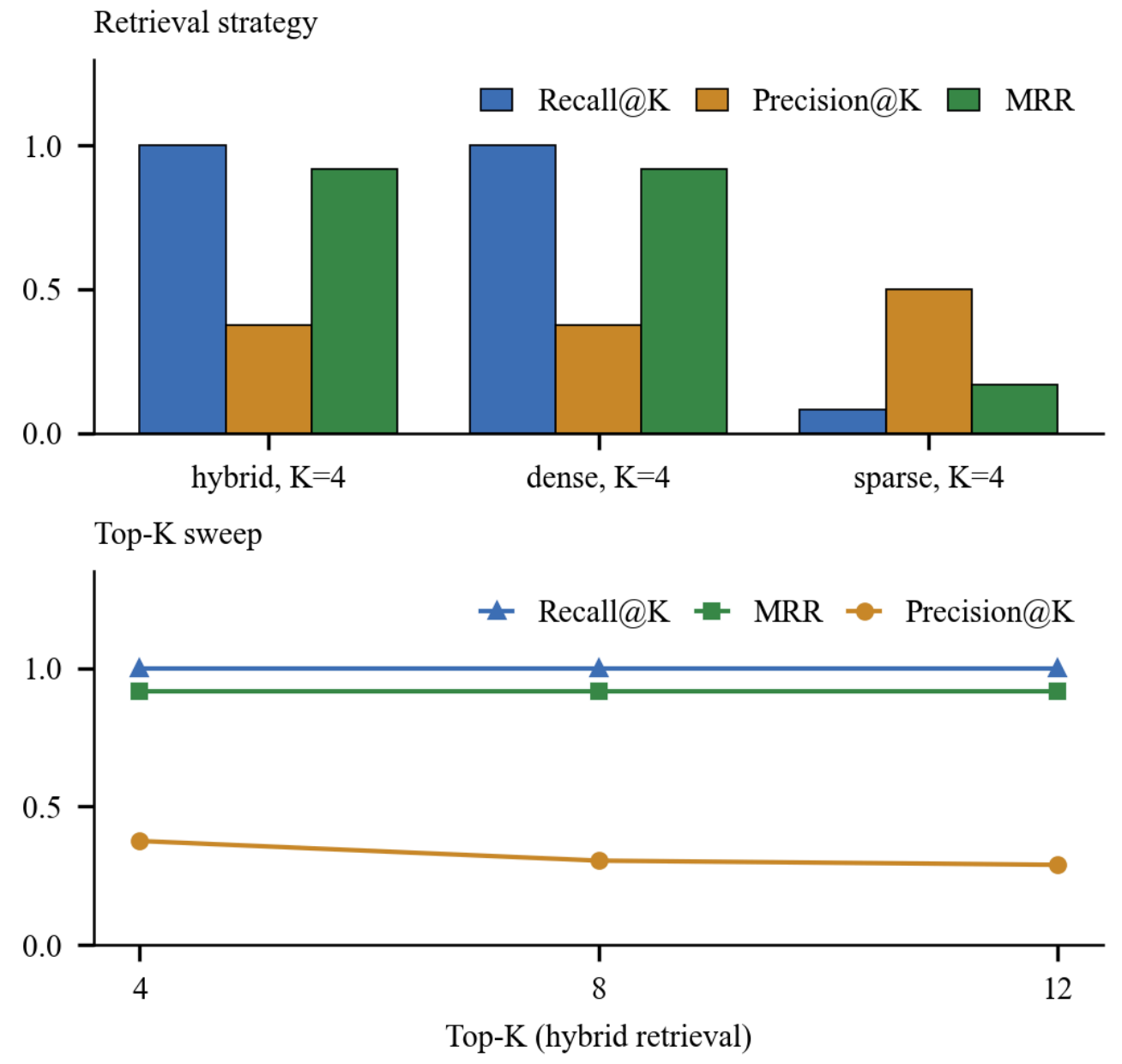

Fig. 22. Retrieval strategy and Top-K experiments (six scenarios).

Nine models were then compared under the same configuration (dense, K = 4, reranker off): GPT-5.6 Sol, GPT-5.6 Terra, GPT-5 mini, Claude Sonnet 5, Claude Opus 5, Claude Haiku 4.5, Gemini 3.8 Flash, Gemini 3.1 Pro and DeepSeek V4 Flash. Three finalists with the strongest balance of quality and latency were repeated: five scenarios times three repeats, 15 runs per model (Table LI, Fig. 23). Terra reached the highest mean correctness (0.995) and the lowest variability (standard deviation 0.013); Haiku was faster (9.8 s). The selected chatbot configuration was dense retrieval, Top-K = 4, reranker off and GPT-5.6 Terra.

TABLE L. Repeated Comparison of the Finalists (15 Runs per Model)

| Model | Pass rate | Mean correctness | Std | Response time |
|---|---|---|---|---|
| GPT-5.6 Terra | 100% | 0.995 | 0.013 | 11.6 s |
| Claude Haiku 4.5 | 100% | 0.989 | 0.017 | 9.8 s |
| Gemini 3.1 Pro | 100% | 0.977 | 0.037 | 14.0 s |

Pass rate applies to the five-scenario subset, not to the full agent campaign.

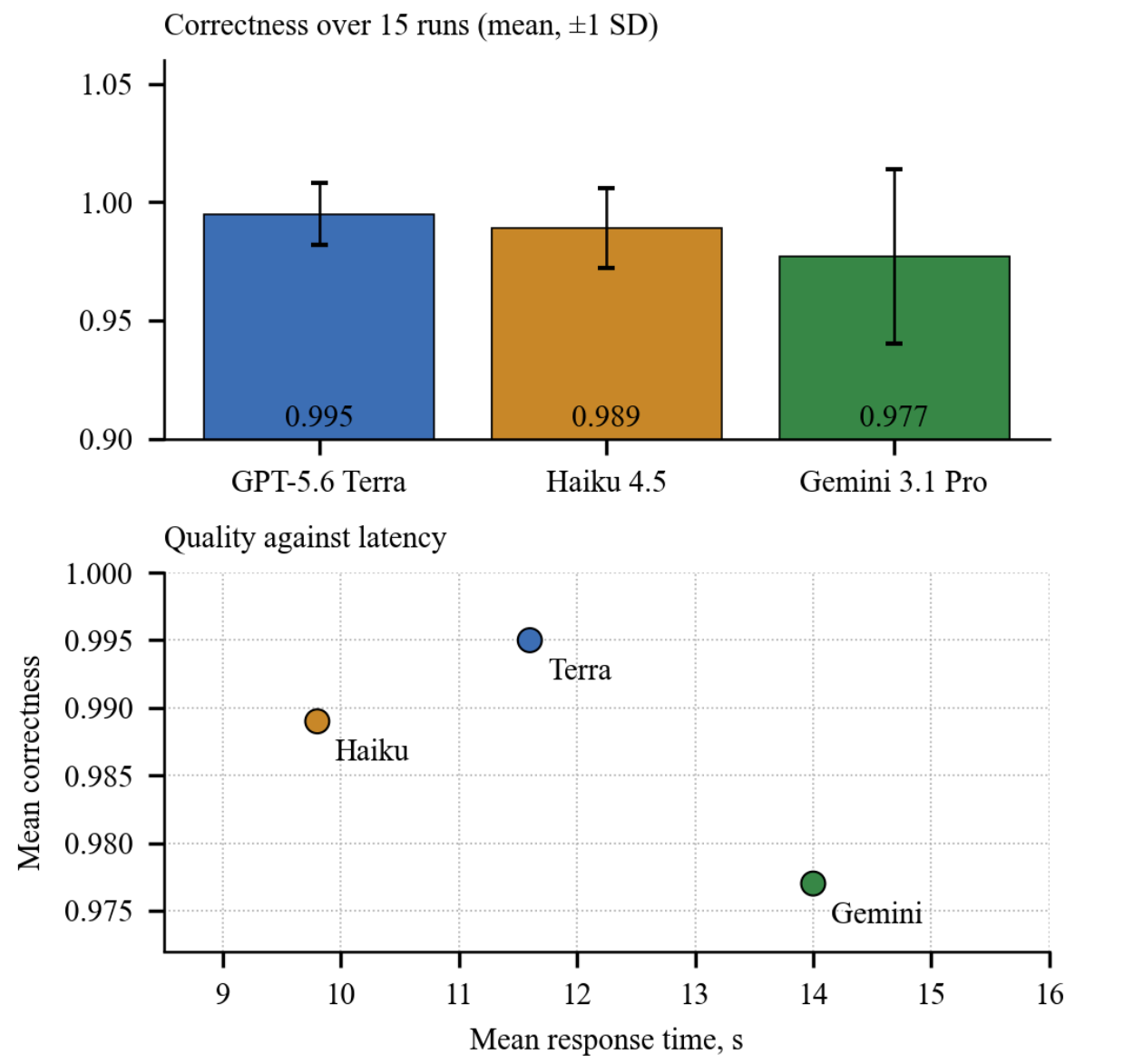

Fig. 23. Finalists: correctness with one standard deviation over 15 runs, and the quality–latency plane.

TABLE LI. Controlled Chatbot Experiments on Retrieval and Generation

| Experiment | Configuration | Result |
|---|---|---|
| Strategy (K = 4) | hybrid / dense / sparse | Recall 1.000 / 1.000 / 0.083; Precision 0.375 / 0.375 / 0.500; MRR 0.917 / 0.917 / 0.167; latency 634 / 487 / 3.5 ms |
| Top-K (hybrid) | K = 4 / 8 / 12 | Precision 0.375 / 0.304 / 0.289; Recall 1.000; MRR 0.917 |
| Reranker | off / on (K = 4) | Precision 0.375 / 0.417; MRR 0.917; latency 762 / 2,960 ms |
| Temperature (GPT-5 mini) | 0.0 / 0.3 / 0.7 | Correctness 0.952 / 0.943 / 0.933; task completion 0.963 / 0.955 / 0.955; faithfulness 1.000 / 1.000 / 0.987 |

Six scenarios per comparison; single run each; dense_top4 fixed for the temperature experiment.

The observed mean correctness difference between Terra and Haiku is 0.006, smaller than their reported standard deviations (0.013 and 0.017). However, the individual repeats and their ranges are unavailable, so the criterion in Equation (8) cannot be evaluated exactly and a formal "tie" is not established. Table LII therefore gives only an illustrative two-objective Pareto comparison of the reported point estimates (correctness maximised and latency minimised). Model selection on these data is provisional: neither superiority in mean correctness nor superiority in stability has been established by a calibrated paired comparison.

The evaluation agent was first run as a diagnostic campaign (v1) on all ten scenarios with the selected configuration (Table LIII, Fig. 24). Five scenarios passed without unnecessary diagnostic calls. Four failed in the agent-and-tools layer because the required search tool call and its arguments were missing; one failed in grounding, because only one of two required source groups was cited, with answer quality also affected.

TABLE LII. Pareto Analysis of the Chatbot Finalists

| Model | Correctness ↑ | Time ↓ | Dominated by | Tie on correctness |
|---|---|---|---|---|
| GPT-5.6 Terra | 0.995 | 11.6 s | — | Haiku |
| Claude Haiku 4.5 | 0.989 | 9.8 s | — | Terra |
| Gemini 3.1 Pro | 0.977 | 14.0 s | Terra, Haiku | — |

Computed from the repeated finalist runs in Table LI. Because individual repeats are unavailable, the exact range-based rule of Equation (8) cannot be applied; the reported standard deviations indicate uncertainty but are not substituted for ranges.

In the first adaptive version the agent had no approved parameter change for the routing failures and tested only Claude Haiku for the grounding failure, which stayed FAIL: one follow-up experiment and no accepted candidate. In v2 the available parameters were extended with the router model and several candidates per failing run were allowed (up to four). Of eight follow-up experiments four candidates were accepted on target-metric improvement; changing only the router to GPT-5.6 Sol improved the tool metrics in all four routing failures and turned two of them into passes, while the other two stayed FAIL because of tool arguments. The four candidates for the grounding failure were all rejected. Four accepted candidates do not mean four passing scenarios, only two additional runs changed to PASS, and nothing was applied automatically to the live chatbot.

TABLE LIII. EVALUATION-AGENT CAMPAIGNS ON TEN SCENARIOS

| Campaign | Outcome |
|---|---|
| v1 diagnostic | 5 PASS; 4 FAIL in agent/tools (search call and arguments missing); 1 FAIL in grounding → answer quality |
| Adaptive v1 | 1 follow-up (Haiku for the grounding failure) rejected; 0 accepted |
| Adaptive v2 baseline | 10 runs: 5 PASS, 5 FAIL |
| Adaptive v2 follow-up | 8 experiments; 4 accepted (target metric +0.02 or PASS); 2 FAIL → PASS |
| Router change | Sol as router: tool metrics improved in 4 of 4; 2 of 4 became PASS |

In the recorded v2 campaign, candidates were provisionally accepted if the target metric increased by at least 0.02 or the scenario became PASS. These single-scenario decisions were not confirmatory noise-aware tests and were not applied to the live system.

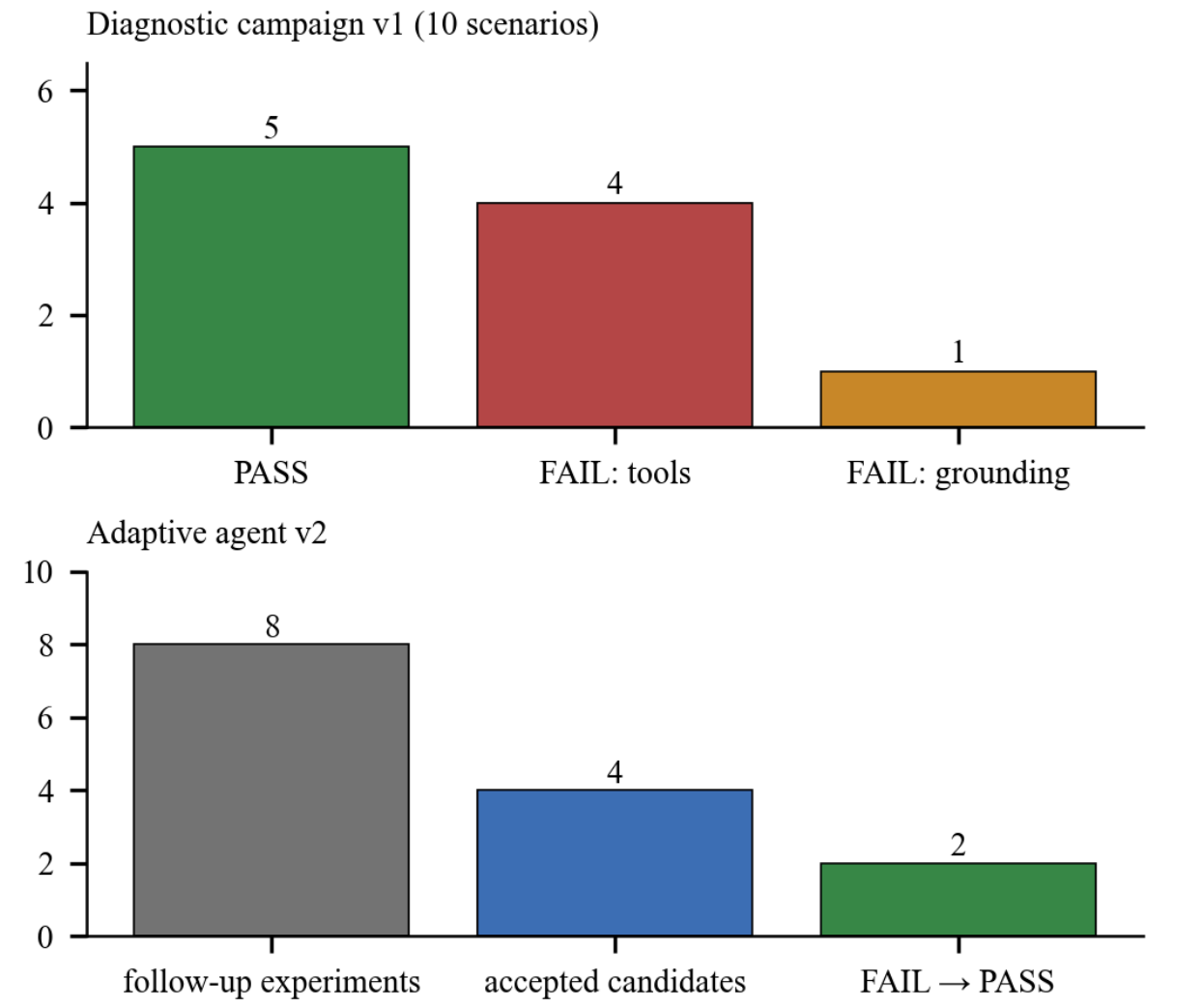

Fig. 24. Evaluation-agent campaigns: diagnostic outcome (v1) and adaptive experiments (v2).

## E. *Case Study: Reaching a Chatbot Configuration*

The chatbot experiments are a complete example of how the testbed workflow turns measurements into a decision, and they also show where that decision rests on weak evidence. The configuration was chosen in six steps, one variable at a time with the others fixed (Fig. 25, Table LIV). At each step we ask whether the evidence meets the noise-floor rule of Equation (8).

TABLE LIV. DECISIONS OF THE CHATBOT CONFIGURATION AND THE STRENGTH OF THEIR EVIDENCE

| Step | Candidates | Driving evidence | Strength |
|---|---|---|---|
| Retrieval strategy | dense, sparse, hybrid | sparse recall 0.083 vs 1.000; dense and hybrid tie, dense faster | sparse rejected clearly; dense vs hybrid by latency |
| Top-K | 4, 8, 12 | precision falls 0.375 → 0.304 → 0.289; recall, MRR tie | monotone trend; single run |
| Reranker | off, on | precision +0.042; MRR tie; latency 762 → 2,960 ms | trade-off; cold start in latency |
| Temperature | 0.0, 0.3, 0.7 | correctness 0.952, 0.943, 0.933 | differences 0.009–0.019; no spread measured |
| Model screening | 9 models | correctness, citations, latency; three finalists | screening, one run |
| Repeats | 3 finalists, 15 runs each | correctness 0.995 ± 0.013, 0.989 ± 0.017, 0.977 ± 0.037 | Terra–Haiku tie on mean; variance and latency decide |

Based on the chatbot experiments summarised in Tables L and LI.

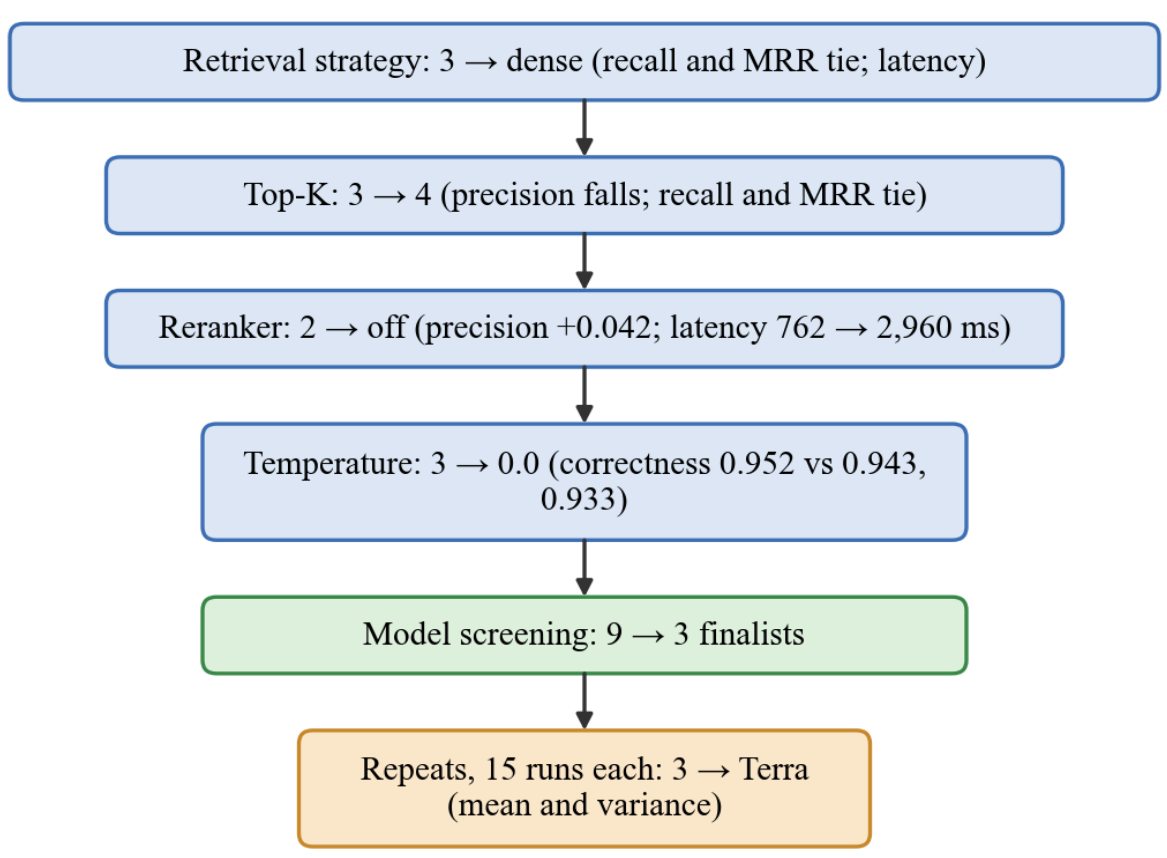

Fig. 25. Decision funnel of the chatbot configuration: one variable per step, with the evidence that drove the decision.

Two observations follow. Clear-cut decisions exist: sparse retrieval is far worse on recall, and retrieving more chunks monotonically lowers precision without compensating gain. But several steps rest on single runs with differences of a few hundredths, and the final model choice is a variance-and-latency decision between models that tie on mean correctness. This does not invalidate the selection, which is a reasonable engineering choice, but it changes how the result should be reported: as a configuration selected by controlled experiments, with the tied alternatives named, rather than as the best configuration. The testbed's role here is to keep every step reproducible, to store each step as a context-tagged measurement and to let a reader recompute the comparison with the noise rule.

The evaluation-agent campaigns add a second type of evidence: a diagnosis rather than a score. Of the ten scenarios, s001–s003 and s009–s010 passed; s004–s007 failed in the tool layer; and s008 failed in grounding with answer quality affected (Table LV). This split is the kind of structure that a single pass-rate would hide, and it is also what makes the router experiment interpretable, because it targets exactly the four scenarios that the diagnosis attributed to routing.

*Breadth.* The first experiment asked whether automatic strategy selection beats a single method (Table LVI, Fig. 26). The agent accepted 1,330 of 1,448 tasks (91.9%), the cheaper model on every task 1,310 (90.5%) and rules alone 205 (14.2%). An accepted task passed every check available for it; most accepted tasks nevertheless stop at static checks, because the fragments do not contain the tables they reference, which is why only 284 of the agent's tasks were installed or executed.

TABLE LV. SCENARIO-LEVEL OUTCOMES OF THE DIAGNOSTIC CAMPAIGN

| Scenarios | Result | Diagnosis |
|---|---|---|
| s001–s003, s009–s010 | 5 PASS | none; no unnecessary diagnostic calls |
| s004–s007 | 4 FAIL | agent/tools: required search call and arguments missing |
| s008 | 1 FAIL | grounding, then answer quality: one of two source groups cited |
| s004, s005 (router → Sol) | FAIL | tool metrics improved; tool arguments still wrong |
| s006, s007 (router → Sol) | PASS | tool metrics improved; scenario passes |

Single campaign on the selected configuration; nothing applied to the live chatbot.

### F. *Migration: Breadth, Depth and the Fix Ledger*

The agent produced 125 exact structural matches with the human reference against 86 for the cheaper model alone (+45.3%), and 426 pairs with similarity at least 0.9 against 379. Rules alone solve a small simple slice accurately (112 exact matches, but only 174 comparable pairs); the agent covers the rest. The token saving relative to the cheaper-model baseline is small, 3.0% fewer tokens and 6.9% fewer calls; repairs and the strong model spend most of what the static path saves. The strong model handled 31 fragments, used 987,364 tokens and produced no exact matches, which is not evidence about the model, since those were the hardest fragments, but it is not evidence that the tier paid for itself. Repair and escalation touched 465 fragments, used 64.4% of the tokens and produced 27 exact matches, while first-attempt fragments used 3.07 million tokens and produced 97. Rules had solved 112 pairs exactly; the agent kept the rules path on 72 of them and, on the other 40, called a model and kept the exact match in only 9, losing 31 matches at a cost of 278,215 tokens.

TABLE LVI. MIGRATION BREADTH: STRATEGY SELECTION ON 1,448 TASKS

| | Agent | Cheap LLM | Rules only |
|---|---|---|---|
| Accepted tasks | 1,330 (91.9%) | 1,310 (90.5%) | 205 (14.2%) |
| Exact structural matches (of 1,000) | 125 | 86 | 112 |
| Similarity ≥ 0.9 (of 1,000) | 426 | 379 | 142 |
| Installed or executed | 284 | 275 | 6 |
| Tokens | 10,020,728 | 10,332,904 | — |

One run per configuration on the same sample (seed 42). Structural match: same normalised syntax tree as the human reference; not proof of equal behaviour.

A second experiment sent each file whole to the same cheaper model with the same prompts and repair loop, without splitting. It produced 99 exact matches against 86 for the split run, a gap that is not distinguishable (McNemar $p = 0.208$). Typical similarity went the other way, 0.674 with the split against 0.520 without, with the drop concentrated in 240 unnamed PL/SQL blocks (0.741 versus 0.179); the whole-file run used 1.37 times more tokens and, in 97 files, used an object before creating it. The split, the dependency order and the per-object repair were removed together, so the run does not isolate splitting by itself.

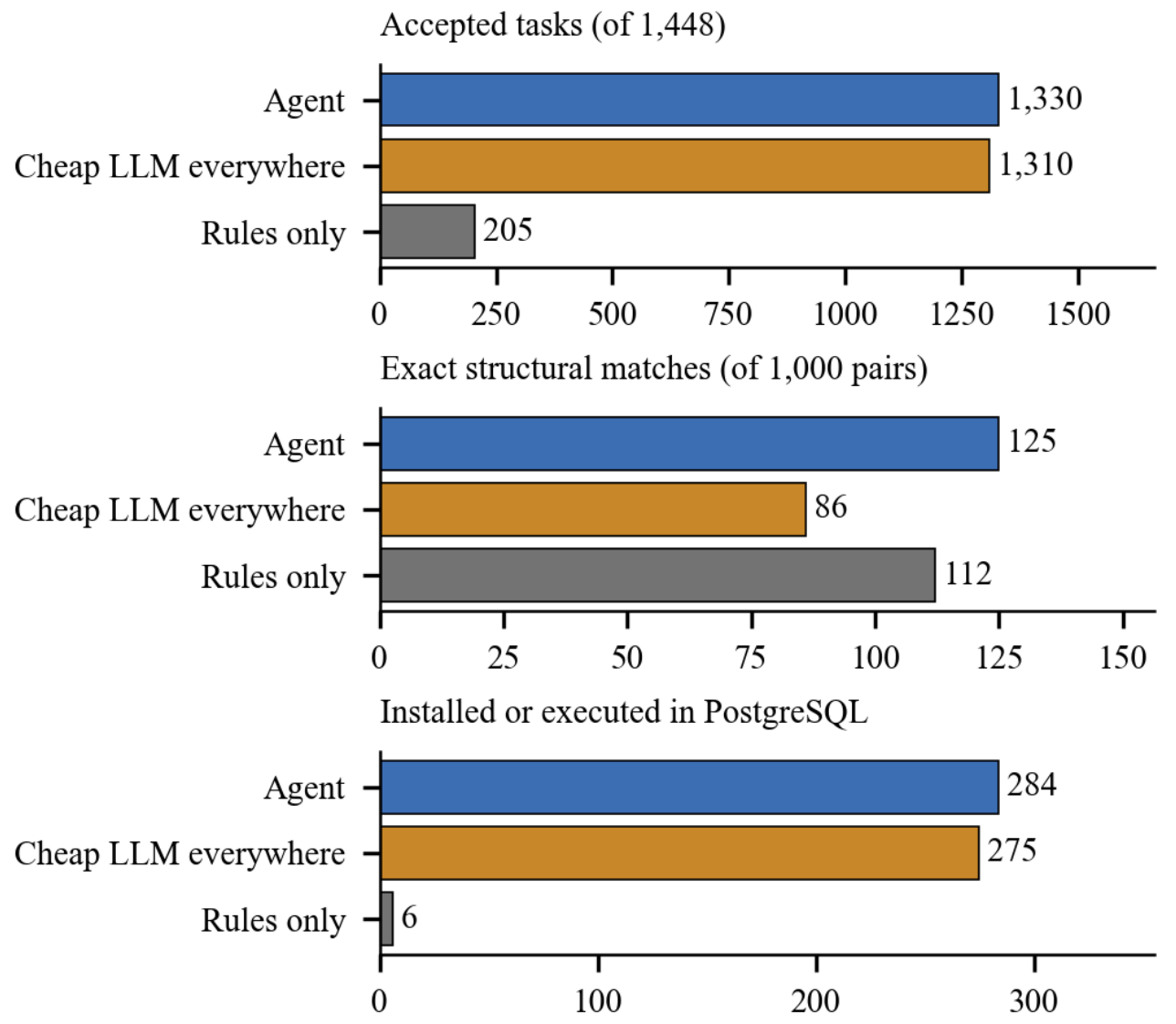

Fig. 26. Migration breadth: accepted tasks, exact matches and executed tasks for the three strategies.

*Component-level evidence: decomposition.* The decomposition stage of the migration agent was itself evaluated before the agent was assembled. Three ways of splitting an Oracle artefact into units were compared on a sample artefact and a 116-file corpus: a static PL/SQL-aware scanner, a grammar-based parser and an LLM splitter (Table LVII). A round trip (split, reassemble, compare with the original) is a cheap deterministic check that a split is information-preserving before any agent is involved. The scanner is the robust default; the grammar-based parser is the most precise but cleanly parses only about 59% of messy real files; the LLM is lossless at corpus scale but limited by token quota and throughput, with boundaries that need structural verification. Splitting a single file is only half the problem, because real dependencies are global (cross-file calls, package specification and body pairs). Lifting the units into one batch-level dependency graph, resolving cross-file references into edges, and using strongly connected components and condensation to handle real cycles gives a dependency-respecting order and parallel phases.

On the 116-file corpus the parser-only run found 1,271 dependencies and 446 internal links, but left 165 units that the grammar could not parse with no dependencies at all. An LLM fallback for exactly those units reduced the number of units without dependencies from 165 to 0, found 496 additional dependencies and raised internal links from 446 to 527 and external links from 825 to 1,182, at the price of about 37 minutes of sequential model calls (Table LVIII, Fig. 27).

TABLE LVII. COMPARISON OF SPLITTING STRATEGIES FOR ORACLE CODE

| Splitter | Coverage and fidelity | Limitation |
|---|---|---|
| static PL/SQL-aware scanner | 100% coverage, lossless, no dependencies | no semantic structure |
| grammar-based parser (ANTLR) | most precise, byte-exact | cleanly parses about 59% of messy real files |
| LLM splitter | lossless at corpus scale (86 of 86 processed files) | token quota, throughput; boundaries need verification |
| sqlglot | plain-SQL cross-check only | not PL/SQL |

Sample artefact and 116-file Oracle corpus.

TABLE LVIII. Dependency Extraction on the 116-File Corpus

| | Parser only | Parser + LLM fallback |
|---|---|---|
| Dependencies found | 1,271 | 1,271 + 496 |
| Internal links | 446 | 527 |
| External links | 825 | 1,182 |
| Units without dependencies | 165 | 0 |
| Added time | — | about 37 min (sequential) |

The corpus contains no cycles; cycle handling was tested on a small dedicated example.

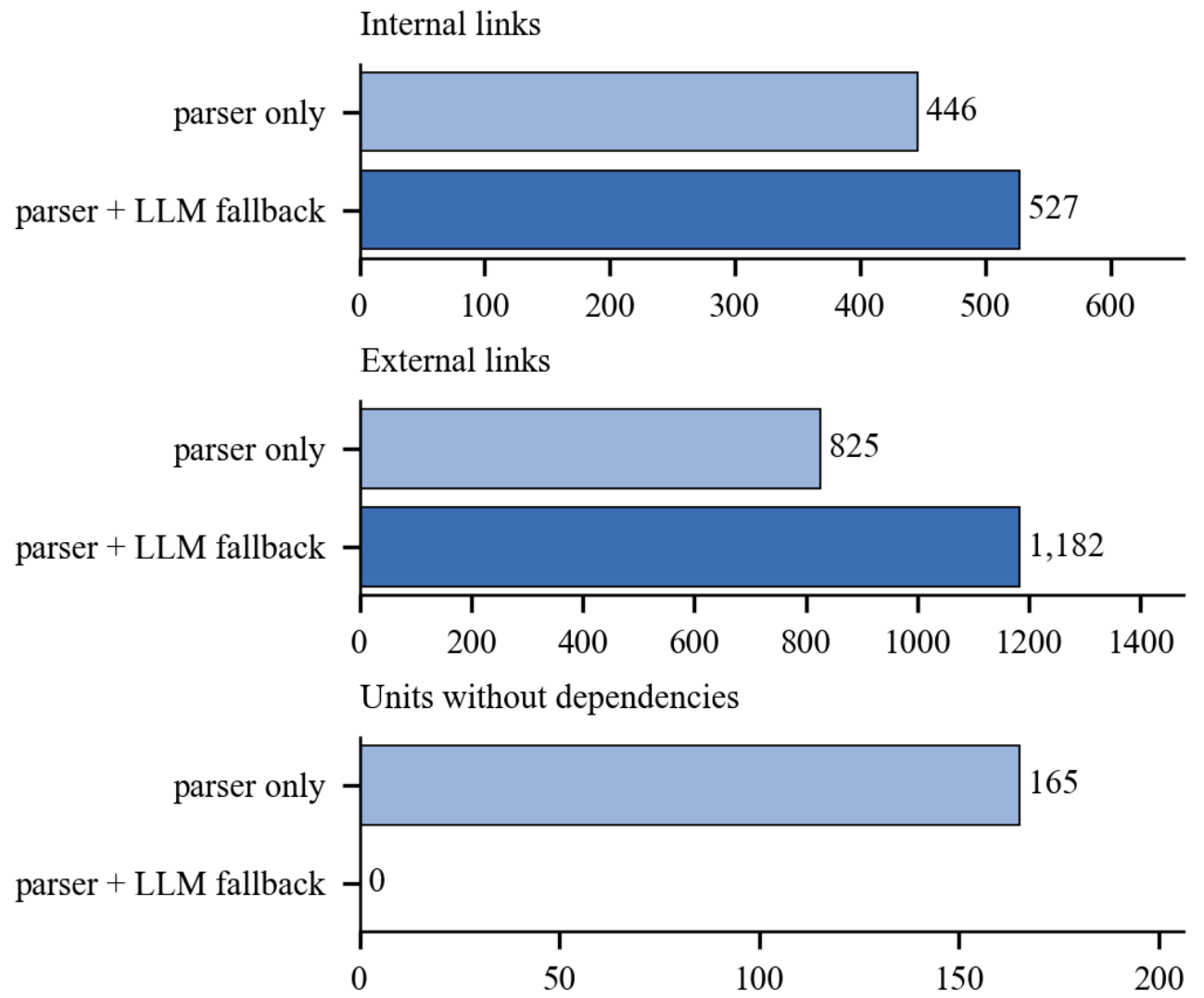

Fig. 27. Effect of the LLM fallback on dependency extraction (116-file corpus).

*Depth.* On 216 migration units executed against a real Oracle instance, two full runs using production routing were compared: migrate/v34 and migrate/v38, with three additional corrections to the evaluation environment (Table LIX, Fig. 28). The agent's own count of ACCEPTED rose by 21, from 93 to 114. Checked against Oracle, the number of demonstrably correct units, defined as ACCEPTED and not contradicted by Oracle's output or errors plus FAILED units that raise exactly what Oracle raises, rose by 10, from 98 to 108. Nine of the 93 earlier and 25 of the 114 later ACCEPTED units are wrong or cannot be verified; most of the extra ones had been BLOCKED before and ran for the first time. Real translation defects rose from 43 to 56 because 32 units that used to be BLOCKED are now tried, and some of them are wrong, which is information that was missing before and not a regression. Examples of wrong answers that the agent had accepted: a collection comparison that says "not equal" where Oracle says "equal"; an error handler that prints NULL where Oracle prints −6502-Value=1000; a bulk update that also touches a row Oracle never updates; a caller that invokes the wrong procedure and returns quietly where Oracle raises.

TABLE LIX. Migration Depth: 216 Units Checked Against Real Oracle

| Category | Before | After |
|---|---|---|
| ACCEPTED (agent's own report) | 93 | 114 |
| Demonstrably correct | 98 | 108 |
| Real translation defects | 43 | 56 |
| BLOCKED (never attempted) | 41 | 9 |
| Not testable by the stand as built | 16 | 23 |
| Other | 18 | 20 |

Runs 20260923-084200-b72cd4ff and 20260923-121730-92079ad3. The ACCEPTED row is a separate, overlapping agent-status indicator. The remaining mutually exclusive evaluation categories sum to 216 units for each run.

*What FAILED contained.* Because the agent's own FAILED status cannot distinguish a wrong translation from a correct one that raises the same error as Oracle, every FAILED unit of the production run was read by hand and, where the judgement rested on Oracle's behaviour, executed on the real Oracle instance (Table LX). Of 61 FAILED units, 27 were real translation defects, 14 were faithful raises, 8 were invalid test inputs that Oracle rejects too, 8 were limitations of the harness and 4 were explained no-ops where the model returned no SQL and said why, for declarations with no PostgreSQL object to create. The distribution of before-and-after unit states is shown in Fig. 28. The migration depth-study status distribution is given in Table LIX.

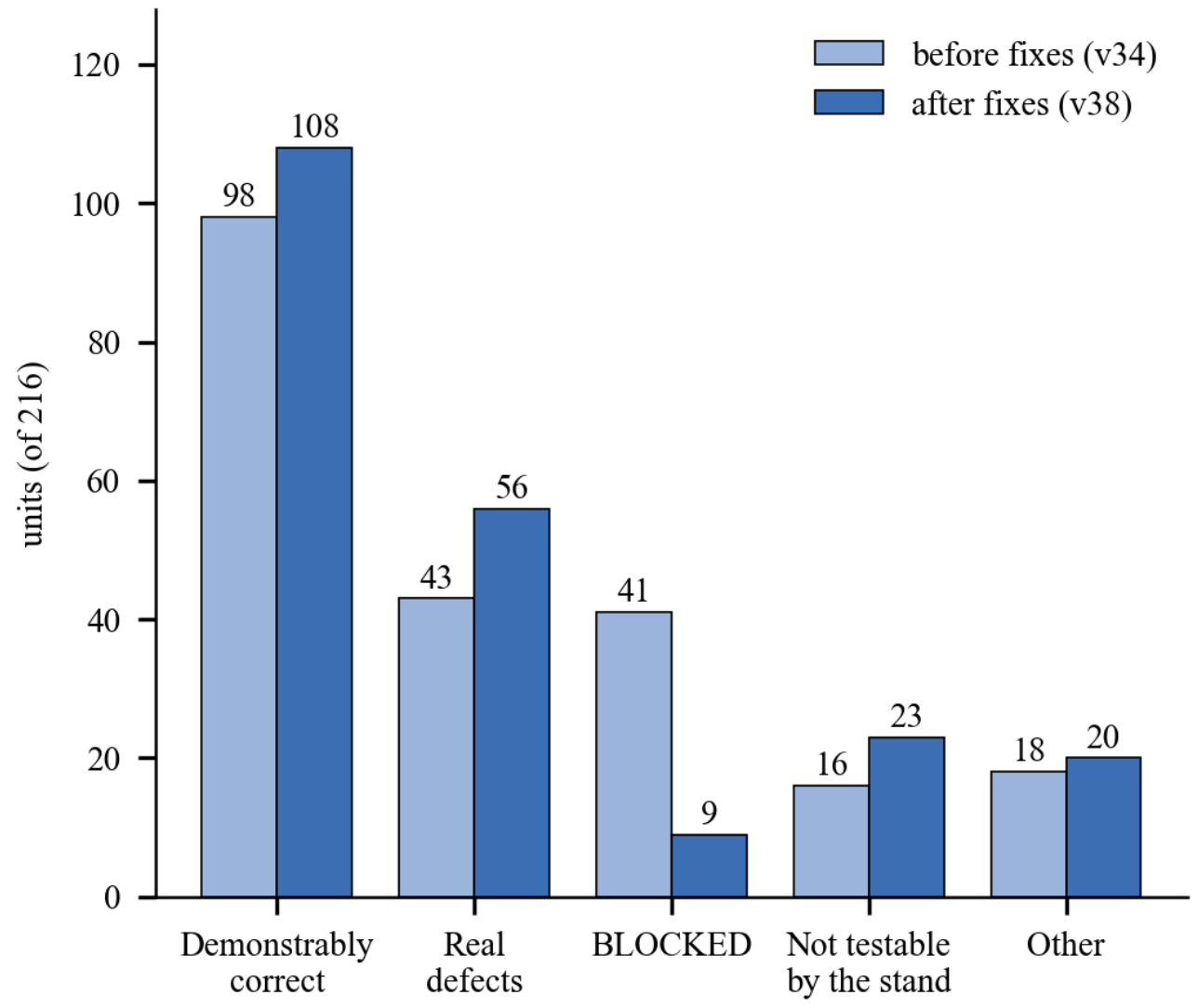

Fig. 28. Depth study: categories before and after the fixes of 2026-09-23.

A first pass that counted every static-stage failure as a defect without reading it had overstated the defect count (32 instead of 27); five units that started as uncertain were all resolved by execution the same day. A similar reading of the BLOCKED units found that about 32 of 41 descended from upstream units that had installed cleanly and failed only at scenario execution; eleven of those upstreams were faithful raises, so that a correct procedure raising NO_DATA_FOUND by design was blocking five other units on its own. The expected consequence was stated in advance: BLOCKED falls and FAILED rises, which read naively looks worse and is not.

TABLE LX. Reclassification of the 61 FAILED Units (Production Routing)

| Class | Units | Meaning |
|---|---|---|
| Real translation defects | 27 | wrong behaviour against Oracle |
| Faithful raises | 14 | raises exactly where Oracle raises, e.g. a numeric overflow that Oracle also raises |
| Invalid test inputs | 8 | inputs that Oracle rejects too (e.g. an injection string as a column name) |
| Harness limitations | 8 | COMMIT/ROLLBACK blocks, hidden dynamic-SQL dependency, cursor argument, non-deterministic cursor name |
| Explained no-ops | 4 | no SQL returned; declaration has no PostgreSQL object |

Method: read the error, the triggering scenario and the Oracle source; execute on real Oracle where the judgement depends on its behaviour.

Two further observations from this reading bear on the testbed. First, a silent-wrong case was caught only because a second model, tried for cost reasons, disagreed with the baseline on one unit and reading both answers showed that the baseline was the wrong one; cross-model disagreement is therefore a cheap *defect detector* that the testbed gives for free whenever two models are run on the same application and context. Second, a correct fix can introduce wrong acceptances: a fix teaching the model that a package-level variable is mutable state moved eleven consumers from 5 ACCEPTED and 6 FAILED to 10 and 1, yet executing the whole file in order on both databases showed that Oracle's collection DELETE leaves a gap which the translation closed up, so that five of the new ACCEPTED units were wrong while every block ran without error. The fix stays partial and the count of demonstrably correct units does not include them. Representative fault classes and associated repairs are itemised in Table LXI.

Model choice and cost also had surprises. A cheaper model on the proxy, about four times cheaper per token, could not complete one full run, because its daily token cap was 2 M against 20 M for the default; on a shared key, quota decides before price. Cost reporting itself was wrong until 2026-09-22, because prices came from a public rate card times a guessed factor; they now come from the proxy's own rates, and two runs in a row matched the proxy's bill to the cent. A full production-routed run costs about USD 3 to 4. The limit of the current numbers is mostly what the stand can verify rather than what the model can translate: about 30 to 40 units are untested for reasons in the stand, and the estimate for the next run with the fixes already committed is 115 to 120 demonstrably correct units, with a ceiling of about 180 that requires fixing statement boundaries in the decomposer and making the test-input generator produce inputs that Oracle accepts. The disposition of confirmed, partial and failed fixes is also summarised in Fig. 29.

TABLE LXI. Examples of Fixes from the Ledger

| Fix | Oracle behaviour taught to the agent |
|---|---|
| FIX-037 | single-row SELECT … INTO raises on 0 or 2+ rows; PostgreSQL needs INTO STRICT |
| FIX-041 | DATE carries a time of day: map to TIMESTAMP(0), not DATE |
| FIX-048 | \|\| treats NULL as an empty string; PostgreSQL's makes the whole string NULL |
| FIX-051 | = between collections compares multisets with their own NULL rules |
| FIX-052 | FORALL … VALUES OF and the total row count across iterations |
| FIX-054 | repairs a regression: TOO_MANY_ROWS translated as DUP_VAL_ON_INDEX |
| FIX-056 | dynamic-SQL :name binds become $1, $2 … by position |

Migration repair outcomes and statuses from the executed fix cycles.

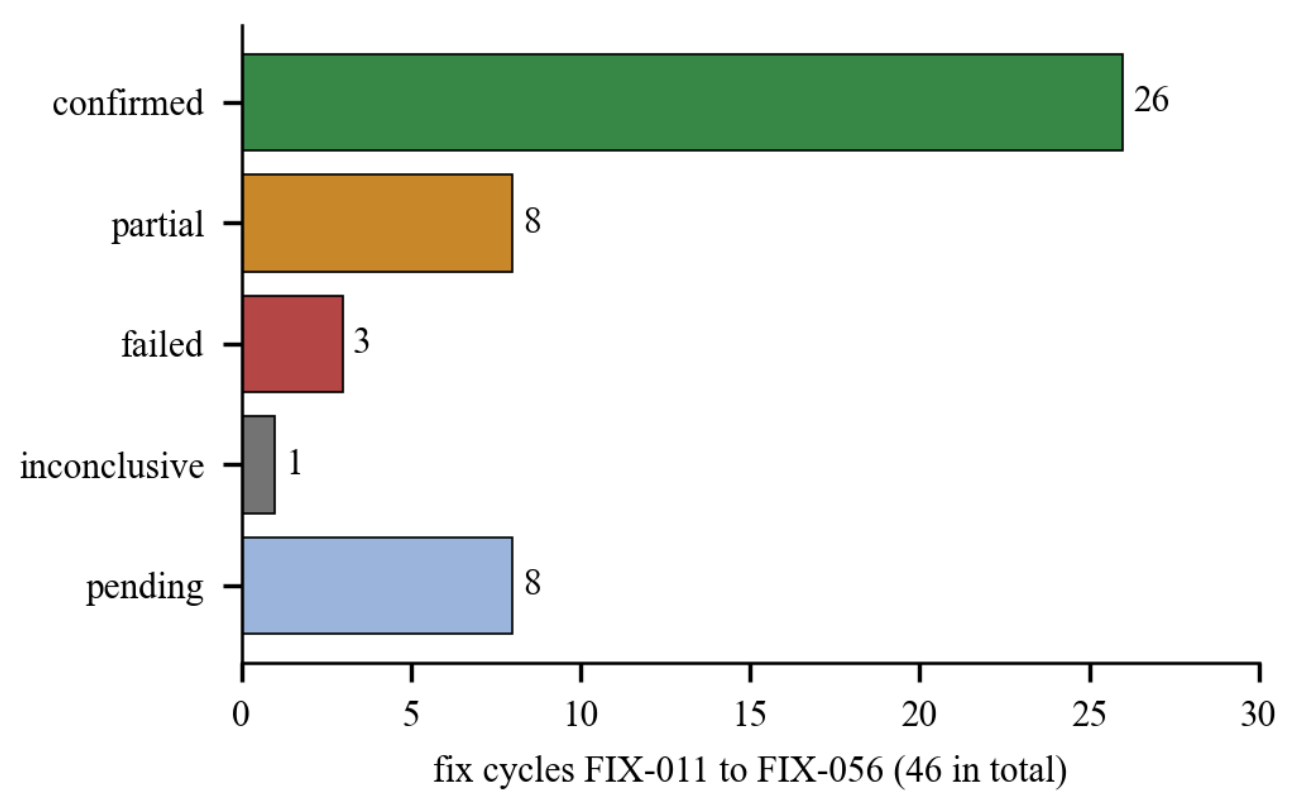

Fig. 29. Verdicts of the 46 fix cycles.

### G. Knowledge Retrieval for Migration: Measurement Pitfalls

The migration agent can retrieve conversion guidance from a vector knowledge base, and the attribute groups of the scheme include "knowledge and retrieval". A series of retrieval experiments on this component, run by colleagues of the programme, is worth reporting briefly because every one of its corrections is an instance of a principle of Section VII. The first proof of concept compared two databases for vector retrieval and found identical results. On re-examination the experiment never exercised a vector index: at 31 chunks both databases fall back to an exact scan, so identical embeddings with identical cosine distance and a full scan can only produce identical rankings, and the finding was arithmetically guaranteed and not measured. The reported latency difference (0.0109 s against 0.0052 s) measured driver overhead. A comparison that cannot come out differently is not a measurement.

Repeated at a realistic scale, the experiment showed that at matched recall the two systems were equivalent in retrieval quality and query latency (about 1.75 ms median at an approximate-search recall near 0.99), while the open-source extension built its indexes 8 to 226 times faster (Table LXII).

TABLE LXII. Vector Index Benchmark: Two Systems at Matched Recall

| Measure | Oracle | pgvector |
|---|---|---|
| HNSW build time, recall ≈ 0.99 | 40.1 s | 4.9 s |
| IVF build time, recall ≈ 0.955 | 249.0 s | 1.1 s |
| Query latency p50 (HNSW) | ≈ 1.75 ms | ≈ 1.75 ms |
| Default search settings, IVF recall | 0.9574 (declarative target) | 0.5975 (probes = 1) |
| Exact search recall@10 / nDCG@10 | 0.4575 / 0.3882 | 0.4575 / 0.3882 |
| Exact scan p50 | 61.6 ms | 30.5 ms |

Oracle AI Vector Search compared with pgvector under matched approximate-recall configurations.

An earlier untuned pass had falsely favoured one system, because the other's defaults silently discarded about 40% of what an exact search would have found (approximate recall 0.5975 at the default probe setting); raising two search parameters closed the gap entirely, so the difference had been the tuning and not the databases. This is a second form of the context rule: a configuration is part of the measurement, and default configurations are not neutral. A third experiment found that fusing a lexical signal into dense retrieval by reciprocal rank fusion [17] made nDCG@10 worse on natural-language queries on both databases (0.3028, 0.3101 and 0.3191 against 0.3882 for dense search alone), and that a native fused index failed to finish building at 86 thousand chunks after more than 90 minutes at full CPU, against 5 seconds for the alternative, because the fused index cannot consume precomputed embeddings. In the same programme a different corpus with identifier-bearing questions favoured hybrid retrieval, and a measured noise floor of the judge (Section VII-F) showed that an earlier reported gap on six cases lay inside the noise. The chatbot, which found dense and hybrid retrieval tied on its own corpus, is consistent with the view that the value of hybrid retrieval depends on the corpus and the queries and must be measured for each application, which is the purpose of the testbed. The measured dense and fused retrieval results appear in Table LXIII.

TABLE LXIII. DENSE VERSUS FUSED RETRIEVAL (NDCG@10)

| Arm | nDCG@10 |
|---|---|
| dense vector search alone | 0.3882 |
| Oracle Text + RRF | 0.3028 |
| BM25 + RRF (PostgreSQL) | 0.3101 |
| tsvector + RRF (PostgreSQL) | 0.3191 |

Natural-language query set; native fused index did not finish building.

### H. Related Measurements from Earlier Studies of the Group

Three earlier studies of the group are relevant for how the testbed treats cost, specifications and judges. *Token optimisation.* The first concerns cost [63]. Twelve token-optimisation strategies for LLM-based Oracle-to-PostgreSQL migration were evaluated on samples of 10 and 100 Oracle SQL queries with Valid Syntax Rate, Exact Match, Semantic Match, CodeBLEU and Token Efficiency. Mild context pruning preserved semantic quality almost at the baseline level (89.75% against 89.80% Semantic Match on 100 queries). Adaptive routing gave the best practical trade-off, reducing input tokens by 8.72% and output tokens by 5.49% while maintaining 88.40% Semantic Match and raising Token Efficiency by 6.67%. Aggressive schema distillation raised Token Efficiency by 132.22% but lowered Semantic Match by 44.50 percentage points (Fig. 30). Token optimisation is therefore not prompt shortening but a multi-objective problem balancing cost, validity and semantic preservation, which is why the metamodel stores cost and quality attributes side by side and why the analytical interface offers a Pareto view instead of a single score.

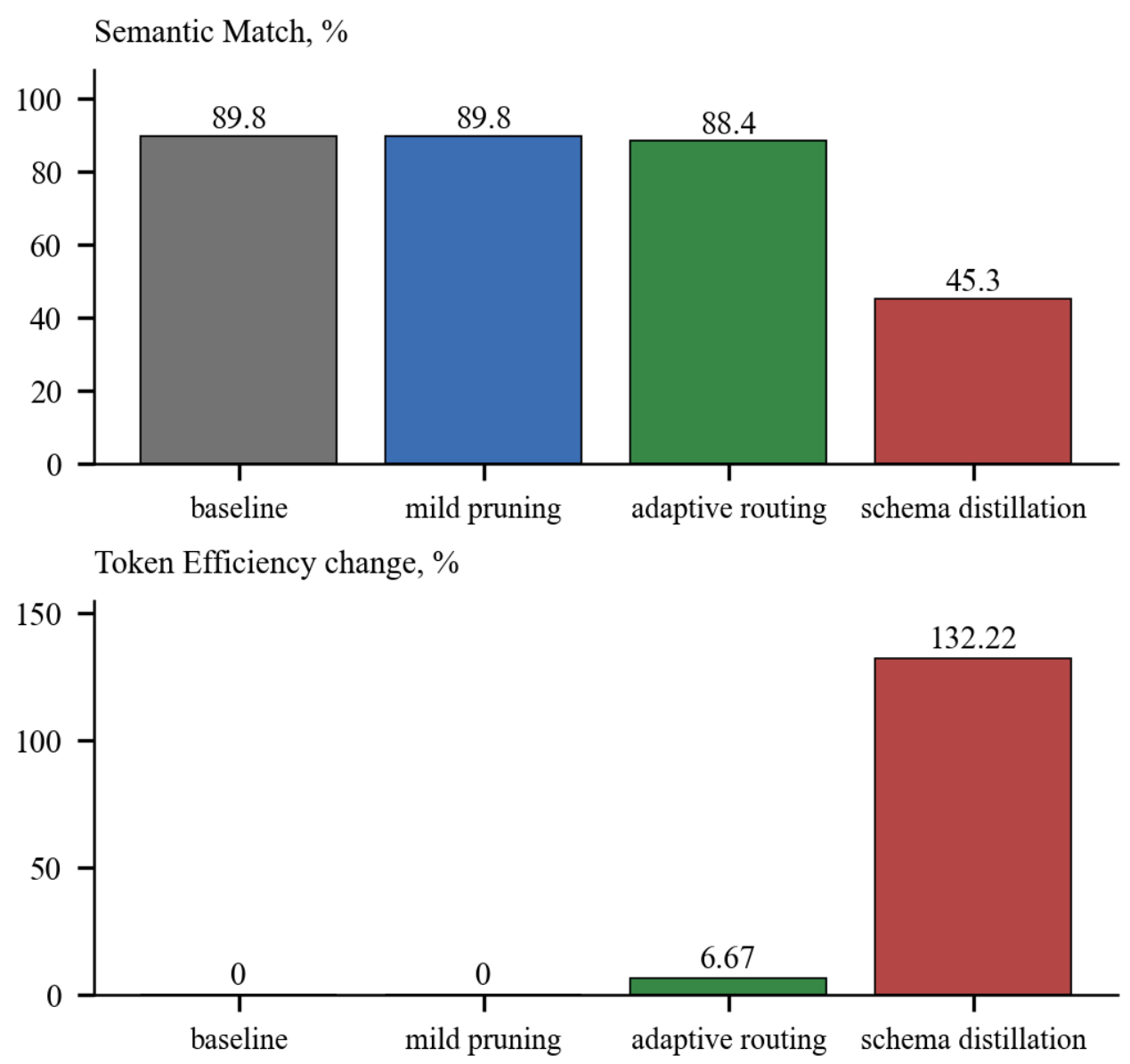

Fig. 30. Token optimisation: semantic match and token efficiency change for four of the twelve strategies on 100 queries (values from [63]; distillation bar derived as 89.80 − 44.50).

*What a specification determines.* The second study introduces a formal semantic-block model of a specification and an execution-judged benchmark with a fixed implementer panel, a mandatory no-specification control arm, and PostgreSQL 16 and a live Oracle instance as deterministic judges [46]. On 75 execution-verified units the full specification raised clean-loading output from 72.0% to 97.3%, whereas aggregate cross-implementer agreement changed only from 83.1% to 83.8%, and structured row-level comparison against live Oracle yielded 19 correct results out of 42 probes in both conditions. The five-layer decomposition reduced the mean per-task context by about 71% through dependency closures. Repeated runs on a 25-unit subsample revealed a median arm-delta spread of 14.4 percentage points. The study therefore supports determinacy as a formal concept but not as a standalone empirical quality metric, and it is the empirical origin of the noise-floor and control-arm rules of Section VII-F.

*Evaluating the evaluators.* The third study builds a leakage-resistant benchmark of 494 labelled source–specification pairs over 19 quality metrics and compares three LLM judges with two trivial baselines on 247 held-out records [45]. The judges reached balanced accuracies of 0.87 (GPT-4o), 0.89 (DeepSeek V4 Pro) and 0.81 (Claude Sonnet 4.5), while a specification-length baseline reached 0.88, and exact paired tests did not separate any judge from that baseline. The judges nevertheless carried information that length does not: the labelled defect category appeared among their findings in 84% to 92% of violated records and the quoted evidence overlapped the labelled span in 92% to 93%. This is the strongest argument we have for the source hierarchy of Table XXXIX: a judge adds typed, evidence-linked findings that help triage, but its binary verdict is not distinguishable from a trivial baseline, so it may explain but not decide. Table LXIV collects the key figures.

TABLE LXIV. KEY FIGURES FROM EARLIER STUDIES OF THE GROUP

| Study | Figure |
|---|---|
| Token optimisation [63] | mild pruning 89.75% vs 89.80% semantic match; adaptive routing −8.72% input tokens at 88.40%; distillation +132.22% efficiency, −44.50 pp |
| Specification benchmark [46] | clean-loading 72.0% → 97.3%; agreement 83.1% → 83.8%; 19/42 live-Oracle probes correct in both arms; context −71% |
| Variability floor [46] | median arm-delta spread 14.4 pp on a 25-unit subsample |
| Judge benchmark [45] | balanced accuracy 0.87 / 0.89 / 0.81 vs 0.88 for length baseline; defect category in 84–92% of findings |

### I. Simulation Study of the Noise Rule

The noise rule of Section VII-F was derived from failures, not from theory, so its operating characteristics had not been studied. We did so by Monte Carlo simulation, the technique used in the authors' earlier work on the evaluation of attribute values under uncertainty [88]. Two arms are simulated with independent normal repeats of equal standard deviation $\sigma$ per run, a true difference $\delta$, and $r$ repeats per arm (40,000 trials per cell, seeded). Three decision rules are compared: the range rule of Equation (8), a Welch test at level 0.05, and a fixed global threshold of two reference deviations applied to every axis. Table LXV reports the minimum detectable differences at 80% statistical power and the probability of sign disagreement between independent replications. Table LXVI presents the proportion of Monte Carlo trials in which an effect is declared under different decision rules, while Fig. 31 illustrates their operating characteristics for five repetitions per experimental arm.

Four findings follow. First, the range rule is a very conservative screen: its false-positive rate is 1.2% with three repeats and below 0.01% with five or more. Second, its sensitivity *falls* as repeats are added, from a minimal detectable difference of 4.65 $\sigma$ with three repeats to 7.17 $\sigma$ with ten, because the range of a sample grows with its size while the standard error shrinks; the Welch test goes the other way (3.51, 2.09 and 1.32 $\sigma$).

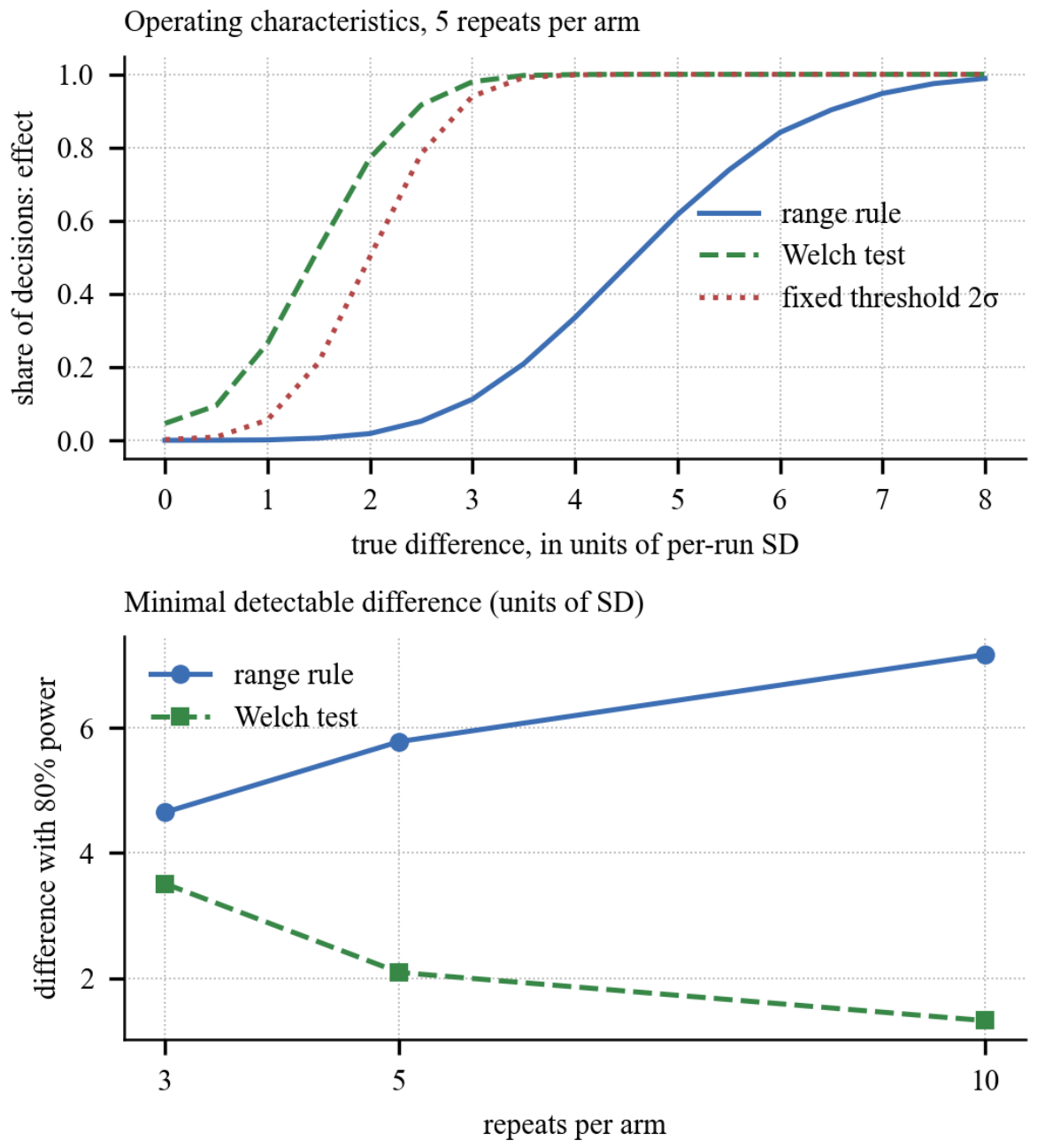

Fig. 31. Operating characteristics of the noise rule (simulation): decision shares against the true difference for five repeats (top), and the minimal difference detected with 80% power against the number of repeats (bottom).

TABLE LXV. MINIMAL DETECTABLE DIFFERENCE AND SIGN INSTABILITY

| Quantity | r = 3 | r = 5 | r = 10 |
|---|---|---|---|
| difference with 80% power, range rule (σ) | 4.65 | 5.78 | 7.17 |
| difference with 80% power, Welch test (σ) | 3.51 | 2.09 | 1.32 |
| sign disagreement of two replications at 0.5 σ | — | 33.7% | — |
| sign disagreement at 1 σ | — | 10.6% | — |
| sign disagreement at 2 σ | — | 0.16% | — |

Sign disagreement: probability that two independent replications of the same experiment give differences of opposite sign.

This is why the verdict of Equation (9) has a second tier. Third, a fixed global threshold is not neutral across axes (Table LXXI): with a threshold of two reference deviations, an axis whose own deviation is twice the reference declares a spurious effect in 11.5% of trials, whereas an axis with half the reference deviation never does, and an effect exactly equal to the threshold is detected in half of the trials on every axis. Fourth, small observed differences cannot be trusted in sign: for a true difference of 0.5 σ, two independent replications disagree on the sign in a third of cases, and at 1 σ in one case in ten, which is consistent with the reports of an axis that changed sign between identical re-runs (+23.3, then −9.3 percentage points). The global-threshold comparison is quantified in Table LXVII.

TABLE LXVI. SHARE OF TRIALS DECLARING AN EFFECT (MONTE CARLO, 40,000 TRIALS PER CELL)

| True difference (units of σ) | Range, r = 3 | Range, r = 5 | Range, r = 10 | Welch, r = 3 | Welch, r = 5 | Welch, r = 10 |
|---|---|---|---|---|---|---|
| 0 (no effect) | 0.012 | 0.000 | 0.000 | 0.035 | 0.044 | 0.047 |
| 1 | 0.049 | 0.001 | 0.000 | 0.121 | 0.265 | 0.557 |
| 2 | 0.182 | 0.021 | 0.000 | 0.366 | 0.771 | 0.989 |
| 3 | 0.415 | 0.110 | 0.002 | 0.677 | 0.981 | 1.000 |
| 4 | 0.673 | 0.330 | 0.030 | 0.884 | 0.999 | 1.000 |
| 6 | 0.953 | 0.837 | 0.469 | 0.994 | 1.000 | 1.000 |
| 8 | 0.997 | 0.989 | 0.929 | 1.000 | 1.000 | 1.000 |

Row 0 is the false-positive rate. Normal i.i.d. repeats; the range rule is Equation (8) with τ = 0, using the means of the repeats.

TABLE LXVII. FIXED GLOBAL THRESHOLD VERSUS PER-AXIS RULES (FIVE REPEATS, THRESHOLD 2 $\Sigma_{REF}$)

| Axis deviation ($\sigma_{ref}$) | True diff. | Global threshold | Range rule | Welch test |
|---|---|---|---|---|
| 0.5 | 0 | 0.000 | 0.000 | 0.043 |
| 0.5 | 2 | 0.498 | 0.330 | 1.000 |
| 1.0 | 0 | 0.002 | 0.000 | 0.045 |
| 1.0 | 2 | 0.502 | 0.020 | 0.768 |
| 2.0 | 0 | 0.115 | 0.000 | 0.043 |
| 2.0 | 2 | 0.500 | 0.002 | 0.264 |

Share of trials declaring an effect; $\sigma_{ref}$ is the reference per-run deviation.

The limits of the simulation are those of its assumptions: normal, independent and identically distributed repeats, equal deviations in both arms, and a single axis. Real axes are bounded rates, repeats of an LLM run may be correlated through shared prompts and caches, and deviations differ between arms. The numbers therefore characterise the rules, not the stands, and they argue for a rule with two tiers and for planning five or more repeats whenever a claim of improvement is intended.

## J. Statistical Re-Analysis of Measured Counts

Some comparisons consist of counts without confidence intervals, and several experimental conditions have only one run. We therefore computed Wilson intervals for binomial proportions, tested available paired discordances when recorded, and otherwise report descriptive approximations without implying that an unpaired test validates a paired design. The migration count intervals and test choices are summarised in Table LXVIII. The re-analysis provides descriptive, not paired-inferential, comparisons.

TABLE LXVIII. WILSON 95% INTERVALS AND UNPAIRED FISHER TESTS FOR MIGRATION COUNTS

| Quantity | Count | Proportion | 95% interval | Comparison, Fisher p |
|---|---|---|---|---|
| demonstrably correct, before fixes | 98 / 216 | 45.4% | 38.9–52.0% | before vs after: 0.39 |
| demonstrably correct, after fixes | 108 / 216 | 50.0% | 43.4–56.6% | |
| accepted but wrong or unverifiable, before | 9 / 93 | 9.7% | 5.2–17.4% | before vs after: 0.023 |
| accepted but wrong or unverifiable, after | 25 / 114 | 21.9% | 15.3–30.4% | |
| exact structural matches, agent | 125 / 1000 | 12.5% | 10.6–14.7% | agent vs cheap: 0.0056 |
| exact structural matches, cheap model | 86 / 1000 | 8.6% | 7.0–10.5% | |
| accepted tasks, agent | 1330 / 1448 | 91.9% | 90.3–93.2% | agent vs cheap: 0.21 |
| accepted tasks, cheap model | 1310 / 1448 | 90.5% | 88.9–91.9% | |

Unpaired tests ignore the matched structure of the two runs and are descriptive approximations; conservativeness is not guaranteed. A McNemar test would require unit-level discordant pairs, which are not reported except for whole-file versus split migration (p = 0.208).

The operational count increased from 98 to 108 of the same 216 migration units; the unpaired Fisher calculation ($p = 0.39$) does not establish either significance or equivalence for these paired data. The share of accepted units subsequently marked wrong or unverifiable increased from 9.7% to 21.9% (unpaired $p = 0.023$), demonstrating a larger observed discrepancy between agent acceptance and external checking, but a paired significance claim requires the cross-classified unit records. Likewise, 125 versus 86 exact structural matches (unpaired $p = 0.0056$) should not be promoted to a paired improvement without the discordant counts. The data support a stronger measurement taxonomy; precise before–after inference awaits the original paired outcomes.

TABLE LXIX. HETEROGENEITY ACROSS FRAMEWORKS AND ACROSS MODELS (AUTOMATION, 49 RUNS PER ARM)

| Quantity | Across | $\chi^2$ (df) | p | Cramér's V |
|---|---|---|---|---|
| effect rate | 4 frameworks | 2.95 (3) | 0.40 | 0.12 |
| effect rate | 6 models | 15.87 (5) | 0.007 | 0.23 |
| unbacked claims | 4 frameworks | 7.92 (3) | 0.048 | 0.20 |
| unbacked claims | 6 models | 35.33 (5) | < 0.0001 | 0.35 |

Effect rate: share of runs with at least one effect; unbacked claims: runs without effect that claim success.

The automation re-analysis identifies heterogeneity within each of the two separate sweeps (Table LXIX). The effect-rate test across four frameworks yielded $p = 0.40$ and Cramér's $V = 0.12$, whereas the test across six model conditions yielded $p = 0.007$ and $V = 0.23$. Unbacked-claim rates also varied (frameworks $V = 0.20$, $p = 0.048$; models $V = 0.35$, $p < 0.0001$). These are within-sweep association statistics, not estimates of causal effects or a directly comparable effect-size decomposition in a crossed factorial design. Illustrative pairwise contrasts include CrewAI versus LlamaIndex effect rates of 57.1% versus 44.9%, Sonnet 4.6 versus gpt-4o of 73.5% versus 44.9%, and gpt-5.5 unbacked claims of 2/49 versus 14/49 for gpt-4o. The results are unreplicated and do not score whether the observed state changes are correct. Uncertainty in the chatbot-finalist ranking is shown in Table LXX.

TABLE LXX. UNCERTAINTY OF THE CHATBOT FINALIST RANKING (15 RUNS PER MODEL, NORMAL APPROXIMATION)

| Comparison or model | Welch t | p | P(highest mean) | P(Pareto) |
|---|---|---|---|---|
| Terra vs Haiku | 1.09 | 0.29 | | |
| Terra vs Gemini 3.1 Pro | 1.78 | 0.093 | | |
| Haiku vs Gemini 3.1 Pro | 1.14 | 0.27 | | |
| GPT-5.6 Terra | | | 0.83 | 0.86 |
| Claude Haiku 4.5 | | | 0.13 | 1.00 |
| Gemini 3.1 Pro | | | 0.03 | 0.03 |

Means and standard deviations from Table LI; 200,000 draws of the mean correctness from a normal distribution with the standard error; latency taken as reported, since its spread was not reported.

For the chatbot finalists, none of the pairwise differences in mean correctness reaches the 5% level (Terra against Haiku $p = 0.29$; Terra against Gemini $p = 0.093$). Under the stated independent normal approximation, Terra ranks first in approximately 83% of the simulated draws, compared with 13% for Haiku and 3% for Gemini. These are model-dependent ranking probabilities rather than calibrated probabilities of real-world superiority. The selection of Terra is thus a reasonable bet with a stated probability, not a statistically established winner, which is the form in which the testbed reports such decisions.

*K. Reference Implementation and Generated Artefacts*

To verify that the formal method is computationally executable, we implemented its core as a Python package, capmap, including typed measurement records, context-aware comparisons, Pareto selection, dashboard specifications and seeded simulations. The code reproduces the deterministic calculations and invariants used here. These verification checks establish implementation consistency, not additional cross-application quality observations.

*Generated metamodel.* A schema-validated model profile was constructed from six measured attributes across three application probes, with seven catalogue attributes explicitly represented as unmeasured (Listing 1). This distinguishes absent evidence from an observed value of zero.

**Listing 1. Excerpt of a generated sparse model profile (measured attributes and explicit missing-value markers; JSON abridged).**

```
{ "metamodels": { "claude": { "claude-haiku-4-5": {
  "attributes": {
    "chat.answer_correctness": [ {
      "value": 0.989, "n": 15,
      "context": "chat_role/dense_top4",
      "dataset": "mvp_scenarios@v6",
      "plan_hash": "7eaad59c7dfe7c28...c1c2",
      "evidence": "L3" } ],
    "aut.unbacked_claim_rate": [ {
      "value": 0.2245, "n": 49,
      "context": "llamaindex/baseline",
      "dataset": "pilot.v4", "evidence": "L1" } ],
    "aut.intent_accuracy": [ { "value": 0.971, "n": 40, ...
    } ],
    ... },
  "not_measured": [
    "chat.citation_validity", "chat.faithfulness",
    "chat.retrieval_mrr", "o2pg.conversion_accuracy",
    "o2pg.judged_quality", "o2pg.refusal_rate",
    "ops.cost_usd" ] } } } }
```

*Generated dashboards.* For the attribute declarations of the four probes (P1 conversion, the chatbot, automation and migration) the provisioner generated one dashboard each (Table LXXI).

TABLE LXXI. DASHBOARDS GENERATED FROM ATTRIBUTE DECLARATIONS

| Probe | Attributes | Panels | Panel types |
|---|---|---|---|
| P1 conversion | 13 | 15 | 10 gauge, 3 bar chart, 1 time series, 1 table |
| P3 chatbot (p2_chatbot) | 22 | 24 | 20 gauge, 1 stat, 1 time series, 1 bar chart, 1 table |
| P4 automation | 12 | 14 | 8 gauge, 2 stat, 2 bar chart, 1 state timeline, 1 table |
| P2 migration | 9 | 11 | 6 stat, 3 bar chart, 1 state timeline, 1 table |

Each dashboard also has three template variables. Text-typed attributes would be omitted.

Every dashboard carries three template variables (context, dataset and model), so that a panel can only show values of one context class; every attribute becomes a panel chosen by its unit (gauge for ratios, stat for counts, time series for latency, bar chart for cost and tokens, state timeline for verdicts), with "not measured" as the no-value text, and two panels common to all dashboards show the spend by role and the run health. The generated documents are valid JSON of the dashboard model; importing them into a running Grafana server and observing live data was not part of this work.

*Property checks.* The three lemmas of Section III-B and two properties of the noise rule were checked on 2,000 random instances each: the plan hash is invariant under permutation of runs; comparability is reflexive, symmetric and transitive; the Pareto set of a finite non-empty set is non-empty; the noise rule is symmetric under exchange of the arms (verdict equal, difference of opposite sign); and it is invariant under a common shift of both arms. All 10,000 checks passed. They do not replace a proof but would detect an implementation that violated the properties on which the mechanisms rely.

### L. Throughput, Latency and Cost of the Testbed Itself

The testbed itself can be evaluated for execution volume, latency and measurement cost (Table LXXII). The P1 evaluation converted 1,802 files with concurrency eight in 48 minutes and scored the resulting outputs in 14 min 38 s. These quantities characterise the tested hardware, gateway and dataset, rather than the throughput of an arbitrary deployment.

TABLE LXXII. MEASURED THROUGHPUT, LATENCY AND VOLUME

| Activity | Figure |
|---|---|
| P1 conversion, 1,802 files, concurrency 8 | 48 min (≈ 38 files/min, derived) |
| P1 evaluation with judge | 14 min 38 s |
| P1 evaluation without judge (from database) | ≈ 7 s |
| Chatbot response time, finalists | 9.8 s / 11.6 s / 14.0 s |
| Intent classification, one repeat of 40 cases | 72 calls; ≈ 37.4 k tokens; 107 s vs 17.1 s control |
| Decomposition LLM fallback, 165 units | ≈ 37 min, sequential |
| Automation run, mean tokens | 1,953 – 9,465 depending on model |

Values measured in the evaluation; throughput ratios are arithmetically derived.

### M. Cross-Application Synthesis: the Evidence Map and the Capability Map

Fig. 32 shows which models have been measured on which application. The matrix is sparse by construction: P1 was measured with one model, automation with six models in one framework, the chatbot with nine models in a screening and three in repeated runs, and the migration agent with the models of its two evidence levels. Sparsity is the reason for the principle that missing is not zero: any ranking that filled the empty cells would be a statement about the filler.

Table LXXIII collects the observations that can be stated per model with the evidence at hand. Each cell is an observation from one application under one configuration; it is not a general property of the model. The comparable statement across applications is structural rather than numerical: the three directions agree that a system's own report of success overstates it. The automation agents claim work they did not do; of the translations the migration agent accepts, 10 to 22% are wrong or cannot be verified against Oracle; and in the chatbot an accepted configuration change is not the same as a passing scenario.

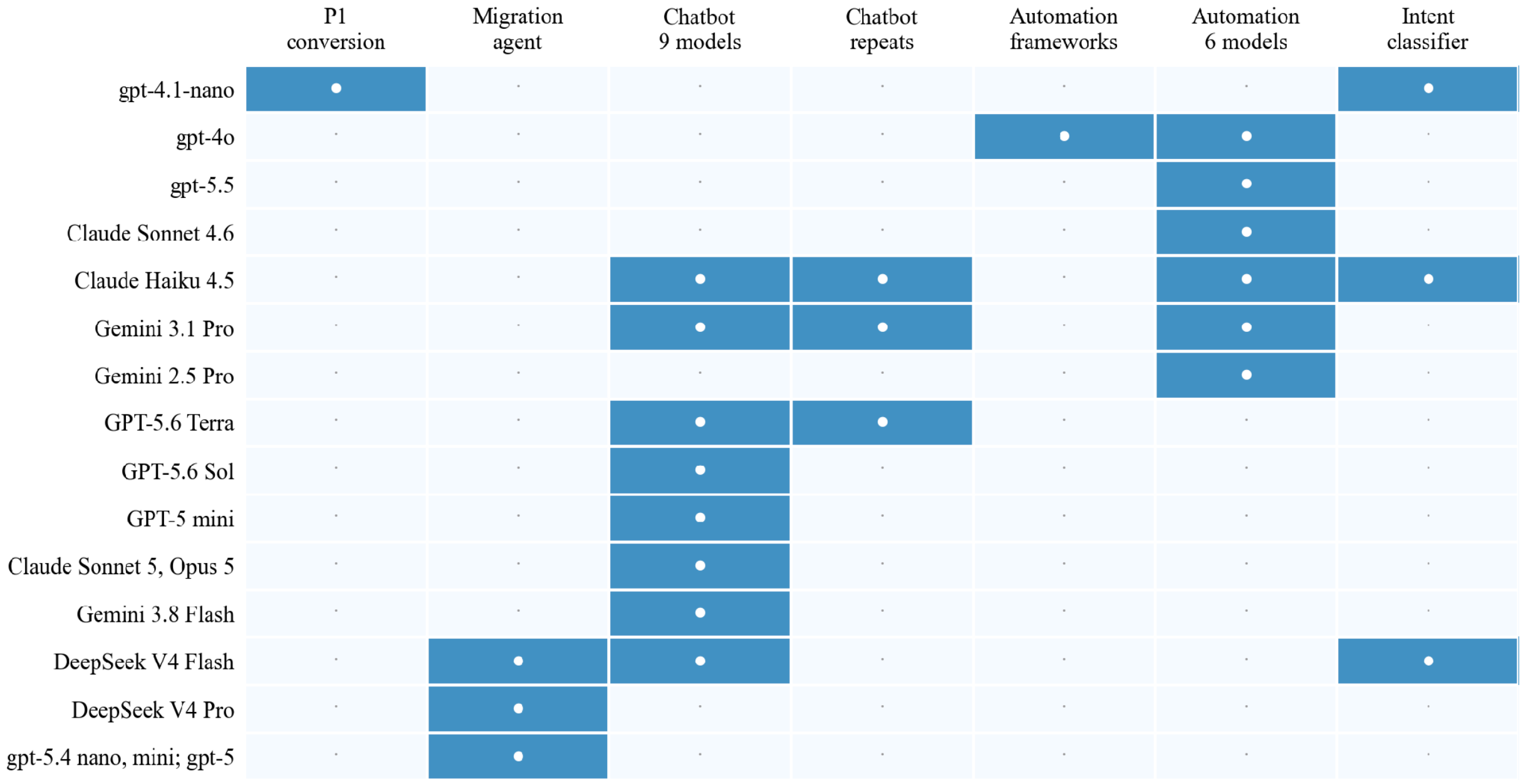

Fig. 32. Evidence coverage of tested model conditions across the reference applications; a missing mark denotes that no measurement exists for the indicated combination.

TABLE LXXIII. CAPABILITY OBSERVATIONS PER MODEL (SINGLE APPLICATION, SINGLE CONFIGURATION)

| Model | Application | Observation | Evidence |
|---|---|---|---|
| gpt-4.1-nano | P1 conversion | abbreviates about 69 large files instead of converting them | full cycle, 1 run |
| gpt-4.1-nano | intent classification | 89.7%; description rewrite +4.6 pp | 5 repeats |
| DeepSeek-V4-Flash | intent classification | 94.3%, zero variance; rewrite −3.4 pp | 5 repeats |
| Claude Haiku 4.5 | intent classification | 97.1%, zero variance; rewrite 0.0 pp | 5 repeats |
| Claude Haiku 4.5 | chatbot | fastest finalist (9.8 s), correctness 0.989 | 15 runs |
| GPT-5.6 Terra | chatbot | highest correctness 0.995, lowest variance | 15 runs |
| Gemini 3.1 Pro | chatbot | faithfulness 1.000 (highest of the finalists); slowest finalist | 5 scenarios |
| Claude Sonnet 4.6 | automation | most thorough (1.8 effects) and costly (9.5 k tokens); unrequested actions; no unbacked claims | 49 samples, 1 run |
| gpt-5.5 | automation | minimal action, but mostly says so (2 unbacked claims) | 49 samples, 1 run |
| Gemini 2.5 Pro | automation | multi-step tasks done; sent no email on email_compose | 49 samples, 1 run |
| gpt-4o, Haiku 4.5, Gemini 3.1 Pro | automation | often report work not done (14, 11, 9 of 49) | 49 samples, 1 run |

Observations are conditional on the specified application, configuration and evidence level; they are not global model rankings.

The two aggregate patterns are worth stating separately. First, within the evaluated automation conditions, observed effects and unbacked success claims varied more across model choices than across framework choices; the limited design does not establish a general causal ordering. A tuning result also depended on the deployment (intent classification). Second, the quality measured by a stage-wise instrument differs from the quality claimed by the system: the metamodel keeps both, as the attributes ACCEPTED and demonstrably correct, effects and unbacked claims, execution success and answer correctness.

For a descriptive test of multi-objective selection, Algorithm 7 was applied to the measured automation aggregates in Table XLVII using mean token count and unbacked success claims as two objectives to minimise. The resulting exact Pareto set contains gpt-5.5 and Gemini 2.5 Pro; all other compared conditions are dominated on those two recorded dimensions (Table LXXIV). This computation must not be interpreted as a deployment recommendation because it omits task correctness and relies on one run per condition.

TABLE LXXIV. Two-Objective Pareto Analysis of Automation Measurements (Token Use and Unsupported Claims)

| Model | Tokens | Unbacked | Status |
|---|---|---|---|
| gpt-5.5 | 1,953 | 2 | Pareto |
| Gemini 2.5 Pro | 2,516 | 0 | Pareto |
| gpt-4o | 2,055 | 14 | dominated by gpt-5.5 |
| Gemini 3.1 Pro | 3,147 | 9 | dominated by gpt-5.5 |
| Haiku 4.5 | 5,384 | 11 | dominated by gpt-5.5 |
| Sonnet 4.6 | 9,465 | 0 | dominated by Gemini 2.5 Pro |

Single-run aggregates from Table XLVII; no spreads available; effects and correctness not used.

# IX. Discussion

## A. Answers to the Research Questions

**RQ1 (architecture).** The testbed treats the application as the unit of extension. An application is a container behind a small contract (health, options, evaluation, attribute declaration, provenance), configured by template variables and, when a variable names an unsupported dataset or instance, by a specification-driven generation hook whose output must pass a smoke test. Onboarding registers a probe, adds owned attributes to the metric catalogue and registers dashboards. The chatbot application was integrated by following this pattern without a chatbot-specific frontend: it appeared in the shared Use Cases, Metrics and Evaluations pages with its own chat.* attributes, and the core was not edited. Reproducibility and cost transparency come from frozen, hashed plans, prompt identity and three separate spends. What we have not shown is the effort required to onboard a fourth application of a very different shape; the placeholder probe of opposite shape that ships with the core keeps the contract honest structurally, but only a second real probe will show it in practice. Table LXXV summarises the answers.

**RQ2 (trust and diagnosis).** The evaluation design separates model, judge and orchestration roles; deterministic metric code prevents a model from supplying its own score. Frozen plans, dataset identities, instrument checks, negative controls and explicit evidence levels prevent several observed measurement errors. The remaining variability and limitations of the source-system oracle are discussed in Section IX-J.

TABLE LXXV. Summary of the Answers to the Research Questions

| Question | Answer | Evidence |
|---|---|---|
| RQ1 architecture | application = container + contract + owned attributes; onboarding registers probe, metrics, dashboards | chatbot integrated without core edit; Fig. 8 |
| RQ2 trust | instrument ≠ subject; code computes; missing ≠ zero; outage ≠ verdict; null-agent check; noise floor | defect tables; null agent beat the agent |
| RQ3 behaviour | separate model/framework sweeps show different heterogeneity; prompt effects are deployment-specific; acceptance overstates correctness | E2–E9 |
| RQ4 decision | Pareto selection with evidence counts and refusal to mix contexts; no ranking claimed | Algorithm 7; Tables LVI, LXXVIII |

**RQ3 (behaviour).** Three preliminary observations emerge. First, model and framework sweeps in the automation stand show different magnitudes of variation in observed side effects and unsupported success claims, without identifying a causal ordering between these factors. Second, rewriting an instruction produced gains, losses or no change across deployments, so prompt effects must be assessed in the specific deployment. Third, reports of completion often exceeded externally observed actions, while migration results also depended on what the measurement stand could test. The chi-square results (model $p = 0.007$; framework $p = 0.40$ for effect rate) describe heterogeneity within separate, one-run sweeps; they do not test a model-by-framework interaction or establish semantic task correctness.

**RQ4 (decision).** The measurement store and deterministic analytical API support task-constrained model filtering, evidence-aware comparisons, Pareto frontiers and the explicit exclusion of models with missing or incompatible measurements. The empirical Pareto calculation in Section VIII-M illustrates these mathematical operations; it is not a universal model ranking or a validated deployment choice.

## B. Implications for Model Selection

Four practical rules follow for teams that select models for applications like ours. (i) Measure the model in the application, in the configuration that will be used; the retrieval settings of the chatbot were chosen by controlled experiments, one variable at a time, and the model was chosen last. (ii) Change the model before changing the words: an instruction tuned on one deployment may do nothing, or harm, on another. (iii) Distrust the system's own success report; compare it with independent evidence such as effects, Oracle's output or the frozen world state. (iv) Treat cost as an objective rather than a footnote, including the cost of measuring: the judge's spend is substantial, a cheaper model can be unusable for lack of quota, and cost reports themselves can be wrong. These rules agree with the observation of HELM that accuracy alone is an incomplete report [3] and with the cost-aware cascades and routers of the model-selection literature [59], [60], but they are drawn from measured defects of our own stands.

## C. Extending the Metamodel: Planned Probes

The scheme names four further attribute groups, P5 to P8, which are not implemented. Table LXXVI sketches what each could measure if onboarded as an application, with the earlier work that would feed it. We include the table to show how the onboarding mechanism would be used; it is a plan and not a result.

TABLE LXXVI. PLANNED PROBES AND CANDIDATE APPLICATIONS

| Probe | Attribute group | Candidate application | Feeds from |
|---|---|---|---|
| P5 | security treatment | security-aware generation and guardrail evaluation | [67], [74] |
| P6 | token efficiency | token-optimisation harness for migration | [63] |
| P7 | required resources | latency, memory and context-window probe | operational set |
| P8 | DevOps | specification-driven DevOps tasks | [45], [78] |
| — | model drift | repeat a frozen plan over time and apply drift detection | [71], [72] |

Two pieces of prior work are natural seeds for P5. A study of security knowledge in specification-driven development found that functional behaviour can be specified explicitly while security behaviour stays implicit, so that generated systems satisfy the visible requirements yet miss ownership checks, administrator safety rules or token-misuse rejection; generic guidance such as ASVS improved generated behaviour, and explicit application-specific security modelling improved it further, most visibly in business-logic and administrator-safety categories [74]. And a guardrail harness around specification-driven workflows returns a structured decision (pass, warning or block) with traceable findings, and ships a deterministic evaluation suite that exits non-zero on unexpected decisions or missing family coverage; its precision and recall would be natural P5 attributes, alongside the offline guardrail measurement reported in Section VIII-C.

### D. Towards Steps 2 and 3: Selecting a Model for a Task

Steps 2 and 3 of the programme, which are out of scope here, are the intended long-term consumers of the metamodel; stating their interface now clarifies what the metamodel must provide. A task is a structured object $T$ = {agents, operations, environments, modules, context, …}. Step 2 turns requirements into a specification $Spec(T, \Psi(m))$ that takes the chosen model's profile into account, for example by adding the safeguards that the profile shows the model needs, and Step 3 splits the specification into parts $T_1$, …, $T_k$ and chooses for each part a model and a configuration. Given a mapping from a task part to the attributes that matter for it, the choice is a constrained multi-objective problem:

$$(m_j, \kappa_j) \in P(O_j) \text{ over } \{(m, \kappa): [\Psi(m)(\sigma) \geq \theta_\sigma \text{ if } \delta_\sigma = \max; \; \Psi(m)(\sigma) \leq \theta_\sigma \text{ if } \delta_\sigma = \min] \text{ for every } \sigma \in \Sigma_j\}. \quad (11)$$

where $\Sigma_j$ is the set of attributes that part $T_j$ depends on, $\theta_\sigma$ are hard thresholds and $\delta_\sigma$ specifies whether each attribute is maximised or minimised, and the Pareto set P is taken over the objectives $O_j$ of Equation (5) in comparable contexts only. The equation makes the dependencies of the later steps explicit. It needs a semantic bridge from a task description to the attributes that matter, which is item 6 on the programme's list of questions; it needs thresholds that are agreed rather than placeholders; and it needs the attributes to be measured in a context that resembles the deployment, which is why the application is part of the measurement. The partiality of the metamodel is a feature here: a model for which a required attribute is absent is excluded with a stated reason, exactly as in the exclusion rule in Algorithm 7.

### E. Relation to Earlier Work of the Group

The testbed continues a line of work on specification-driven development and monitoring. The comparative analysis of frameworks for the same three application classes [78] provided the architectures; the unified architecture metamodel [77] supplied the idea that a model or system can be described by a versioned, structured record; the benchmark on what a specification determines [46] and the leakage-resistant evaluation of specification judges [45] supplied the discipline of execution-judged evaluation and of sealed ground truth; the token-optimisation study [63] showed the multi-objective nature of cost. The monitoring publications of co-author V. Lyashkevych [69]–[73] investigate ontology-driven dashboards and drift in evolving information systems, and a natural extension of the testbed is to apply their drift detection to the metamodel itself, so that a change in a hosted model, a gateway or a dataset version is detected as drift rather than discovered by a surprised user. The security and risk perspective [74], [75] is the basis for the planned security attributes (P5), and the orchestration of agents with dynamic involvement of new agents [76] relates to the planner of the core.

### F. Generalisability and Alternative Designs

The probe-per-application design has competitors. A single monolithic harness that runs all applications in one process would have been simpler to start and would have given one language, one dependency set and one test runner. We rejected it because the applications are built by different teams with different stacks (the migration agent depends on a parser generator and a PostgreSQL container, the chatbot on a document index and a graph runtime, the automation agent on four agent frameworks), and because the frameworks themselves conflict: two frameworks that both want to run the steps cannot be assumed to nest. The price of isolation is a network boundary and a contract to maintain. A second alternative is to adopt an existing evaluation library as the definition of a score. We use such libraries as implementations of individual axes, but not as the answer to what a score is, because they supply implementations and not a hierarchy of sources, a significance threshold or a control arm. A third alternative, a single leaderboard, is simply a different product.

Three conditions limit generalisation. The design assumes that an application can expose an evaluation API and declare its attributes; an application whose evaluation requires a human in the loop at every step would need a different probe type. It assumes a single gateway for model access, which is convenient for cost attribution but couples the testbed to the gateway's deployment catalogue, including its quotas and the identifiers it returns. And it assumes that the programme's model of an application, a dataset plus a configuration space plus an evaluator, is adequate; applications whose behaviour depends on long-lived state that cannot be reset between runs would violate the fresh-state assumption on which the repeat logic rests.

### G. Lessons Learned

Table LXXVII lists the lessons that the authors consider transferable to teams that build similar infrastructure. Most of them are about instruments, not models.

TABLE LXXVII. TRANSFERABLE LESSONS

| Lesson | Origin in this work |
|---|---|
| Score the agent that does nothing before trusting a metric. | an empty migration outscored the agent on every published headline metric |
| A verdict is only as strong as the weakest level at which it was checked. | about 78% of 'ok' verdicts proved nothing; 108 of 216 demonstrable |
| Read the failures before fixing the model. | FAILED contained 14 faithful raises and 8 invalid inputs |
| Expect a correct fix to expose the next defect, and to move the headline the wrong way. | ACCEPTED fell after a correctness fix; a FIX_PARTIAL is usually a correct fix |
| Change the model before changing the words. | a rewrite worth +4.6, −3.4 and 0.0 points on three deployments |
| Separate the price of measuring from the property measured. | judge spend exceeded conversion spend in P1 |
| An outage must not look like a bad agent. | 109 FAILED units from a network outage |
| Pin the ruler. | two linter versions in one stand; metric libraries frozen |
| Put the denominator in the record. | a blank denominator passed a presence check |
| Where two models disagree, read the unit. | first silently wrong accepted translation found by disagreement |

Each rule is supported by the measured incidents and verification results described in Sections VII and VIII.

## H. *Practical and Ethical Considerations*

The application datasets are private or synthetic, so no personal data enter the store; the automation dataset in particular is a closed simulation with synthetic content, which avoids side effects on real tickets and emails. The testbed spends money on every run, and in the experiments the spend was shared by several users of one key: the migration team recorded that the shared key stood at about USD 331 of a USD 500 rolling budget at one point. The design decision that the caller's key pays for the call, and that no service stores it, is therefore also an ethical decision about who bears the cost of measuring. Human approval gates remain in the loop where a change would reach a live system (publication of a knowledge snapshot) or where spending is unattended (pre-registration sign-off and permission to spend are two explicit gates in the fix cycle). Finally, because the metamodel can be used to decide which model handles which task, it can be misused to over-trust a profile; the partial metamodel, evidence counts and context warnings are the countermeasures built into the interface, and we recommend that downstream consumers keep them.

## I. *Threats to Validity and Limitations*

The testbed implementation, automated verification and empirical measurements in Sections V–VIII provide the evidence base for the claims in this work. The limitations concern generalisability, statistical sensitivity, unmeasured dimensions, and transfer to new applications rather than the existence of the implemented artifacts. The complete set of threats is tabulated in Table LXXVIII.

Two further limitations concern the design. The caller's key crosses a process boundary in the task message and is protected operationally, not cryptographically. And ground-truth leakage is controlled mechanically only in the specification-evaluation work [45]; for the chatbot and the automation agent it is controlled by construction (expected answers and tools are used only by the evaluation engine), which is weaker. Finally, public benchmark contamination [49], [50] is not a concern for the application datasets, which are private or synthetic, but it would become one if the testbed were used on public data.

## J. *Validated Uses, Boundaries of Inference, and Transfer Conditions*

**Interpretation principle.** The capability profile is evidence about a model operating inside an explicitly configured application and measurement procedure, not a context-free estimate of model intelligence. The testbed can record which application, dataset version, prompt variant, evaluator and deployment produced a value; attach evidence and sample counts; keep unmeasured attributes absent; and expose comparisons under controlled conditions. These functions are implemented and exercised in the reported stands. Their correctness as software mechanisms must not be conflated with the much stronger proposition that any resulting score predicts performance on an unseen production workload. The former is established by the implemented pipeline and its tests; the latter is a transfer claim that needs additional data.

TABLE LXXVIII. THREATS TO VALIDITY AND LIMITATIONS

| Threat | Where | Status / mitigation |
|---|---|---|
| single runs, small samples | E1–E3, E5, E7 | reported as single run; noise rule applied; repeats planned |
| P1 noise floor unmeasured | P1 conversion | model numbers withheld; repeated identical plans planned |
| automation not scored against ground truth | E2, E3 | claims and effects reported as such; scoring pending |
| applicability thresholds are placeholders | automation Stage 0 | screening not executable until agreed |
| synthetic, partly unreachable dataset | automation pilot.v4 | 15 samples need an unreadable address; known |
| self-authored labels; one set | intent classification | upper bound; control arm is a floor |
| judge instability and self-preference | chatbot, P1 | 16.2% churn at T = 0; judge separated; cross-vendor judge open |
| migration headline predates last fixes | E8 | estimate 115–120 awaits next full run |
| token counts not comparable across vendors | E3 | stated; own tokenizers |
| shared quota and key | E7, E8 | full runs depend on a daily cap |
| Grafana registration not evaluated | Section V-H | mechanism described; no quantitative result |
| generation hook not evaluated | Section V-F | contract and smoke test only |
| models of different periods and gateways | all | context recorded; no cross-experiment ranking |
| simulation assumes normal i.i.d. repeats, one axis | Section VIII-I | characterises the rules, not the stands; repeats may be correlated |
| unpaired tests on paired runs; no multiplicity correction | Section VIII-J | descriptive; paired inference unavailable without unit-level records |
| generated dashboards not imported into a Grafana server | Section VIII-K | validated as JSON only; live evaluation pending |

The applicability restrictions are derived from the actual measurement designs and are not claims about untested models or applications.

**What a model-selection decision can support.** For an application represented by a reference probe, an engineer can filter models by measured hard constraints, examine separate quality, effect, latency and cost attributes, and construct a Pareto set over objectives evaluated in a common context (Sections III-A, V-H and VIII-M). The interface also reveals missing attributes, context mismatches and evidence counts; it can decline a comparison instead of silently imputing missing scores. This is a defensible use of the profile as a decision aid or shortlist. The system does not establish that the leading model is universally best, that an unmeasured constraint has been satisfied, or that a Pareto-optimal candidate will remain optimal after changes in the workload, provider or evaluation method. A selection must be revisited whenever the target context differs materially from the recorded context.

**Reproducibility versus repeatability.** Frozen plans, hashes, dataset identifiers and registered prompts make the intended experiment and its provenance recoverable. Replaying a frozen plan avoids asking an LLM planner to reproduce its earlier choices; it does not force a hosted model, judge, external API or concurrent runtime to return byte-identical outputs. Thus plan-level reproducibility is supported, whereas numerical repeatability is an empirical property to be measured separately for each attribute, model and deployment. The reported Monte Carlo experiments characterise the operating behaviour of the range-based and three-state decision rules under their stated assumptions; they are not a measurement of run-to-run noise in every hosted application. With single runs, observed contrasts are valid descriptive results but cannot carry a claimed confidence interval for a deployment-level mean.

**Migration capability and semantic equivalence.** The Oracle-to-PostgreSQL direction can decompose and order migration units, route them to conversion strategies, run bounded repairs and verify PostgreSQL execution. Its stronger evidence comes from differential checks against a real Oracle system. The reported 108 demonstrably correct units out of 216 refer to the exercised corpus and the behaviours tested there. They do not imply equivalence on every input, independence from schema state, complete support for the Oracle PL/SQL language or correctness of a complete enterprise migration. In particular, installation alone is weaker than behavioural equivalence; even L4 observations cover only the tested cases and source conditions. A new database estate therefore needs dependency coverage, fixture representativeness, representative input scenarios and re-execution against its own Oracle reference before a production-quality claim is made.

**Chatbot capability and the limits of semantic judging.** The chatbot probe can isolate execution, tool choice, retrieval, citation structure, grounding and judged answer quality; controlled retrieval experiments and the integrated smoke run demonstrate that these attributes can be produced and persisted. This does not make retrieval recall a property of the language model when the retrieval configuration is varied, nor does a well-formed citation prove that the answer is factually supported. Some expected labels in the scenario dataset remain provisional, and an LLM judge can disagree with itself or favour a related model family. The current semantic scores are therefore advisory unless independently validated against the intended task criteria. A high-stakes application needs a curated held-out set, a documented adjudication protocol, negative and refusal cases, and checks of both retrieved evidence and the final answer.

**Automation capability: effects, claims and correctness.** The enterprise-support stand uses production-shaped synthetic events and a frozen state to observe tool requests, accepted arguments, actual side effects, missing operations and unsupported success claims. This permits a grounded comparison of observable action behaviour in the evaluated conditions. However, an action occurring is not the same as the intended business outcome being correct or authorised: an agent can change state and still choose the wrong issue, page, recipient or sequence. The pilot set contains 49 samples and known cases whose required address cannot be retrieved by any available tool. Accordingly, the model-versus-framework differences reported for side effects and unbacked claims are descriptive within separate sweeps; they are not a factorial causal estimate and do not establish end-to-end task accuracy. That stronger claim requires scored final states, matched model-by-framework cells and repeats over independent scenarios.

**Evidence levels and instrument dependence.** The evidence ladder is an explicit safeguard against treating all PASS labels alike. L0 self-reports, static checks, installation, reference comparisons and differential execution answer different questions. The testbed can report the strongest available verification and its denominator, distinguish a failure of the evaluation infrastructure from a failure of the tested model, and apply the null-agent check before promoting an aggregate to a headline metric. Nonetheless, an evaluator can overlook a defect when its tests are incomplete, and a metric can reward an unintended shortcut even when the evaluation code executes correctly. The operational conclusion is not that the recorded measurements are unreliable, but that every downstream claim should remain bounded by the specific property, test coverage and evidence level actually observed.

**Statistical sensitivity and cross-context comparison.** The range-based noise guard favours avoiding unsupported effect claims, but its sensitivity depends on the number and spread of repeats; the simulation indicates that simply increasing the number of repeats need not improve its power. The Welch-based second tier provides a suggestive signal under its modelling assumptions and does not replace a prespecified paired analysis where runs share cases. The existing unpaired re-analysis cannot recover unit-level transition counts absent from the reported aggregates, and several comparisons are unadjusted for multiplicity. Likewise, token counts from different providers are not automatically a common unit of computational effort. Cross-model comparisons therefore require an agreed metric definition, fixed dataset and evaluator versions, recorded pricing assumptions, and uncertainty assessment on the same experimental unit.

**Software integration and operational boundary.** MAS Core, the deterministic gate, the P1 execution path, the shared analytical interface and the chatbot integration have implementation and test evidence in this work. The migration and automation systems also have running evaluation stands, but have not yet been onboarded end to end as ordinary probes in the common core. The dashboard provisioner generates valid dashboard definitions, while automatic import into a live Grafana instance and its onboarding cost are not assessed. The specification-driven adapter-generation hook is described and subject to a smoke-test contract; the general quality of generated adapters has not been evaluated. These distinctions delimit the presently demonstrated level of application extensibility: the architecture and one integrated application show the mechanism, but integration effort and maintenance cost across a broader class of applications remain open empirical questions.

**Security, privacy and deployment constraints.** Per-caller model credentials, role-separated accounting, isolated evaluation state and explicit approval gates reduce several operational risks, but they do not constitute a comprehensive production security assurance. The caller key still crosses a worker/broker boundary without end-to-end cryptographic protection in the described design. No general guarantee is provided against prompt injection in arbitrary input documents, disclosure through a provider, unauthorised live side effects, or safety failures on untested tools; the security probe P5 remains planned. Availability and economics are also environment-dependent: shared quotas, deployment catalogues, gateway behaviour, pricing changes and service outages can limit scale or invalidate cost comparisons. Production deployment consequently requires its own secrets handling, access control, threat assessment, rate-limit testing and observation of live drift.

**Conditions for transfer and future evaluation.** The method is best transferred to a new application when (i) a representative, versioned task set and a defensible target outcome can be defined; (ii) the application can expose a resettable or otherwise controlled evaluation boundary; (iii) the probe declares metric ownership, units, directions and evidence sources; (iv) tested models and judges are assigned distinct roles; (v) configuration, software versions and prompt identities are retained; and (vi) the evaluator is challenged with null agents, negative cases and repeated controls. When one of these conditions is absent, the architecture can still collect diagnostic traces, but an evidence-based selection claim should be withheld or explicitly qualified. The next validation step is therefore not to invent a broader universal leaderboard: it is to connect additional real probes, measure onboarding effort and live dashboard operation, score automation ground truth, quantify within-context repeatability and test whether recommendations hold on new datasets and model deployments.

### *K. Engineering Status*

Table LXXIX lists the implementation status of the components at the time of writing. We include it to avoid ambiguity between what has been built and what is designed.

TABLE LXXIX. IMPLEMENTATION STATUS OF THE COMPONENTS

| Component | Status |
|---|---|
| MAS Core: planning, gate, freezing, workers, aggregation | implemented; verified end to end on P1 with gpt-4.1-nano |
| P1 conversion probe | measures for real; noise floor pending |
| Chatbot probe p2_chatbot | integrated; smoke run and interactive launch verified |
| Migration and automation probes | own stands; integration as probes planned |
| Analytical UI (Explore, Analytics, Launch, Agent) | implemented; analytics deterministic |
| Grafana connection to the metric store | connected; dashboard ownership separate |
| Automatic dashboard registration at onboarding | provisioner implemented and run on four probes; generated JSON not imported into a live Grafana |
| Reference implementation of the computational core (capmap) | implemented; property checks and simulations reproducible |
| Specification-driven generation hook | designed; not evaluated |
| Security, tokens, resources, DevOps probes (P5–P8) | planned |
| Steps 2 and 3 of the programme | planned |

Measured implementation coverage of the application probes; planned functions are labelled separately.

## X. CONCLUSION AND FUTURE WORK

We presented a method for application-specific capability profiling of LLMs and the containerised testbed that realises it. The method profiles LLMs through reference applications and accumulates the measurements into a versioned, sparse metamodel of each model. The application is the unit of extension: onboarding registers a probe and owned attributes in the metric catalogue and prepares dashboard definitions, so that the set of measurable LLM features grows with the set of applications. The testbed enforces a small set of principles (instrument is not subject, agents decide and code computes, reproducibility by frozen plans, missing is not zero, outage is not a verdict) and two checks, the null-agent check and the noise-floor rule, each justified by a defect in the authors' own stands. It diagnoses the instrument before the system. A simulation showed that the noise rule is a conservative screen whose sensitivity falls as repeats are added, which led to a three-state verdict with a calibrated second tier.

The three reference applications—a migration agent, a chatbot and an automation agent—showed context-dependent model behaviour, deployment-specific effects of instruction tuning, and gaps between reported success and externally observed outcomes. In the automation conditions tested, variation across models exceeded variation across frameworks on selected measured axes; the current experimental design does not justify generalising this ordering. The first result of the migration agent measured against real Oracle was 108 of 216 demonstrably correct units, with a ceiling limited by what the stand can verify. These findings are offered as the first measured structure of the metamodel, not as a ranking of models. The practical scope and non-guarantees of these measured capabilities, together with the conditions required to transfer them to new applications, are set out in Section IX-K.

Future work follows the open items. The immediate item is to characterise the run-to-run variation of P1 with repeats of one frozen plan, since everything downstream of "which prompt is better" waits on it. Next, the migration and automation stands will be connected as probes, which will prove the probe contract on a second and third real application and allow the evidence map of Fig. 32 to be filled by planned runs instead of by history. The automation results need ground-truth scoring and agreed thresholds for the screening stage. The quality of generated adapters and the effort of onboarding, as well as the dashboard registration, should be evaluated quantitatively, for example by the time and defects of onboarding a fourth application. Finally, with stable profiles, Steps 2 and 3 of the programme can be undertaken: generating specifications tailored to a chosen model, decomposing them and deciding which model and configuration handles each part.

## APPENDIX: ATTRIBUTE CATALOGUE OF THE CHATBOT PROBE

Table LXXX lists the attributes owned by the probe p2_chatbot as shown on the Metrics page of the interface, with their group and direction. The operational attributes with the prefix ops. are common to all probes. Directions follow the interface: an upward arrow means that higher is better.

TABLE LXXX. Attributes of the Chatbot Probe

| Attribute | Group | Better |
|---|---|---|
| chat.agent_iterations | performance | lower |
| chat.answer_correctness | quality (judge) | higher |
| chat.answer_relevance | quality (judge) | higher |
| chat.citation_completeness | grounding | higher |
| chat.citation_compliance | grounding | higher |
| chat.citation_source_match | grounding | higher |
| chat.citation_validity | grounding | higher |
| chat.faithfulness | grounding (judge) | higher |
| chat.forbidden_tool_free_rate | reliability | higher |
| chat.helpfulness | quality | higher |
| chat.metadata_complete | retrieval | higher |
| chat.refusal_appropriateness | quality | higher |
| chat.retrieval_empty_correct | retrieval | higher |
| chat.retrieval_mrr | retrieval | higher |
| chat.retrieval_precision_at_k, r_precision, recall_at_k | retrieval | higher |
| chat.task_pass_rate | quality | higher |
| chat.tool_argument_correctness, tool_selection_correctness | quality | higher |
| chat.tool_execution_success | reliability | higher |
| chat.turn_latency_ms | performance | lower |
| ops.tokens_per_case | cost | lower |
| ops.completion_rate, ops.format_compliance | reliability | higher |

Attribute names, categories, and optimization directions are defined by the chatbot probe's registered metric catalogue. Operational attributes are shared across application probes.

The metrics that the evaluation phase of the P1 stand collects and stores, and that serve as the data source for the dashboards, are the NLP aggregate and per-metric values (recall, BLEU, chrF), the four judge criteria, the linter error rate and warnings, the file efficiency and refusal rate of the annotators, and the run totals of tokens and cost. An inventory of the stored values was compiled from the code and feeds the discussion of the final metric set, which includes the observation that two linter-derived entries of the selectable NLP list duplicate the validation stage.

## Data and Code Availability

Data and code availability—The results are derived from implemented software, recorded application datasets and executed evaluations. Code artefacts, versioned configurations, test datasets and run identifiers are retained by the authors. Public access, licensing and the availability of restricted enterprise or Oracle datasets are subject to the distribution arrangements specified by the submitting authors; independent replication outside those arrangements is not asserted.

## Declaration on Generative AI

During the preparation of this work, the authors used Generative AI tools for language editing and editorial refinement. The authors reviewed and verified the resulting text and take full responsibility for the content of the publication.